\documentclass{article}
\usepackage{arxiv}
\usepackage[utf8]{inputenc} 
\usepackage[T1]{fontenc}    
\usepackage{hyperref}       
\usepackage{url}            
\usepackage{booktabs}       
\usepackage{amsfonts}       
\usepackage{amssymb}        
\usepackage{nicefrac}       
\usepackage{microtype}      
\usepackage{graphicx}
\usepackage{amsmath}
\usepackage{subcaption}
\usepackage{longtable,tabularx}
\usepackage[switch]{lineno}
\usepackage{placeins}      

\usepackage[table]{xcolor}
\renewcommand{\arraystretch}{1.5} 
\title{Sharp-interface VOF method for phase-change simulations on unstructured meshes}

\author{
	Jan Kren \\
	Laboratory for Simulation and Modelling\\
	Paul Scherrer Institute\\
	Villigen 5234, Switzerland \\
	\texttt{jan.kren@psi.ch} \\
	\And
	Bojan Ničeno \\
	Laboratory for Simulation and Modelling\\
	Paul Scherrer Institute\\
	Villigen 5234, Switzerland \\
	\texttt{bojan.niceno@psi.ch} \\
	\And
	Yohei Sato \\
	Laboratory for Simulation and Modelling\\
	Paul Scherrer Institute\\
	Villigen 5234, Switzerland \\
	\texttt{yohei.sato@psi.ch} \\
}

\begin{document}
	\maketitle
	\begin{abstract}
		Unstructured meshes represent one of the most versatile and straightforward approaches for capturing non-canonical geometries in fluid dynamics simulations. Despite efforts in the CFD community, most high-fidelity first-principles phase-change models are proposed and applied to structured meshes. In this paper, we present a method for phase-change simulations that works on unstructured meshes. We combine the algebraic Volume-of-Fluid (VOF) technique with geometric interface reconstruction to capture phase-change phenomena accurately, and implement it in an in-house open-source CFD code. Phase-change rates are computed from local temperature gradients evaluated at the geometrically reconstructed interface, without empirical closure models, using an interface reconstruction procedure that operates on arbitrary polyhedral cells. Since the method relies on standard finite-volume framework, it can be integrated into other cell-centred finite-volume codes that support unstructured meshes. Our simulation method is validated against the one-dimensional Stefan and Sucking problems and the three-dimensional Scriven bubble growth on both structured hexahedral and polyhedral meshes. The results show good agreement with analytical solutions for all three cases. A detailed analysis of the Scriven problem reveals that the interface-modified least-squares gradient stencil on Cartesian meshes overestimates the interfacial temperature gradient, which produces a persistent overshoot of the analytical bubble radius, and introduces a coherent four-fold anisotropy that elongates the bubble along the grid diagonals. On polyhedral meshes, the irregular face orientations eliminate both effects, that yields isotropic growth and monotonic convergence. We demonstrate the capabilities of the framework with turbulent upward co-current annular boiling flow, where the early transient results are qualitatively consistent with a previous LES study and experimental observations of wave-modulated evaporation.
	\end{abstract}
	
	
	\clearpage
	\section*{Nomenclature}
	
	{\renewcommand\arraystretch{1.0}
		\noindent\begin{longtable}{@{}lll@{}}
			\caption{Nomenclature}
			\label{tab:nomenclature}\\
			\hline
			\textbf{Symbol} & \textbf{Definition} & \textbf{Unit}\\
			\hline
			\endfirsthead

			\multicolumn{3}{@{}l}{\small\slshape (continued from previous page)}\\
			\hline
			\textbf{Symbol} & \textbf{Definition} & \textbf{Unit}\\
			\hline
			\endhead

			\hline
			\multicolumn{3}{r}{\small\slshape (continues on next page)}\\
			\endfoot

			\hline
			\endlastfoot

			\multicolumn{3}{@{}l}{\textit{Latin symbols}}\\
			\hline
			
			$a_f$                  & Interface heat transfer coefficient & W/K \\
			$a_{PQ}$               & Pressure-coupling coefficient & m$^3$/(Pa$\cdot$s) \\
			$A_f$                  & Face area (scalar) & m$^2$ \\
			$\mathbf{A}_f$         & Face-area vector & m$^2$ \\
			$A_\mathrm{int}$       & Interfacial area (cell) & m$^2$ \\
			$A_{\mathrm{int},f}$   & Interface area projected onto face & m$^2$ \\
			$\mathbf{b}_P$         & Right-hand side vector component in cell $P$ & N \\
			$c$                    & Wave phase velocity & m/s \\
			$c_p$                  & Specific heat & J/(kg$\cdot$K) \\
			$C_f$                  & Face-centred coefficient & m$^3$/(Pa$\cdot$s) \\
			$d$                    & Distance from cell centre to interface intersection & m \\
			$d_f$                  & Normal distance from cell centre to interface & m \\
			$\mathbf{d}_f$         & Vector connecting adjacent cell centres & m \\
			$dS$                   & Surface-integration element & m$^2$ \\
			$dV$                   & Volume-integration element & m$^3$ \\
			$\mathbf{f}_\mathrm{body}$ & Body forces vector & N/m$^3$ \\
			$\mathbf{f}_\mathrm{face}$ & Face-centred body force vector & N/m$^3$ \\
			$F_f$                  & Mass flux through face $f$ & kg/s \\
			$\mathbf{F}_{\sigma}$  & Surface-tension force & N/m$^3$ \\
			$\mathbf{g}$           & Gravity vector & m/s$^2$ \\
			$\mathbf{G}_P$         & Least-squares gradient matrix in cell $P$ & m$^2$ \\
			$h$                    & Cell spacing & m \\
			$h_{lv}$               & Latent heat of vaporisation & J/kg \\
			$k$                    & Thermal conductivity & W/(m$\cdot$K) \\
			$L$                    & Axial domain length & m \\
			$\dot{m}$              & Volumetric mass transfer rate & kg/(m$^3\cdot$s) \\
			$M_P$                  & Momentum matrix coefficient & kg/s \\
			$\mathbf{n}$           & Unit normal vector & -- \\
			$p$                    & Pressure & Pa \\
			$\mathcal{P}_k$        & Isopolygon & -- \\
			$q$                    & Heat transfer rate & W \\
			$\dot{Q}_\mathrm{phase}$ & Latent heat source/sink & W/m$^3$ \\
			$\mathbf{r}_{PQ}$      & Displacement vector from cell $P$ to cell $Q$ & m \\
			$\mathbf{r}_{Pf}$      & Shortened displacement vector to interface intersection & m \\
			$R$                    & Pipe inner radius / bubble radius & m \\
			$R_0$                  & Initial bubble radius & m \\
			$R_\mathrm{Scriven}$   & Analytical Scriven bubble radius & m \\
			$\mathbf{s}_f$         & Unit vector connecting cell centres & -- \\
			$T$                    & Temperature & K \\
			$T_\mathrm{far}$       & Far-neighbour cell temperature & K \\
			$T_\infty$             & Bulk liquid temperature & K \\
			$T_\mathrm{wall}$      & Wall temperature & K \\
			$t_m$                  & Time-weighted scaling factor & -- \\
			$\Delta U$             & Relative velocity between phases & m/s \\
			$\mathbf{u}$           & Velocity vector & m/s \\
			$U_{y,v}$, $U_{y,l}$   & Mean axial vapour / liquid velocity & m/s \\
			$u_\mathrm{flux}$      & Volume flux through face $f$ & m$^3$/s \\
			$v_m$                  & Ratio $V_P/M_P$ & m$^3$s/kg \\
			$\mathbf{v}_i^k$       & Vertices of isopolygon $\mathcal{P}_k$ & m \\
			$V_P$                  & Cell volume & m$^3$ \\
			$w_f$                  & Geometrical face weight factor & -- \\
			$w_{\hat{\alpha}, i}$  & Colour-based weighting factor & -- \\
			$w^{o}$, $w^{oo}$      & Time-discretisation weights & -- \\
			$\mathbf{x}_P$         & Cell centre position & m \\
			$\mathbf{x}_f$         & Front intersection point & m \\
			$x_\gamma$             & Interface position & m \\
			$x_b$                  & Domain outlet position & m \\
			
			\\
			\multicolumn{3}{@{}l}{\textit{Greek symbols}}\\
			\hline
			$\alpha$               & Liquid volume fraction & -- \\
			$\hat{\alpha}$         & Smoothed (convoluted) volume fraction & -- \\
			$\alpha_l$             & Liquid thermal diffusivity & m$^2$/s \\
			$\alpha_v$             & Vapour thermal diffusivity & m$^2$/s \\
			$\beta$                & Scriven growth constant & -- \\
			$\Gamma$               & Diffusion coefficient & variable dependent \\
			$\delta$               & Local film thickness & m \\
			$\delta_0$             & Initial film thickness & m \\
			$\bar{\delta}$         & Mean film thickness & m \\
			$\delta_s$             & Dirac delta function & 1/m \\
			$\delta_T$             & Thermal penetration depth & m \\
			$\Delta T$             & Superheat, $T_\infty - T_\mathrm{sat}$ or $T_\mathrm{wall} - T_\mathrm{sat}$ & K \\
			$\Delta s$             & Distance between cell centres & m \\
			$\Delta t$             & Time step increment & s \\
			$\Delta t_\sigma$      & Capillary time step constraint & s \\
			$\varepsilon$          & Small regularisation constant & -- \\
			$\eta$                 & Interface perturbation from mean & m \\
			$s_\eta$               & Std.\ dev.\ of interface perturbation & m \\
			$\theta_P$             & Cell-centre superheat, $T_P - T_\mathrm{sat}$ & K \\
			$\kappa$               & Curvature & 1/m \\
			$\lambda$              & Stencil shortening ratio, $d/h$ & -- \\
			$\mu$                  & Dynamic viscosity & Pa$\cdot$s \\
			$\mu_t$                & Turbulent (eddy) viscosity & Pa$\cdot$s \\
			$\phi$                 & Generic transported variable & -- \\
			$\psi$                 & Scheme-dependent blending factor & -- \\
			$\rho$                 & Density & kg/m$^3$ \\
			$\sigma$               & Surface tension coefficient & N/m \\
			$\tau$                 & Pseudo-time & -- \\
			$\omega_f$             & Non-orthogonal correction scaling factor & -- \\
			$\xi$                  & Stefan growth constant & -- \\
			\\
			\multicolumn{3}{@{}l}{\textit{Subscripts}}\\
			\hline
			$_A$                   & Acceptor (downwind) cell & \\
			$_D$                   & Donor (upwind) cell & \\
			$_f$                   & Cell face & \\
			$_\mathrm{int}$        & Interface (iso-surface) & \\
			$_\mathrm{iso}$        & Isovalue for interface extraction & \\
			$_l$, $_v$             & Liquid, vapour & \\
			$_P$                   & Primary cell & \\
			$_Q$                   & Neighbouring cell & \\
			$_\mathrm{sat}$        & Saturation & \\
			$_U$                   & Far-upwind cell & \\
			$_\mathrm{wall}$       & Wall & \\
			$_\infty$              & Bulk / far-field & \\
			\\
			\multicolumn{3}{@{}l}{\textit{Superscripts}}\\
			\hline
			$^{(o)}$               & Previous time-step value & \\
			$^{(oo)}$              & Second previous time-step value & \\
			$^{*}$                 & After Choi's unsteady correction & \\
			$^{**}$                & Before corrections (from momentum equation) & \\
			\\
			\multicolumn{3}{@{}l}{\textit{Abbreviations}}\\
			\hline
			BiCG                   & Bi-Conjugate Gradient method & \\
			CFD                    & Computational Fluid Dynamics & \\
			CG                     & Conjugate Gradient method & \\
			CICSAM                 & Compressive Interface Capturing Scheme for Arbitrary Meshes & \\
			Co                     & Courant number & \\
			CSF                    & Continuum Surface Force & \\
			GFM                    & Ghost Fluid Method & \\
			IDW                    & Inverse-Distance Weighting & \\
			Ja                     & Jakob number & \\
			LES                    & Large Eddy Simulation & \\
			LSQ                    & Least-Squares (gradient reconstruction) & \\
			MTR                    & Mass Transfer Rate & \\
			PETSc                  & Portable, Extensible Toolkit for Scientific Computation & \\
			PISO                   & Pressure-Implicit with Splitting of Operators & \\
			SIMPLE				   & Semi-Implicit Method for Pressure Linked Equations & \\
			Ste                    & Stefan number & \\
			VOF                    & Volume-Of-Fluid method & \\
			WALE                   & Wall-Adapting Local Eddy-viscosity & \\
	\end{longtable}}
	\clearpage
	\section{Introduction}
	Boiling is one of the most effective modes of heat transfer, which enables the removal of large heat fluxes in many industrial systems. For example, nuclear power reactors deliberately operate in the nucleate boiling regime to achieve high surface heat fluxes for cooling while avoiding the burnout beyond the critical heat flux \cite{nishida_front_2023, lim_exploring_2022}. Besides nuclear energy, boiling heat transfer is integral to processes in chemical manufacturing, power generation, and electronics cooling \cite{sohel_murshed_critical_2017}, where efficient thermal management is important. However, predicting boiling phenomena---such as bubble nucleation, growth, and departure---is challenging yet essential for the safe and optimal design of such equipment. Direct experimentation on boiling flows can be difficult due to measurement constraints and the complex interactions inherent in two-phase flows \cite{kharangate_review_2017, hu_comprehensive_2025}. This motivates the development of high-fidelity computational simulations that can capture the detailed fluid mechanics and heat transfer in boiling, which provides insight into the underlying physics and helps to design better thermal systems \cite{mb_numerical_2024, chen_review_2024}.

	Accurate boiling simulation requires resolving the liquid--vapour interface as bubbles form and evolve. Numerical methods for this purpose are broadly classified into two categories: the interface-tracking and interface-capturing approaches. In interface-tracking methods, the interface is explicitly represented by a moving mesh or marker points, as in the front-tracking method, arbitrary Lagrangian-Eulerian (ALE) methods, or marker-and-cell (MAC) approaches. Front-tracking achieves high geometric fidelity and early implementations demonstrated the feasibility of simulating boiling in simple geometries \cite{juric_computations_1998, esmaeeli_front_2004}, but it requires complex algorithms for topological changes such as bubble merging or breakup.

	In contrast, interface-capturing methods describe the interface implicitly through a scalar field defined on a fixed grid, such as a colour function, volume fraction, or level-set function. Phase-field (or diffuse-interface) methods represent the interface as a finite-thickness region governed by free-energy-based equations such as the Cahn--Hilliard or Allen--Cahn formulations \cite{anderson-diffuse1998, roccon_boiling_2025}, which naturally handles topological changes but requires care to ensure mass conservation and sharp-interface recovery.

	The level-set method represents the interface implicitly as the zero level set of a continuous signed-distance function \cite{sussman_level_1994}. This approach automatically handles complex interface deformations and topological changes by advecting the level-set function with the flow. Level-set methods enable straightforward computation of geometric properties---curvature and interface normals---essential for surface tension and phase-change modelling. Although traditional level-set methods are not inherently mass-conserving, recent developments have improved on this limitation and made them viable for boiling simulations \cite{tanguy_level_2007, sahut_numerical_2021, fan_high-fidelity_2022, iskhakova_interface_2023}.

	The focus of this paper, however, is the volume-of-fluid (VOF) approach, which uses a colour function or volume fraction to indicate the proportion of each computational cell occupied by liquid versus vapour \cite{hirt_volume_1981}. Early VOF implementations used algebraic methods \cite{zhang_refined_2014, saincher_efficient_2022} that struggled with phenomena such as bubble disintegration, coalescence, and interfacial heat and mass transfer. The accuracy of VOF methods improved substantially with geometric interface capturing, in which the interface is first reconstructed (e.g.\ via Piecewise Linear Interface Calculation) and then advected geometrically \cite{bures_piecewise_2021, long_edge-based_2024}. Despite these advances, implementing geometric interface capturing on arbitrary meshes remains challenging. Such methods have consequently been applied mostly on structured grids \cite{sato_sharp-interface_2013, bures_direct_2021, poblador-ibanez_momentum_2025}.

	In all cases, capturing the physics of phase change requires solving the energy transport (heat equation) alongside the hydrodynamics and applying the correct jump conditions for temperature and heat flux at the interface. Techniques like the Ghost Fluid Method (GFM) \cite{fedkiw_non-oscillatory_1999, tanguy_level_2007, gibou_level_2007, rueda_villegas_ghost_2016} are often used in level-set or front-tracking frameworks to impose discontinuities (e.g. enforcing the interface at saturation temperature and accounting for latent heat) without unphysically smearing those jumps. Interface-capturing CFD can now resolve boiling at the scale of individual bubbles, which gives access to local heat transfer processes that are difficult to measure experimentally.

	Although interface-capturing methods have matured, most developments and validations were performed on structured grids. Structured grids simplify interface advection and achieve high accuracy with aligned interfaces; however, they struggle to conform to the complex geometries typical of industrial equipment. Beyond geometric flexibility, there is growing evidence that mesh topology itself influences accuracy in ways that go beyond simple resolution effects. Mari\'{c} et al.~\cite{Maric2020} showed that algebraic VOF advection on Cartesian meshes deforms a sphere into an octahedron due to mesh-aligned numerical diffusion. Herrmann~\cite{Herrmann2008} demonstrated that balanced-force algorithms achieve machine-zero spurious currents on structured grids but fail on unstructured grids due to loss of mesh symmetry. Nishikawa~\cite{Nishikawa2018} proved that Green--Gauss gradients are second-order accurate only on structured meshes, degrading on general unstructured grids, and Syrakos et al.~\cite{Syrakos2017} showed that least-squares gradient accuracy depends critically on stencil symmetry. These findings suggest that structured meshes, while convenient, may introduce systematic biases when the gradient stencil is modified to account for interface position---an issue that has not been examined in the context of phase-change simulations.

	Unstructured grids, with their ability to represent arbitrary polyhedral cells, offer geometric flexibility for practical applications but introduce numerical challenges in: (i)~interface reconstruction, where computing interface geometry (e.g., normal vectors and curvature) is more complex due to irregular, non-orthogonal cells; (ii)~gradient calculation, where accurately evaluating temperature gradients on either side of a curved interface is needed for resolving phase-change rates; and (iii)~numerical stability and conservation, where maintaining the sharpness of the interface during phase change is particularly challenging. Recent efforts have begun to address these challenges: Sahut et al.\ \cite{sahut_numerical_2021} extended coupled level-set/ghost-fluid approaches to arbitrary meshes, while Giustini and Issa \cite{giustini_method_2021} coupled algebraic VOF with a Lee-like empirical model \cite{lee1980pressure} for mass transfer. Scheufler and Roenby~\cite{Scheufler2023} developed the TwoPhaseFlow framework for OpenFOAM, which combines MULES algebraic transport with an $\alpha = 0.5$ isosurface reconstruction for phase-change calculations; however, their approach is limited to first-order interface geometry and has been validated primarily on structured grids.
	
	This paper presents an interface-capturing framework for boiling on unstructured grids with the volume-of-fluid method. The mass transfer rate is computed directly from the heat fluxes in the liquid and vapour phases at the geometrically reconstructed interface, without empirical closure models. The interface is represented by polygons computed using the isoap library \cite{lopez_isoap_2021}, which reconstructs the interface from the volume fraction field $\alpha$ in arbitrary polyhedral cells. Unlike the $\alpha = 0.5$ isosurface approach of Scheufler and Roenby~\cite{Scheufler2023}, the isoap reconstruction provides a piecewise-linear representation of the interface within each cell, yielding accurate face-area vectors and normal orientations that are essential for computing heat fluxes at the interface on cells of arbitrary shape.

	The main contributions of this work are: (i)~a direct heat-flux-driven mass transfer model on general unstructured meshes, combining CICSAM algebraic VOF transport with isoap geometric interface reconstruction for the phase-change calculation; (ii)~a systematic analysis of the interaction between the interface-modified least-squares gradient stencil and mesh topology, revealing a coherent four-fold temperature gradient anisotropy on structured hexahedral meshes that is absent on polyhedral meshes---a finding that connects and extends the observations of Nishikawa~\cite{Nishikawa2018}, Syrakos et al.~\cite{Syrakos2017}, and Mari\'{c} et al.~\cite{Maric2020} on mesh-induced numerical biases to the specific context of interface-capturing phase-change simulations; and (iii)~demonstration of the framework on turbulent annular boiling flow in a geometry using unstructured meshes.

	The paper is structured as follows. In Section~\ref{sec:methodology}, we introduce the methodology, from the interface capturing method and governing equations to discretisation and solving. In Section~\ref{sec:validation}, we validate the method against benchmark problems and present a detailed analysis of the temperature gradient anisotropy on structured versus polyhedral meshes. Section~\ref{sec:annular} demonstrates the framework on an upward co-current annular boiling flow case. In Section~\ref{sec:conclusions}, we conclude with implications and future research directions. \ref{app:cicsam} and~\ref{app:LeastSQGR} provide details of the CICSAM advection scheme and the least-squares gradient reconstruction, respectively. \ref{app:gradient_magnitude} derives the gradient overestimate mechanism on hexahedral meshes, and~\ref{app:computational_cost} reports the computational cost breakdown.
	
	\FloatBarrier
	\section{Methodology}
	\label{sec:methodology}
	This section details the governing equations, numerical methods, and solver implementation. T-Flows \cite{DelNovTFlows} is an incompressible flow solver written in Fortran 2008 and using Message Passing Interface (MPI) for parallelisation, developed by the CFD group at Paul Scherrer Institute, Switzerland. It supports unstructured meshes and is therefore suitable for simulations involving complex geometries in engineering applications. It has recently been upgraded to handle two-phase flow problems. T-Flows takes advantage of open-source libraries: PETSc, the Portable, Extensible Toolkit for Scientific Computation \cite{petsc-user-ref}, is used for solving matrix systems, and the isoap library \cite{lopez_isoap_2021} has been integrated for geometric reconstruction of the interface needed for phase-change problems.
	
	\subsection{Interface capturing method}
	A two-phase vapour–water system has been modelled using the one-fluid formulation \cite{tryggvason_direct_2011} of the Navier–Stokes equations and the volume-of-fluid (VOF) approach for interface capturing. In this approach, \(\alpha(\mathbf{x}, t)\) denotes the volume fraction of the liquid phase  (\(0 \leq \alpha \leq 1\)). The volume fraction of the vapour phase is then \((1 - \alpha)\).
	The advection equation for \(\alpha\), including mass transfer source terms  \(\dot{m}\), can be written as:
	
	\begin{equation}
		\frac{\partial \alpha}{\partial t} 
		+ \nabla \cdot \bigl(\alpha\,\mathbf{u}\bigr)
		= -\frac{\dot{m} }{\rho_l},
		\label{eq:vof_equation}
	\end{equation}
	where \(\mathbf{u}\) is the mixture velocity, \( \rho_l\) is the liquid density and $\dot{m}$ is the volumetric mass transfer rate, positive for vaporisation and negative for condensation. The derivation of Eq. \eqref{eq:vof_equation} is given in Appendix A of \cite{sato_sharp-interface_2013}.
	
	Volume-averaged mixture properties (e.g., density \(\rho\) and viscosity \(\mu\)) within a computational cell are defined as an arithmetic mean:
	\begin{align}
		\phi(\alpha) &= \alpha\,\phi_l + (1 - \alpha)\,\phi_v, \label{eq:mixture_rho}
	\end{align} 
	where $\phi$ is any above-mentioned property. Specific heat $c_p$ and thermal conductivity $k$ are replaced by their phase-specific values ($c_{p,l}$, $k_l$ or $c_{p,v}$, $k_v$) in cells and at faces where one phase dominates ($\alpha > 0.5$ or $\alpha < 0.5$).
	
	The Compressive Interface Capturing Scheme for Arbitrary Meshes (CICSAM) developed by Ubbink and Issa (1999) \cite{ubbink_method_1999} is a high-resolution volume-of-fluid (VOF) method for tracking sharp fluid-fluid interfaces on arbitrary meshes. It solves a scalar transport equation for the volume fraction in a conservative finite-volume framework. The scheme blends a compressive differencing scheme (Hyper-C), which maintains interface sharpness, with a high-order downwind scheme (Ultimate Quickest) that prevents interface wrinkling. The blending factor depends on the local orientation of the interface relative to the mesh faces to minimise numerical diffusion. Temporal discretisation follows an implicit, unsplit, predictor--corrector approach that maintains boundedness and second-order accuracy without explicit clipping of the solution. A brief mathematical formulation of the method can be found in \ref{app:cicsam}; for more details, the reader is referred to the original paper by Ubbink and Issa~\cite{ubbink_method_1999}.
	
	To reconstruct the interface from an $\alpha$ field, the \texttt{isoap} library \cite{lopez_isoap_2021} has been integrated into the T-Flows code, which extracts the isosurface corresponding to a prescribed threshold value $\alpha_{\mathrm{iso}}$ (typically $0.5$) on arbitrary polyhedral cells. Let $\alpha(\mathbf{x})$ denote the scalar field defined at the vertices of a polyhedral cell. The interface is defined as the surface where $\alpha = 0.5$.
	
	For an edge connecting vertices $i$ and $j$, with coordinates $\mathbf{x}_i$ and $\mathbf{x}_j$ and corresponding scalar values $\alpha_i$ and $\alpha_j$, the intersection point $\mathbf{x}_{ij}$ is determined by linear interpolation:
	\begin{equation}
		\mathbf{x}_{ij} = \mathbf{x}_i + \lambda_{ij} (\mathbf{x}_j - \mathbf{x}_i),
		\quad \text{with} \quad \lambda_{ij} = \frac{\alpha_{\mathrm{iso}} - \alpha_i}{\alpha_j - \alpha_i}.
		\label{eq:edge_intersection}
	\end{equation}
	A valid intersection is computed only when $\alpha_i$ and $\alpha_j$ straddle the isovalue $\alpha_{\mathrm{iso}}$.
	
	The \texttt{isoap} algorithm proceeds as follows:
	\begin{enumerate}
		\item \textbf{Edge intersections:} For each face of the polyhedral cell, identify all edges where the scalar values at the vertices straddle $\alpha_{\mathrm{iso}}$. Compute the intersection points using Eq.~\eqref{eq:edge_intersection}.
		\item \textbf{Isosurface assembly:} The computed intersection points are then assembled into one or more \emph{isopolygons} that represent the reconstructed interface within the cell. Each isopolygon $\mathcal{P}_k$ is defined as an ordered set of vertices:
		\[
		\mathcal{P}_k = \{\mathbf{v}_1^k, \mathbf{v}_2^k, \dots, \mathbf{v}_{n_k}^k\},
		\]
		where the vertices are arranged in a consistent order (e.g., anticlockwise when viewed from the continuous phase) to accurately capture the interface geometry.
	\end{enumerate}
	
	This reconstructed interface is subsequently used to evaluate mass transfer rates across the interface as described in the following sections. By coupling the advection of the $\alpha$ field via the CICSAM scheme with the interface extraction capabilities of the \texttt{isoap} library \cite{lopez_isoap_2021}, the solver combines high-resolution advection of the volume fraction with geometric interface reconstruction on arbitrary meshes. 
	
	\subsection{Continuity and momentum equation}
	
	For an incompressible two-phase mixture with phase-change, the continuity equation (assuming constant phase density) takes the form \cite{sato_sharp-interface_2013}:
	\begin{equation}
		\nabla \cdot \mathbf{u} = \dot{m} \Big( \frac{1}{\rho_v} - \frac{1}{\rho_l}\Big).
		\label{eq:continuity}
	\end{equation}
	A single momentum equation is solved for the mixture velocity \(\mathbf{u}\):
	\begin{equation}
		\frac{d}{dt}\int_{V_P} \rho\,\mathbf{u}\,dV
		+ \int_{\partial V_P} \rho\,\mathbf{u}\otimes\mathbf{u}\cdot\mathbf{n}\,dS
		= -\int_{\partial V_P} p\,\mathbf{n}\,dS + \int_{\partial V_P} (\mu + \mu_t) \left(\nabla\mathbf{u} + \left(\nabla\mathbf{u}\right)^T\right)\cdot\mathbf{n}\,dS
		+ \int_{V_P} \rho\,\mathbf{g}\,dV + \int_{V_P} \mathbf{F}_\sigma\,dV,
		\label{eq:momentum}
	\end{equation}
	where \(V_P\) is the cell volume, \(p\) is the pressure (shared by both phases), \(\rho\) and \(\mu\) are the mixture density and viscosity from Eq.~\eqref{eq:mixture_rho}, \(\mathbf{g}\) is the gravitational acceleration vector and \(\mathbf{F}_\sigma\) is the surface tension force. For turbulent simulations the turbulent viscosity $\mu_t$ is introduced to model the unresolved subgrid-scale stresses in the LES framework, and is added to the molecular viscosity to form the effective viscosity $\mu_\mathrm{eff} = \mu + \mu_t$.
	
	\subsection{Energy equation with phase-change}
	The energy equation is solved for temperature \(T\), and we assume a single energy equation for the mixture. Using the divergence theorem, the convective and diffusive terms are transformed into surface integrals over the cell boundary \(\partial V_P\):
	\begin{equation}
		\int_{V_P} \rho\,c_p\,\frac{\partial T}{\partial t}\,dV 
		+\int_{\partial V_P}  \rho\,c_p\, T\,(\mathbf{u}\cdot\mathbf{n})\,dS 
		= \int_{\partial V_P} k\,(\nabla T\cdot\mathbf{n})\,dS 
		+ \int_{V_P} \dot{Q}_{\mathrm{phase}}\,dV.
		\label{eq:energy}
	\end{equation}
	
	where $k(\alpha)$ is the mixture thermal conductivity and $c_p (\alpha)$ is the specific heat capacity, $\dot{Q}_{\mathrm{phase}}$ is the latent heat source/sink due to condensation/evaporation. 
	\subsection{Discretisation}
	
	\subsubsection{Convective term}
	
	The convection term appears in all transport equations and is discretised by integrating over the cell volume using the Gauss theorem and the midpoint rule:
	\begin{equation}
		\int_{V_P} \rho u_j \frac{\partial \phi}{\partial x_j} \, dV 
		\approx \sum_f F_f \, \phi_f,
	\end{equation}
	where $F_f = \rho_f (\mathbf{u}_f \cdot \mathbf{n}_f) A_f$ is the mass flux through face $f$ and $\phi_f$ is the face value of the transported variable. Here $\phi$ represents any transported quantity: velocity components $u_i$ in the momentum equations, temperature $T$ in the energy equation, or scalar concentrations. The accuracy and stability of the convective discretisation depends on how $\phi_f$ is evaluated.
	
	We denote by $D$ the upwind (donor) cell and by $A$ the downwind (acceptor) cell, assigned according to the sign of the mass flux. The face value is expressed in the general form
	\begin{equation}
		\phi_f = \phi_D + \tfrac{1}{2}\,\psi \, (\phi_A - \phi_D),
	\end{equation}
	where $\psi$ is a scheme-dependent limiter function. For \emph{upwind differencing} (UD), $\psi = 0$ and the face value equals the upwind cell value, which gives a first-order accurate, unconditionally bounded scheme that introduces significant numerical diffusion. For \emph{central differencing} (CD), $\psi = 1$ on a uniform mesh (more generally $\psi = 2 f_j$, where $f_j$ is the geometric interpolation factor). This results in a second-order accurate scheme that may exhibit nonphysical and unbounded oscillations.

	To combine the stability of UD with the accuracy of higher-order schemes, \emph{total variation diminishing} (TVD) limiters compute $r$ as a function of the gradient ratio
	\begin{equation}
		r = \frac{\phi_D - \phi_U}{\phi_A - \phi_D},
	\end{equation}
	where $\phi_U$ is the far-upwind value. On unstructured meshes, $\phi_U$ is not directly available and must be estimated by gradient extrapolation from the upwind cell:
	\begin{equation}
		\phi_U = \phi_A - 2 \, \nabla\phi|_D \cdot \mathbf{d}_f,
	\end{equation}
	where $\mathbf{d}_f$ is the vector connecting cell centres $D$ and $A$. The gradient of $\phi$ at cell centre is computed with least-squares method as described in \ref{app:LeastSQGR}. This extrapolated value is limited to lie within local bounds $[\phi_{\min}, \phi_{\max}]$ computed from neighbouring cells.   Available TVD schemes in T-Flows include MINMOD, SUPERBEE, SMART, and the Gamma scheme. Given the gradient ratio $r$, the limiter function $\psi$ for each scheme is:
  \begin{align*}
		\text{MINMOD:}    \quad \psi &= \max\bigl(0,\, \min(r,\, 1)\bigr), \\
		\text{SUPERBEE:}  \quad \psi &= \max\bigl(0,\, \min(2r,\, 1),\, \min(r,\, 2)\bigr), \\
		\text{SMART:}     \quad \psi &= \max\Bigl(0,\, \min\bigl(4r,\, \tfrac{3+r}{4},\, 2\bigr)\Bigr).
	\end{align*}
	These schemes range from the most diffusive (MINMOD) to the most compressive (SUPERBEE). The Gamma scheme \cite{jasak_high_1999} operates in the Normalised Variable Framework rather than through a limiter function: it blends between upwind
	and central differencing based on the normalised face value $\tilde{\phi}_f$, with a blending parameter $\beta_m = 0.1$ controlling the transition.
	
	To ensure numerical stability, a deferred correction approach is used. The upwind scheme is implemented implicitly to guarantee diagonal dominance of the coefficient matrix, while the higher-order correction is added explicitly to the right-hand side:
	\begin{equation}
		F_f \, \phi_f = \underbrace{F_f \, \phi_D}_{\text{implicit}}
		+ \underbrace{F_f \left( \phi_f - \phi_D \right)^{(o)}}_{\text{explicit correction}},
	\end{equation}
	where the superscript $(o)$ denotes evaluation using values from the previous iteration. At convergence, the explicit correction vanishes and the solution corresponds to the selected higher-order scheme.
	\subsubsection{Diffusion term}
	
	The diffusion term is discretised using a deferred-correction approach to handle 
	mesh non-orthogonality. For scalar transport equations, such as energy, the 
	diffusion term takes the form $\nabla \cdot (\Gamma \nabla \phi)$, where $\Gamma$ is the diffusion coefficient. For the 
	momentum equations, diffusion involves the viscous stress tensor 
	$\boldsymbol{\tau} = \mu (\nabla \mathbf{u} + (\nabla \mathbf{u})^T)$, which 
	couples velocity components through the symmetric gradient. However, since 
	T-Flows solves the momentum equations component by component, the non-orthogonal 
	correction for face gradients is applied in the same manner for both cases. 
	We illustrate the approach for scalar diffusion. Applying the Gauss theorem:

	\begin{equation}
		\int_{V_P} \nabla \cdot (\Gamma \nabla \phi) \, dV
		= \oint_{\partial V_P} \Gamma_f \nabla \phi \cdot \mathbf{n}_f \, dS
		\approx \sum_f \Gamma_f (\nabla \phi \cdot \mathbf{n}_f) A_f,
	\end{equation}

	where $\Gamma_f$ is the diffusion coefficient at face $f$, $\mathbf{n}_f$ is the outward unit normal vector, and $A_f$ is the face area.

	For orthogonal meshes with linear scalar variation, the face-normal gradient is approximated using central differencing:
	\begin{equation}
		\nabla \phi|_f \cdot \mathbf{n}_f = \frac{\phi_Q - \phi_P}{\Delta s},
	\end{equation}
	where $\phi_P$ and $\phi_Q$ are the values in the primary cell $P$ and its neighbouring cell $Q$, and $\Delta s = |\mathbf{x}_Q - \mathbf{x}_P|$ is the distance between them.
	
\begin{figure}[htbp]
		\centering
		\includegraphics[width=\linewidth]{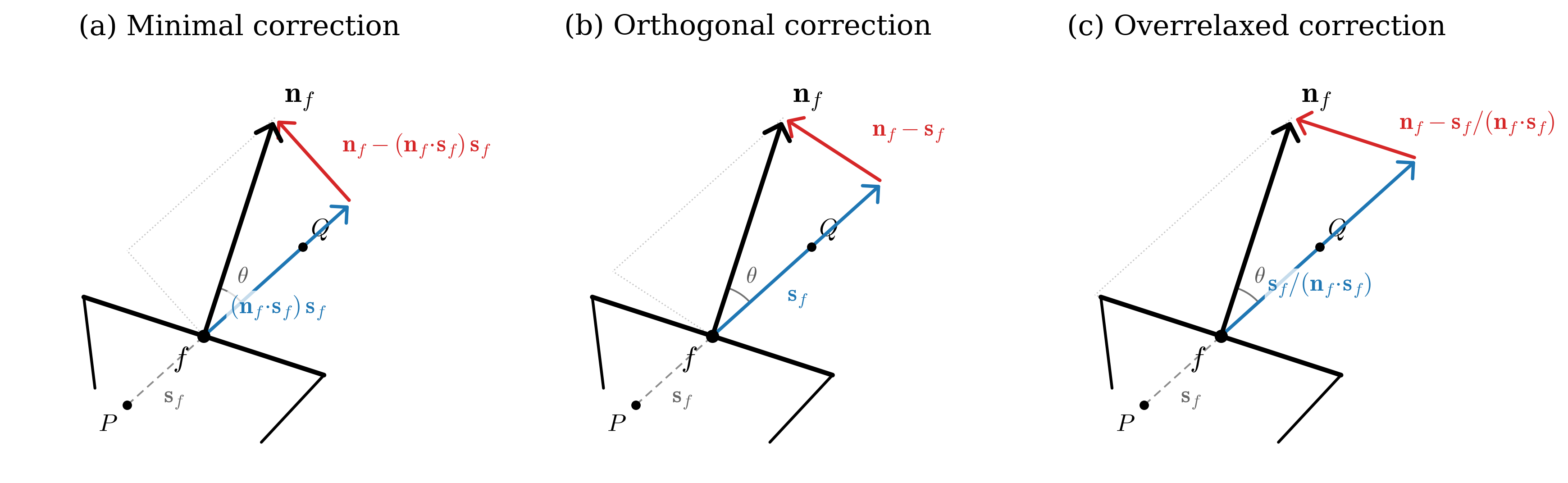}
		\caption{Non-orthogonal correction methods at face $f$ of a non-orthogonal two-dimensional mesh. The face normal $\mathbf{n}_f$ (black) is decomposed into an implicit contribution along $\mathbf{s}_f$ (blue) and an explicit correction (red). From (a) to (c), the implicit weight $\omega_f$ increases, which improves stability on distorted meshes.}
		\label{fig:overrelaxed}
	\end{figure}

	However, orthogonality is rare in unstructured meshes. To maintain stability and accuracy, a deferred correction approach decomposes the face-normal gradient into orthogonal and non-orthogonal contributions. We define:
	\begin{itemize}
		\item $\mathbf{n}_f$ -- the unit normal to face $f$,
		\item $\mathbf{s}_f = (\mathbf{x}_Q - \mathbf{x}_P)/|\mathbf{x}_Q - \mathbf{x}_P|$ -- the unit vector connecting cell centres.
	\end{itemize}
	On orthogonal meshes $\mathbf{n}_f = \mathbf{s}_f$; on non-orthogonal meshes they differ (see Fig.~\ref{fig:overrelaxed}).
	
	The gradient decomposition reads:
	\begin{equation}
		\nabla \phi|_f \cdot \mathbf{n}_f
		= \underbrace{\omega_f \frac{\phi_Q - \phi_P}{\Delta s}}_{\text{orthogonal (implicit)}}
		+ \underbrace{\overline{\nabla \phi}|_f \cdot (\mathbf{n}_f - \omega_f \mathbf{s}_f)}_{\text{non-orthogonal (explicit)}},
		\label{eq:difforth}
	\end{equation}
	where $\omega_f$ is a scaling factor and the overbar denotes interpolation from cell-centred gradients:
	\begin{equation}
		\overline{\nabla \phi}|_f = (1 - w_f) \nabla \phi|_P + w_f \nabla \phi|_Q,
	\end{equation}
	with $w_f$ being the geometric interpolation weight. The non-orthogonal term is evaluated using gradients from the previous iteration, hence the deferred correction.
	
	Three choices for the scaling factor $\omega_f$ are commonly used \cite{denner_balanced-force_2013}:
	\begin{itemize}
		\item \textbf{Minimal correction} ($\omega_f = \mathbf{n}_f \cdot \mathbf{s}_f$): minimises the explicit correction term and gives the highest theoretical accuracy but is prone to instability at high non-orthogonality.

		\item \textbf{Orthogonal correction} ($\omega_f = 1$): maintains a fixed implicit contribution regardless of mesh geometry, but diverges when the angle between $\mathbf{n}_f$ and $\mathbf{s}_f$ exceeds approximately $38^\circ$.

		\item \textbf{Overrelaxed correction} ($\omega_f = (\mathbf{n}_f \cdot \mathbf{s}_f)^{-1}$): increases the implicit contribution with growing non-orthogonality and provides the highest numerical stability on distorted meshes.
	\end{itemize}
	
	Given the focus on unstructured meshes with potentially significant local non-orthogonality, the overrelaxed correction is adopted throughout this work to ensure stable convergence \cite{mathur_pressure-based_1997}.
	
	The complete diffusion flux through face $f$ is:
	\begin{equation}
		\Gamma_f (\nabla \phi \cdot \mathbf{n}_f) A_f
		= \Gamma_f A_f \left[ \omega_f \frac{\phi_Q - \phi_P}{\Delta s}
		+ \overline{\nabla \phi}|_f^{(o)} \cdot (\mathbf{n}_f - \omega_f \mathbf{s}_f) \right],
	\end{equation}
	where the superscript $(o)$ denotes values from the previous iteration. The orthogonal term enters the coefficient matrix implicitly, which couples cells $P$ and $Q$, while the non-orthogonal correction is added to the right-hand side.
	
	The orthogonal contribution provides numerical robustness through implicit treatment, while the explicit non-orthogonal correction recovers accuracy on distorted meshes. At convergence, the iterative corrections vanish and the scheme gives a consistent approximation of the full diffusion operator.
	
	\subsubsection{Momentum specific terms}
	\textbf{Surface tension term}
	
	In T-Flows, surface tension is modelled using the continuum surface force (CSF) method introduced by Brackbill et al. \cite{brackbill_continuum_1992} and subsequently improved by Denner \cite{denner_balanced-force_2013}. In this approach, the surface tension force is distributed as a volumetric force over the interface region.
	
	To reduce errors in the interface normal and curvature calculation due to the discontinuous nature of the volume fraction field, the volume fraction is first smoothed by iterative face-area-weighted averaging. The smoothed volume fraction in the primary cell $P$, $\hat{\alpha_P}$, is defined as:
	\begin{equation}
		\hat{\alpha}_P = \frac{\sum_f \hat{\alpha}_f \, A_f}{\sum_f A_f},
		\quad
		\hat{\alpha}_f = (1 - w_f)\,\alpha_P + w_f\,\alpha_Q,
	\end{equation}
	where $A_f$ is the face area, $\alpha_Q$ in the neighbour cell, and $w_f$ is the scaling factor \cite{denner_balanced-force_2013} defined in the previous section. The procedure is repeated for a prescribed number of smoothing iterations. The unit normal vector $\mathbf{n}_P$ to the interface in cell $P$, oriented from the vapour phase towards the liquid phase, is then defined as
	\begin{equation}
		\label{eq:unit_normal}
		\mathbf{n}_P = \frac{\nabla \hat{\alpha}_P}{ \lVert \nabla \hat{\alpha}_P \rVert + \varepsilon},
	\end{equation}
	where $\varepsilon = 10^{-15}$ is a small regularisation constant that prevents division by zero in pure-phase cells.

	The raw curvature $\tilde{\kappa}_P$ of the interface in cell $P$ is computed as the negative divergence of the unit normal,
	\begin{equation}
		\tilde{\kappa}_P = -\nabla \cdot \mathbf{n}_P.
	\end{equation}

	This raw curvature is then smoothed in two successive steps.
	In the first step, $\tilde{\kappa}_P$ is averaged with the $\hat{\alpha}$-based weighting factor
	\begin{equation}
		w_{\hat{\alpha},i} \;=\; \bigl[\,1 - 2 \,\bigl|\tfrac12 - \hat{\alpha}_i \bigr|\bigr]^8,
	\end{equation}
	to give an \emph{intermediate} curvature $\kappa^*_P$:
	\begin{equation}
		\kappa^*_P
		\;=\;
		\frac{
			\tilde{\kappa}_P\,w_{\hat{\alpha},P} \;+\; \sum_{Q}\,\tilde{\kappa}_Q\,w_{\hat{\alpha},Q}
		}{
			w_{\hat{\alpha},P} \;+\; \sum_{Q} w_{\hat{\alpha},Q} \;+\; \varepsilon
		},
	\end{equation}
	where $\varepsilon$ prevents division by zero in pure-phase cells where all weights vanish.
	
	In the second step, $\kappa^*_P$ is further smoothed in the direction of the interface normal $\mathbf{n}_P$ from Eq.~\eqref{eq:unit_normal}.
	For every face-sharing neighbour \(Q\in\mathcal N(P)\) we define
	\begin{equation}
		\label{eq:sQ}
		\mathbf s_Q
		\;=\;
		\frac{\mathbf x_Q-\mathbf x_P}
		{\lVert \mathbf x_Q-\mathbf x_P \rVert},
	\end{equation}
	with \(\mathbf x_P\) and \(\mathbf x_Q\) denoting the cell–centre coordinates.
	The directional weight for the contribution from neighbour $Q$ to cell $P$ is then:
	\begin{equation}
		w_{n,Q}
		\;=\;
		\bigl|\mathbf{n}_P \,\cdot\, \mathbf{s}_Q \bigr|^8
	\end{equation}
	
	The final smoothed curvature is:
	\begin{equation}
		\kappa_P 
		\;=\; 
		\frac{
			\kappa^*_P\,w_{\hat{\alpha},P}\ \;+\;
			\sum_{Q}\,\kappa^*_Q\,w_{\hat{\alpha},Q}\,w_{n,Q}
		}{
			w_{\hat{\alpha},P} \;+\;
			\sum_{Q} w_{\hat{\alpha},Q}\,w_{n,Q}
		}.
	\end{equation}
	
	In the CSF method, the singular surface tension force $\sigma \kappa \, \delta_s \, \mathbf{n}$ is regularised by replacing the Dirac delta $\delta_s$ with $|\nabla \alpha|$, which gives the volumetric force:
	\begin{equation}
		\mathbf{F}_\sigma = \sigma \kappa_P \, \nabla \alpha.
	\end{equation}
	Note that the force uses the gradient of the sharp (unsmoothed) volume fraction $\alpha$, not $\hat{\alpha}$: the curvature benefits from smoothing to reduce spurious currents, but the force gradient must remain sharp to maintain the balanced-force property with the pressure gradient in the momentum equation~\cite{denner_balanced-force_2013}.

	\textbf{Rhie and Chow correction}

	Face velocities are computed using the Rhie--Chow interpolation~\cite{rhie_numerical_1983} with the unsteady correction of Choi~\cite{choi1999, choi_use_2003} and the body-force treatment following the formulation of Mencinger and \v{Z}un~\cite{mencinger_finite_2007}. This procedure prevents pressure--velocity decoupling on collocated grids by reconstructing face mass fluxes from interpolated cell velocities and face-centred pressure gradients~\cite{yu_discussion_2002, majumdar_role_1988, papageorgakopoulos_enhancement_2000}.

	The pressure difference between adjacent cells $P$ and $Q$ approximates the face-normal pressure gradient:

	\begin{equation}
		\left(\nabla p \cdot \mathbf{n}\right)_f \approx \frac{p_P - p_Q}{\Delta s},
	\end{equation}
	\noindent where $\Delta s$ is the distance between the two cell centres. For VOF simulations, the balanced-force formulation subtracts the surface tension contribution:

	\begin{equation}
		\left(\nabla p \cdot \mathbf{n}\right)_f \approx \frac{(p_P - p_Q) - \sigma \kappa_f(\alpha_P - \alpha_Q)}{\Delta s},
	\end{equation}
	\noindent where $\alpha_P$ and $\alpha_Q$ are the volume fractions in cells $P$ and $Q$, and $\kappa_f = (1 - w_f)\,\kappa_P + w_f\,\kappa_Q$ is the face-interpolated curvature.

	Choi's unsteady correction removes the contribution of previous time-step velocities:

	\begin{equation}
		\mathbf{u}_P^{*} = \mathbf{u}_P^{**}
		- \frac{V_P \rho_P}{\Delta t \, M_P}
		\left(w^{o} \mathbf{u}_P^{(o)}
		+ w^{oo} \mathbf{u}_P^{(oo)}\right),
	\end{equation}

	\noindent where $\mathbf{u}_P^{**}$ is the velocity from the momentum equation for cell~$P$, $V_P$ the cell volume, $\rho_P$ the density, $M_P$ the central coefficient of the discretised momentum equation, superscripts $(o)$ and $(oo)$ denote values from the previous and second-previous time levels, and $w^{o}$, $w^{oo}$ are the corresponding time-discretisation weights ($w^{o} = 1$, $w^{oo} = 0$ for first-order backward Euler; $w^{o} = 2$, $w^{oo} = -1/2$ for the second-order backward differentiation formula). The body-force correction accounts for gravity and surface tension:

	\begin{equation}
		\mathbf{u}_P = \mathbf{u}_P^{*}
		- \frac{V_P}{M_P}\mathbf{f}_\mathrm{body}.
	\end{equation}

	Face velocities are obtained by linear interpolation of the corrected cell velocities:

	\begin{equation}
		\mathbf{u}_f = (1 - w_f)\,\mathbf{u}_P
		+ w_f\,\mathbf{u}_Q.
	\end{equation}

	The volume flux across face $f$ is then:

	\begin{multline}
		u_\mathrm{flux} = (\mathbf{u}_f \cdot \mathbf{A}_f)
		+ a_{PQ}(p_P - p_Q)
		+ C_f\,(v_m \nabla p)_f \cdot \mathbf{d}_f \\
		+ \underbrace{\left[((1 - w_f)\,t_{m,P}
		+ w_f\,t_{m,Q})\,(w^{o}\,u_\mathrm{flux}^{(o)}
		+ w^{oo}\,u_\mathrm{flux}^{(oo)})\right]
		}_{\text{Choi's correction}} \\
		+ \underbrace{\left[((1 - w_f)\,v_{m,P}
		+ w_f\,v_{m,Q})\,(\mathbf{f}_\mathrm{face}
		\cdot \mathbf{A}_f)\right]
		}_{\text{body-force correction}},
	\end{multline}
	where $a_{PQ}$ is the pressure-coupling coefficient, $C_f$ is the face-centred coefficient from the discretised momentum equation, and $\mathbf{d}_f$ is the vector connecting the cell centres $P$ and $Q$. The ratio $v_m = V_P/M_P$ characterises each cell's volumetric response to applied forces, $t_m = V_P \rho_P / (\Delta t\,M_P)$ is the time-weighted scaling factor, and $\mathbf{f}_\mathrm{face}$ denotes face-centred body forces (gravity, surface tension). The first pressure term, $a_{PQ}(p_P - p_Q)$, removes the pressure gradient already embedded in the interpolated cell velocities using a compact two-cell stencil. The second term, $C_f\,(v_m \nabla p)_f \cdot \mathbf{d}_f$, restores the face-centred pressure gradient using cell-centred gradients computed from the extended neighbourhood. On orthogonal meshes these terms partially cancel; on non-orthogonal or unstructured meshes they differ because the cell-centred gradient $\nabla p$ accounts for pressure variations in all directions, not solely along the cell-connecting vector.
	
	\subsubsection{Energy specific terms}

	\paragraph{Mass transfer rate.}
	For cells containing the interface, the interfacial area $A_\mathrm{int}$ and outward unit normal $\mathbf{n}_\mathrm{int}$ are computed using the isoap library \cite{lopez_isoap_2021}, which provides the exact geometry of the reconstructed isopolygon. The heat transfer rates from each phase to the interface are:
	\begin{equation}
		q_l = k_l (\nabla T_l \cdot \mathbf{n}_\mathrm{int})\, A_\mathrm{int}, \qquad
		q_v = -k_v (\nabla T_v \cdot \mathbf{n}_\mathrm{int})\, A_\mathrm{int},
	\end{equation}
	where $k_l$ and $k_v$ are the thermal conductivities and the extrapolated gradients $\nabla T_l$ and $\nabla T_v$ represent the temperature gradients at the interface from the liquid and vapour sides, respectively. Both heat transfer rates are defined as positive when directed towards the interface.

	The volumetric mass transfer rate is determined from the energy imbalance via the Rankine--Hugoniot jump condition:
	\begin{equation}
		\dot{m} = \frac{q_l + q_v}{h_{lv}\, V_P}.
		\label{eq:mass_transfer}
	\end{equation}
	It should be recalled that positive and negative $\dot{m}$ correspond to evaporation and condensation, respectively. This formulation directly links phase change to local thermal gradients at the interface without relying on empirical correlations.

	\paragraph{Interface-modified temperature gradient.}
	The temperature field is discontinuous across the liquid--vapour interface: the interface is held at the saturation temperature $T_\mathrm{sat}$, while bulk temperatures on each side vary smoothly. Computing accurate gradients near the interface therefore requires accounting for this discontinuity. In the standard least-squares gradient reconstruction (\ref{app:LeastSQGR}), the gradient at cell $P$ is obtained by fitting a linear field to the values at neighbouring cell centres $Q$. For cells adjacent to the interface, however, a neighbouring cell on the opposite side of the phase boundary would introduce a value from a different phase, corrupting the gradient estimate.

\begin{figure}[htbp]
		\centering
		\includegraphics[width=0.65\linewidth]{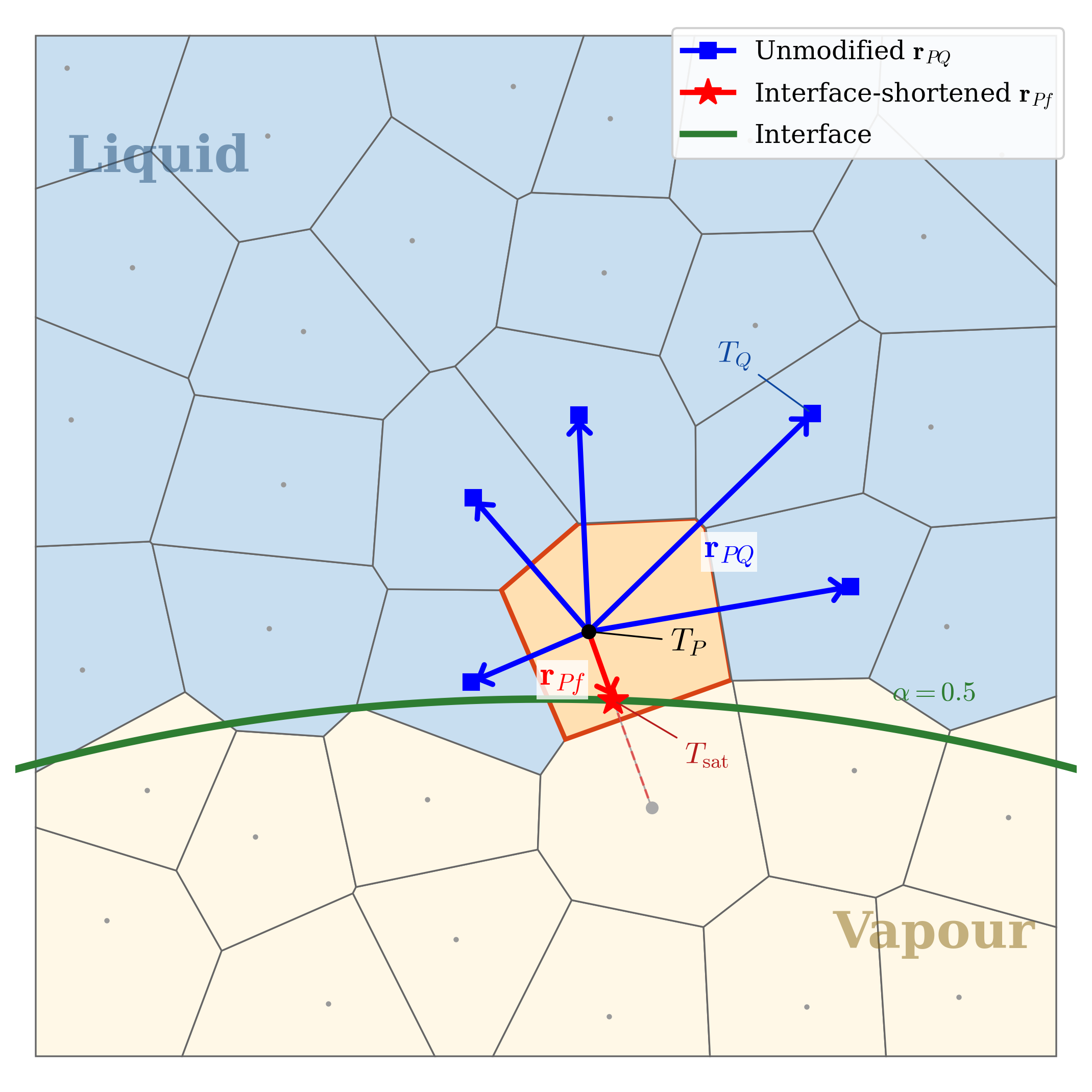}
		\caption{Interface-modified gradient stencil on a polyhedral mesh. For the highlighted cell~$P$, displacement vectors to same-phase (liquid) neighbours (blue arrows, $\mathbf{r}_{PQ}$) remain unmodified. Where the cell-centre connection crosses the interface, the vector is shortened to the intersection point $\mathbf{x}_f$ (red arrow, $\mathbf{r}_{Pf}$) and the neighbour's temperature is replaced by $T_\mathrm{sat}$.}
		\label{fig:gradient_stencil}
	\end{figure}

	To address this, the gradient stencil is modified at the interface. When the volume fraction field indicates that a face between cells $P$ and $Q$ is crossed by the interface---that is, $(\alpha_P - 0.5)(\alpha_Q - 0.5) < 0$---the displacement vector $\mathbf{r}_{PQ}$ is replaced by a shortened vector $\mathbf{r}_{Pf} = \mathbf{x}_f - \mathbf{x}_P$, where $\mathbf{x}_f$ is estimated by linear interpolation of the volume fraction along the cell-centre connection:
	\begin{equation}
		\mathbf{x}_f = \mathbf{x}_P + \frac{|\alpha_Q - 0.5|}{|\alpha_P - \alpha_Q|} \, (\mathbf{x}_Q - \mathbf{x}_P).
		\label{eq:front_interpolation}
	\end{equation}
	The scalar value at this point is set to $T_\mathrm{sat}$, replacing the cross-interface neighbour's temperature $T_Q$. This modification ensures that the least-squares gradient is computed using only same-phase information, with the interface temperature imposed as a boundary condition. Figure~\ref{fig:gradient_stencil} illustrates this stencil modification: blue arrows denote unmodified displacement vectors to same-phase neighbours, while the red arrow shows the shortened vector to the interface intersection point where $T = T_\mathrm{sat}$.

	\paragraph{Temperature gradient extrapolation.}
	Although a single temperature field is solved, the mass transfer rate requires
	temperature gradients from each phase evaluated at the interface. These are
	obtained by extrapolating bulk temperature gradients towards the interface
	along the normal direction, which is done by solving an advection equation in
	pseudo-time:
	\begin{equation}
		\frac{\partial \phi}{\partial \tau}
		+ \mathrm{sgn}(\alpha - 0.5) \, (\mathbf{n}_\mathrm{int} \cdot \nabla \phi) = 0,
		\label{eq:extrapolation}
	\end{equation}
	where $\phi$ is the temperature gradient component being extrapolated, $\tau$
	is pseudo-time, and $\mathbf{n}_\mathrm{int}$ is the interface unit normal
	pointing from vapour to liquid. The sign function ensures that information
	propagates from each bulk phase towards the interface. This procedure is
	applied to each gradient component separately to obtain $\nabla T_l$ and
	$\nabla T_v$ at interface cells. The equation is discretised using a first-order upwind scheme and advanced for up to 120 pseudo-time iterations. The pseudo-time increment $\Delta\tau$ is set to the minimum face-to-face distance across the entire mesh, $\Delta\tau = \min_f |\mathbf{d}_f|$, which ensures stability of the explicit integration. The iteration terminates early when the maximum change between successive iterations falls below $10^{-3}$~K/m (absolute tolerance, corresponding to less than $0.01\%$ of the typical interface gradient of ${\sim}10^4$~K/m), which is typically achieved within 10--30 iterations for well-resolved interfaces. The results are insensitive to the maximum iteration count provided it exceeds the number of cells spanning the thermal boundary layer (typically 5--15 cells).

	\paragraph{Heat source/sink.}
	When liquid transitions to vapour, the process absorbs energy proportional to the mass transfer rate and latent heat of vaporisation $h_{lv}$. Conversely, condensation releases energy. The volumetric energy source due to phase change is:
	\begin{equation}
		\dot{Q}_{\mathrm{phase}} = -\dot{m}\, h_{lv},
		\label{eq:latent_source}
	\end{equation}
	where $\dot{m}$ is the mass transfer rate per unit volume, positive for evaporation and negative for condensation. The mass transfer rate is computed only in cells where the isoap library identifies a non-zero interfacial area ($A_\mathrm{int} > 0$), and is set to zero elsewhere. This naturally restricts the phase-change source to a narrow band of 1--2 cells around the interface, consistent with a sharp-interface approach. The source term enters the VOF equation (Eq.~\eqref{eq:vof_equation}) and the continuity equation (Eq.~\eqref{eq:continuity}) explicitly: $\dot{m}$ is evaluated from the temperature field at the current time level and added to the right-hand side.

	In the energy equation, however, the latent heat exchange is treated implicitly through an interface condition rather than a volumetric source. For each face that crosses the interface, the standard diffusive coupling between the two neighbouring cells is removed (the face is ``detached''), and each cell is instead coupled to the saturation temperature through an interface heat transfer coefficient:
	\begin{equation}
		a_f = \frac{k_\phi \, A_{\mathrm{int},f}}{d_f},
		\label{eq:interface_htc}
	\end{equation}
	where $k_\phi$ is the thermal conductivity of the phase in which the cell resides, $A_{\mathrm{int},f}$ is the interface area projected onto the face, and $d_f$ is the normal distance from the cell centre to the $\alpha$-interpolated interface position (Eq.~\eqref{eq:front_interpolation}), consistent with the interface intersection used in the gradient stencil modification. This coefficient is added to the matrix diagonal and contributes $a_f \, T_\mathrm{sat}$ to the right-hand side, which creates an implicit Robin-type condition that pulls the cell temperature toward~$T_\mathrm{sat}$. The coupling strength scales as $1/d_f$: cells close to the interface experience a strong pull toward saturation, while cells farther away are only weakly affected. This self-adjusting mechanism naturally maintains interface-adjacent cell temperatures close to~$T_\mathrm{sat}$, a consequence that is analysed further in~\ref{app:gradient_magnitude}.
	
	\subsection{Solution algorithm}
	\label{sec:solution_algorithm}
	At each time step, the discrete operators for the governing equations are assembled into sparse matrices. Face-based contributions (diffusive and convective fluxes) and volumetric source terms (body forces, phase-change sources, boundary conditions) are collected into a coefficient matrix and right-hand side vector for each equation. The solution procedure splits into two stages: the VOF equation is solved first to update the interface and material properties, then momentum, pressure, and energy are solved iteratively until convergence.
	
	In practice, a VOF-based boiling solver executes the following steps at each time step:
	\begin{enumerate}
		\item \textbf{VOF advection and interface reconstruction.}
		The liquid–vapour volume fraction field is updated by solving the VOF transport equation with interfacial mass-transfer source terms. An isosurface of the interface is then reconstructed (using the \texttt{isoap} library) to provide accurate interface curvature, surface-tension forces, and geometric data for phase-change modelling.
		
		\item \textbf{Mixture-property update.}
		All cell-centred mixture properties---density, viscosity, specific heat, thermal conductivity---are recomputed from the updated volume fraction.
		
		\item \textbf{Outer iteration loop.}
		Momentum, pressure, and energy are solved together in an outer iteration loop~\cite{ferziger_computational_2020}. Each iteration proceeds as follows:
		\begin{enumerate}
			\item The momentum equations are assembled and solved implicitly (optionally with under-relaxation) to obtain a provisional velocity field incorporating viscous, body, and surface-tension forces.
			\item Face mass fluxes are reconstructed via Rhie–Chow interpolation, a pressure-correction equation is assembled and solved, and both the pressure field and cell velocities are corrected.
			\item If the PISO algorithm is selected, a fixed number of additional corrector steps follow: the momentum equations are re-assembled and solved explicitly (without invoking the linear solver or applying under-relaxation), the pressure-correction equation is re-solved, and the velocity is corrected again. This makes PISO well suited for transient flows with a small time increment.
			\item The mass transfer rate is recomputed from the current temperature gradients at the interface, and the energy equation is solved with the updated latent-heat source term and the latest velocity field.
		\end{enumerate}
		For the SIMPLE algorithm, the outer loop repeats steps (a)--(d) until momentum, continuity, and energy residuals fall below prescribed tolerances. For PISO, the same loop is used, with the additional correctors in step~(c), which accelerates convergence of the pressure--velocity coupling within each outer iteration.
	\end{enumerate}

	Within each outer iteration, the linear systems arising from the discretised momentum, energy, and pressure-correction equations are solved by preconditioned Krylov subspace methods (inner iterations). The conjugate gradient (CG) method is used for the symmetric pressure-correction system, and the bi-conjugate gradient (BiCG) method for the non-symmetric momentum and energy systems, both with incomplete Cholesky preconditioning. The inner iterations continue until the normalised residual falls below a prescribed tolerance or a maximum iteration count is reached.

	\paragraph{Time step.} A fixed time step $\Delta t$ is used throughout each simulation. Since the surface tension force is treated explicitly in the CSF method, $\Delta t$ should be chosen to satisfy the capillary time step constraint~\cite{brackbill_continuum_1992}:
	\begin{equation}
		\Delta t \;<\; \min\!\bigl(\Delta t_\sigma,\, \Delta t_\mathrm{CFL}\bigr), \qquad
		\Delta t_\sigma = \sqrt{\frac{(\rho_l + \rho_v)\, \Delta x^3}{4\pi\sigma}},
		\label{eq:capillary_dt}
	\end{equation}
	where $\Delta x$ is the characteristic cell size and $\Delta t_\mathrm{CFL}$ is the Courant--Friedrichs--Lewy limit. The capillary constraint ensures that capillary waves are resolved in time and prevents their spurious amplification; it is not enforced automatically by the solver and must be verified by the user when selecting $\Delta t$. For the VOF advection, the solver computes the interface Courant number at each time step. When this number exceeds a prescribed threshold (0.25 by default), the VOF equation is automatically subcycled with smaller internal time steps while the remaining equations advance with the full $\Delta t$.
	
	\FloatBarrier
	\section{Validation and verification cases}
	\label{sec:validation}
	The solver is validated against three canonical phase-change benchmarks \cite{tanguy_benchmarks_2014}: the Stefan problem, Sucking problem, and Scriven problem. All three cases use water--steam thermophysical properties at atmospheric pressure, summarised in Table~\ref{tab:material_properties}.

	\begin{table}[htbp]
		\centering
		\caption{Thermophysical properties used in the benchmark cases (water--steam at atmospheric pressure).}
		\label{tab:material_properties}
		\begin{tabular}{llll}
			\hline
			\textbf{Property}           & \textbf{Vapour}          & \textbf{Liquid}          & \textbf{Unit}        \\
			\hline
			Density, $\rho$             & 0.597                    & 958.4                    & kg/m$^3$             \\
			Dynamic viscosity, $\mu$    & $1.22 \times 10^{-5}$    & $2.82 \times 10^{-4}$    & Pa$\cdot$s           \\
			Thermal conductivity, $k$   & 0.025                    & 0.679                    & W/(m$\cdot$K)        \\
			Specific heat, $c_p$        & 2030                     & 4216                     & J/(kg$\cdot$K)       \\
			Latent heat, $h_{lv}$       & \multicolumn{2}{c}{$2.26 \times 10^{6}$}              & J/kg                 \\
			Surface tension, $\sigma$   & \multicolumn{2}{c}{0.059}                              & N/m                  \\
			\hline
		\end{tabular}
	\end{table}

	\subsection{One-dimensional benchmark problems: Stefan and Sucking}

\begin{figure}[htbp]
		\centering
		\includegraphics[width=0.7\linewidth]{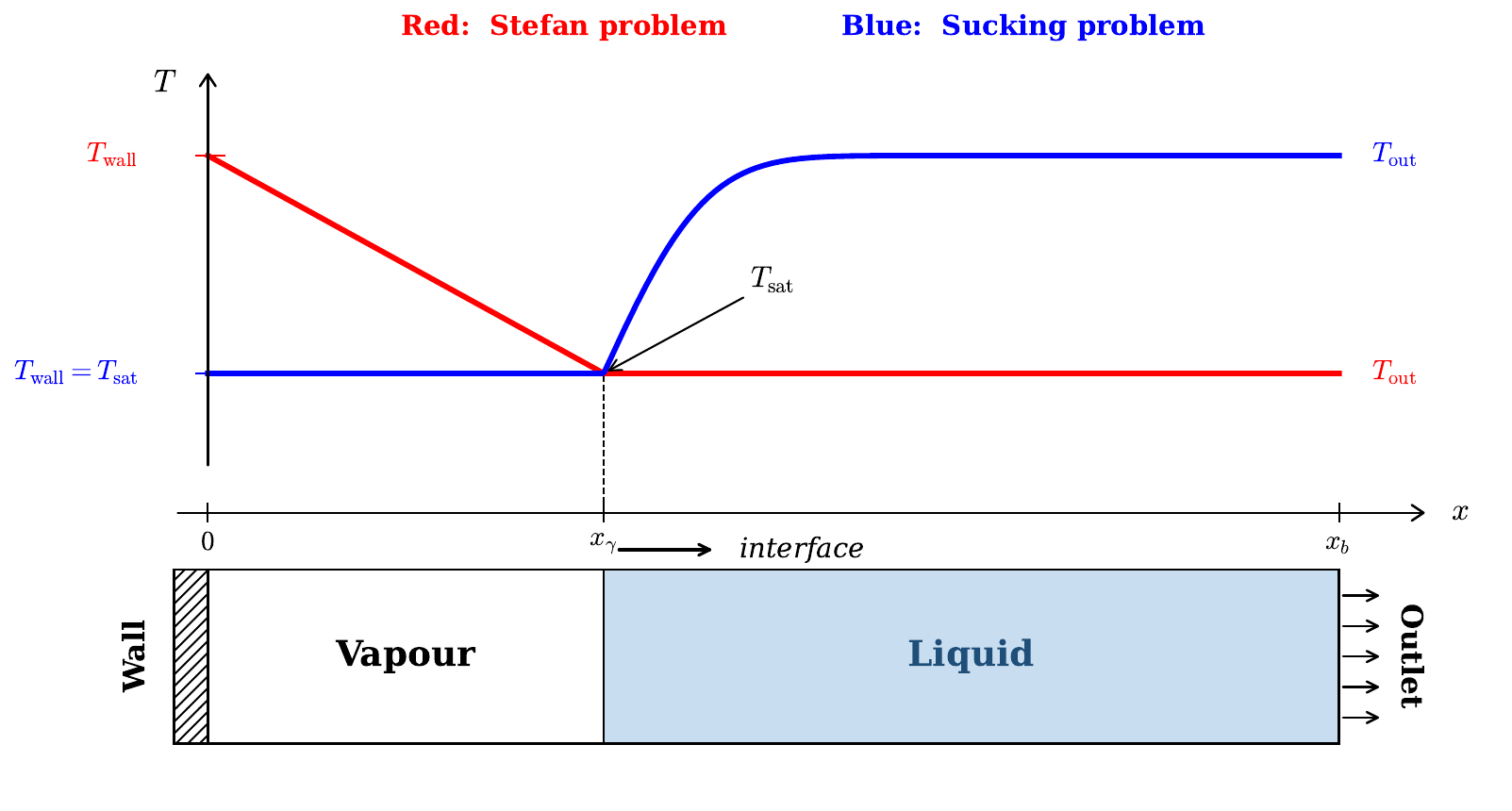}
		\caption{Schematic of the one-dimensional Stefan and Sucking problems. The red curve shows the temperature profile for the Stefan problem (linear in the vapour phase, constant in the liquid), while the blue curve shows the Sucking problem (constant in the vapour phase, increasing in the liquid). Both profiles meet at $T_\mathrm{sat}$ at the interface.}
		\label{fig:stefan_sucking_schematic}
	\end{figure}

	Figure~\ref{fig:stefan_sucking_schematic} illustrates the one-dimensional configuration shared by both the Stefan and Sucking problems. In each case, a planar vapour--liquid interface is located at position $x_\gamma$, separating the vapour phase (adjacent to the wall) from the liquid phase (extending to the outlet at $x_b$). In the Stefan problem, the wall is superheated ($T_\mathrm{wall} > T_\mathrm{sat}$), and a linear temperature gradient in the vapour drives phase change at the interface, where the temperature equals $T_\mathrm{sat}$. In the Sucking problem, the wall is held at $T_\mathrm{sat}$ and the liquid temperature increases from $T_\mathrm{sat}$ at the interface towards the outlet.
	
	The Stefan problem is a classical benchmark for conduction-driven phase change. A superheated wall drives heat conduction through a vapour layer to a planar liquid--vapour interface at saturation temperature; the heat arriving at the interface supplies the latent heat of vaporisation, causing the interface to advance into the quiescent liquid. Because the liquid remains at $T_\mathrm{sat}$, the problem reduces to vapour-phase conduction and admits the self-similar solution $x_\gamma(t) = 2\xi\sqrt{\alpha_v t}$, where $\alpha_v = k_v/(\rho_v c_{p,v})$ is the vapour thermal diffusivity. The growth constant $\xi$ is obtained from the transcendental equation $\xi\,\exp(\xi^2)\,\mathrm{erf}(\xi) = \mathrm{Ste}/\sqrt{\pi}$, with the Stefan number $\mathrm{Ste} = c_{p,v}\,\Delta T / h_{lv}$. The computational domain spans 2~mm in the streamwise ($x$) direction with a cross-section of $10 \times 10$~mm (periodic in $y$ and $z$), discretised using $101 \times 6 \times 6$ equidistant cells. The wall at $x = 0$ is held at a superheat of $\Delta T = 10$~K above $T_\mathrm{sat}$, the outlet at $x = x_b$ has zero heat flux, and periodic conditions are applied in the transverse directions. The initial vapour layer thickness is 0.5~mm, with the liquid initialised at $T_\mathrm{sat}$. The simulation runs for 1800 time steps with $\Delta t = 5 \times 10^{-3}$~s.
	
	The Sucking problem extends the Stefan problem by introducing a coupled velocity field induced by the phase change itself. Here the wall is held at $T_\mathrm{wall} = T_\mathrm{sat}$ (no wall superheat), while the liquid far from the interface is superheated to $T_\infty = T_\mathrm{sat} + 5$~K. Heat conducts from the superheated liquid to the interface, driving evaporation. The large density ratio between liquid and vapour ($\rho_l/\rho_v \approx 1605$) causes the newly formed vapour to expand, which induces a velocity field that draws liquid towards the interface---hence the name. This coupling between the thermal and velocity fields makes the Sucking problem considerably more challenging than the purely conductive Stefan problem: the solver must simultaneously resolve the temperature boundary layer in the liquid, the vapour expansion, and the moving interface. The computational domain is 8~mm long with a cross-section of $1 \times 1$~mm ($401 \times 6 \times 6$ cells, periodic in $y$ and $z$). The wall at $x = 0$ is held at $T_\mathrm{sat}$, the outlet at $x = 8$~mm has zero heat flux, and the initial conditions comprise a vapour layer extending 2.2~mm from the wall at $T_\mathrm{sat}$, with the liquid initialised at $T_\infty$. The simulation uses $\Delta t = 2.5 \times 10^{-5}$~s for 24\,000 time steps (0.6~s physical time).
	
\begin{figure}[htbp]
		\centering
		\includegraphics[width=\linewidth]{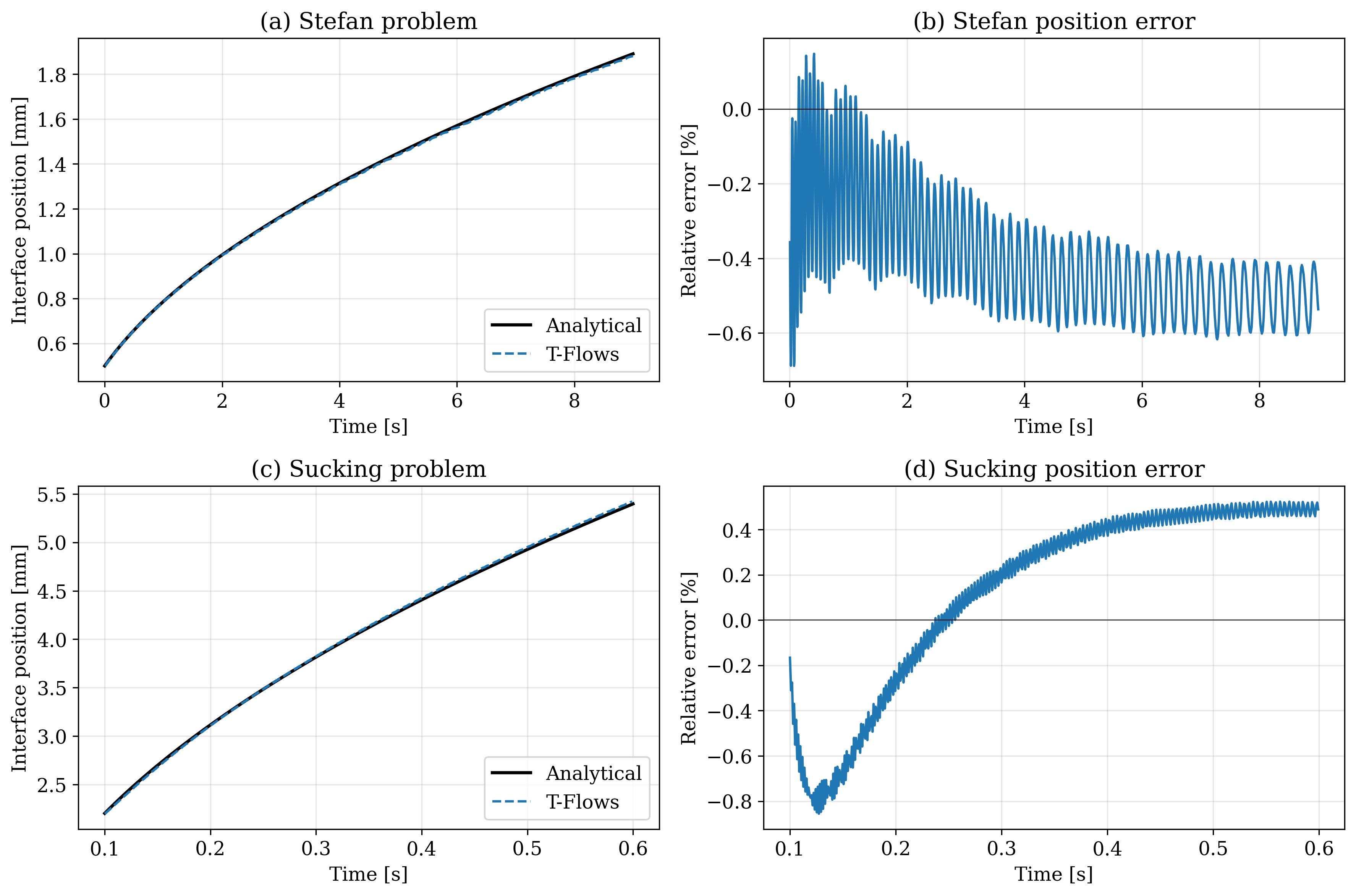}
		\caption{One-dimensional evaporation benchmarks: interface position and relative error for (a,~b)~the Stefan problem (101 cells) and (c,~d)~the Sucking problem (401 cells). The relative error remains below $1\%$ for both cases throughout the simulation.}
		\label{fig:stefan_sucking_results}
	\end{figure}

	Figure~\ref{fig:stefan_sucking_results} compares the analytical solutions with the T-Flows simulation results for both problems. The interface position shows good agreement for both cases, and the relative error remains below $1\%$ throughout the simulation, with mean absolute errors of $0.4\%$ for both problems. Because the domain is one-dimensional, the vapour volume is directly proportional to the interface position. The small, drift-free error therefore confirms that mass is accurately conserved over the full integration time.
	
\begin{figure}[htbp]
		\centering
		\includegraphics[width=\linewidth]{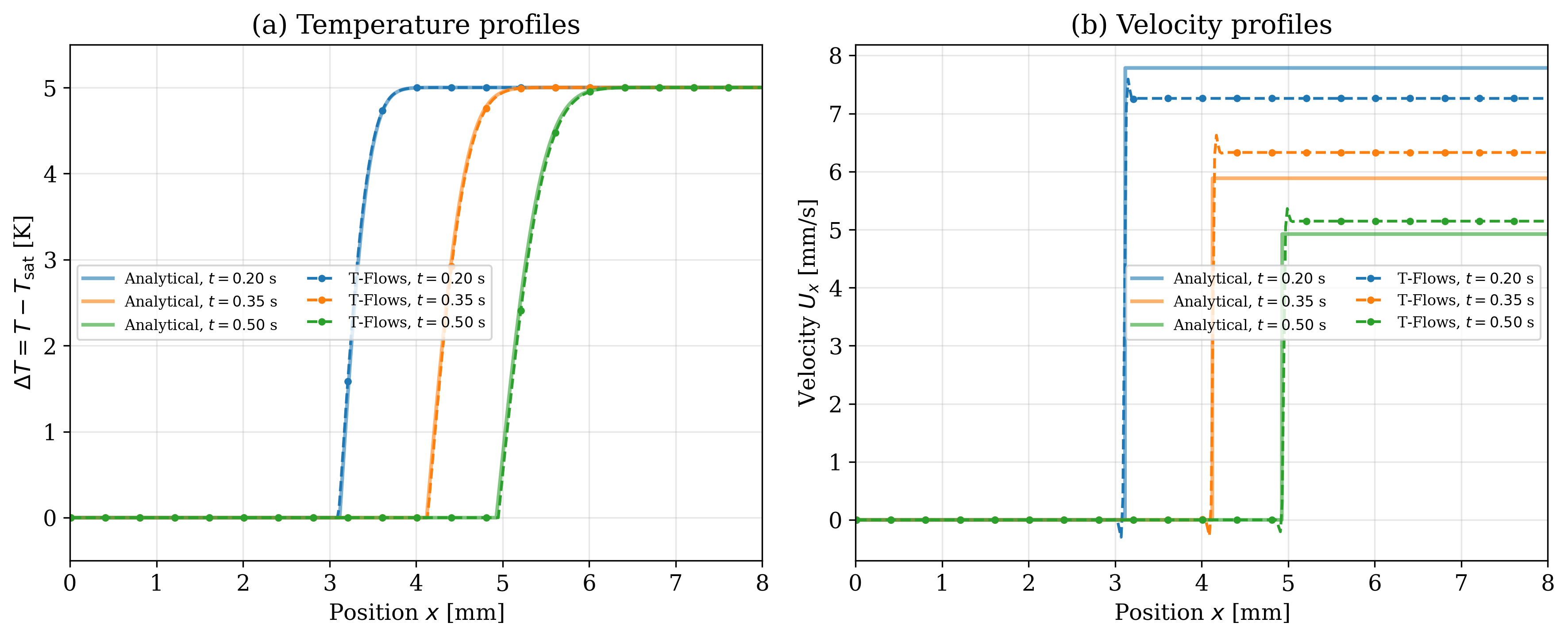}
		\caption{Sucking problem: instantaneous (a)~temperature and (b)~velocity profiles at $t = 0.2$, $0.35$, and $0.5$~s, comparing T-Flows results (dashed) with analytical solutions (solid).}
		\label{fig:sucking_profiles}
	\end{figure}

	To further validate the Sucking problem beyond interface position, Figure~\ref{fig:sucking_profiles} presents instantaneous temperature and velocity profiles at three time instants ($t = 0.2$, $0.35$, and $0.5$~s). The temperature 
	profiles exhibit the expected error-function shape in the liquid thermal boundary layer, with $T = T_\mathrm{sat}$ in the vapour region and a smooth transition to $T_\infty$ in the bulk liquid. The velocity profiles confirm that   
	the liquid ahead of the interface is pushed at a uniform velocity $u_l = \dot{x}_\mathrm{int}(1 - \rho_v/\rho_l)$, while the vapour remains nearly stationary due to the wall boundary condition. The temperature field shows
	near-perfect agreement with the analytical solution, whereas the velocity field exhibits a small discrepancy in the immediate vicinity of the interface. This difference arises because the temperature is continuous across the phase 
	boundary (both phases are at $T_\mathrm{sat}$ at the interface), which makes it inherently smooth and well-resolved. The velocity, by contrast, undergoes a sharp discontinuity at the interface due to the large density ratio $\rho_l /    
	\rho_v \approx 1605$: the analytical solution prescribes a step function from $u = 0$ in the vapour to a finite value in the liquid. Although the interface is geometrically reconstructed as a sharp plane within each cell, the      
	velocity field is represented by cell-centred values and the jump is necessarily spread over the cells straddling the reconstructed interface. Despite this finite-resolution smoothing of the velocity jump, the overall agreement    
	confirms accurate coupling between the thermal and velocity fields.

	\subsection{Three-dimensional benchmark problem}

	The Scriven problem~\cite{scriven_dynamics_1959} describes the growth of a spherical vapour bubble in an infinite superheated liquid. Unlike the one-dimensional Stefan and Sucking problems, the Scriven problem is inherently three-dimensional and involves a curved interface with associated surface tension effects. The analytical solution predicts that the bubble radius grows proportionally to the square root of time, $R(t) = 2\beta\sqrt{\alpha_l t}$, where $\beta$ is the growth constant determined by the Jakob number and $\alpha_l$ is the liquid thermal diffusivity. It therefore tests spherical interface geometry, curvature computation, and three-dimensional heat transfer simultaneously.

\begin{figure}[htb!]
		\centering
		\includegraphics[width=0.85\linewidth]{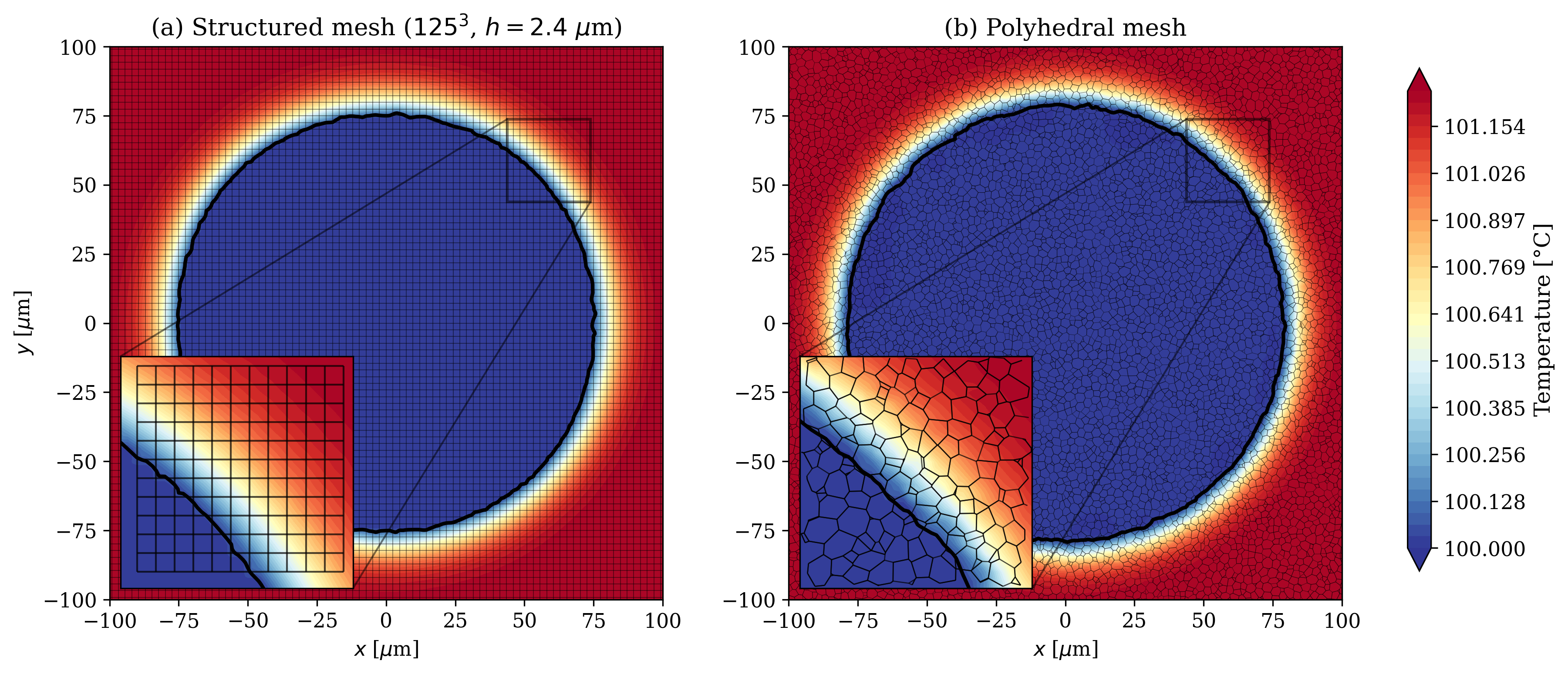}
		\caption{ Scriven problem at $t = 0.63$~ms: temperature field and mesh cross-sections through the bubble centre for (a)~the structured hexahedral ($125^3$) and (b)~the polyhedral mesh. The solid black line shows the numerical interface ($\alpha = 0.5$). Insets magnify the interface region, showing the thermal boundary layer resolved by approximately 4--6 cells in the radial direction. The full computational domain is a cube with a side length $300$ $\mu$m.}
		\label{fig:scriven_meshes}
	\end{figure}

	The computational domain is a cube of side length $300~\mu$m centred at the origin, with all boundaries set as outflow. A spherical vapour bubble with an initial radius $R_0 = 50~\mu$m and a temperature of  $T_\mathrm{sat} = 100$~°C is placed at the centre of the domain, and the liquid is initialised with the analytical Scriven temperature profile corresponding to $\Delta T = T_\infty - T_\mathrm{sat} = 1.25$~K of superheat. The initial thermal boundary layer thickness around the bubble is 13.5~$\mu$m. The Jakob number $\mathrm{Ja} = \rho_l c_{p,l} \Delta T / (\rho_v h_{lv}) \approx 3.7$ measures the ratio of sensible heat in
	the superheated liquid to the latent heat of vaporisation and governs the thermal boundary layer thickness relative to the bubble radius, $\delta_T / R \sim 1/\mathrm{Ja}$. The moderate value of Ja in the present case creates a boundary layer resolved by approximately 4--6 cells on the $125^3$ mesh (Figure~\ref{fig:scriven_meshes}). To assess mesh sensitivity, simulations were performed on four structured hexahedral meshes of $75^3$, $100^3$, $125^3$, and $150^3$ cells (cell sizes of $4.0$, $3.0$, $2.4$, and $2.0~\mu$m, respectively), as well as four polyhedral meshes of $75^3$, $100^3$, $125^3$, and $150^3$ cells, obtained by dualising tetrahedral meshes generated with GMSH~\cite{geuzaine_gmsh_2009}.
	Temporal sensitivity is first assessed on the polyhedral $75^3$ mesh at $\Delta t = 1$, $2$, and $10~\mu$s (Section~\ref{sec:temporal}), establishing that $\Delta t \leq 2~\mu$s is temporally converged while $\Delta t = 10~\mu$s introduces a uniform $8$--$10\%$ error from the sequential operator splitting in the solution algorithm (Section~\ref{sec:solution_algorithm}). The polyhedral mesh convergence study accordingly uses $\Delta t = 2~\mu$s. For the structured meshes, all four refinement levels completed only at $\Delta t = 10~\mu$s; at smaller time steps, the gradient overestimate mechanism (Section~\ref{sec:struct_hex}) produces a mild instability on the finer meshes. Moreover, the operator-splitting bias at $\Delta t = 10~\mu$s partially compensates the gradient overshoot, yielding the closest agreement with the analytical solution. The temporal analysis confirms this bias is mesh-independent and does not affect convergence rates.
	Figure~\ref{fig:scriven_meshes} shows cross-sections of the structured hexahedral and polyhedral meshes used for the Scriven problem, together with the initial bubble interface. The polyhedral mesh, generated by dualising a tetrahedral mesh, produces cells of varying size and irregular shape, which presents a more challenging test for the interface reconstruction and curvature computation algorithms.

\begin{figure}[htb!]
		\centering
		\includegraphics[width=\linewidth]{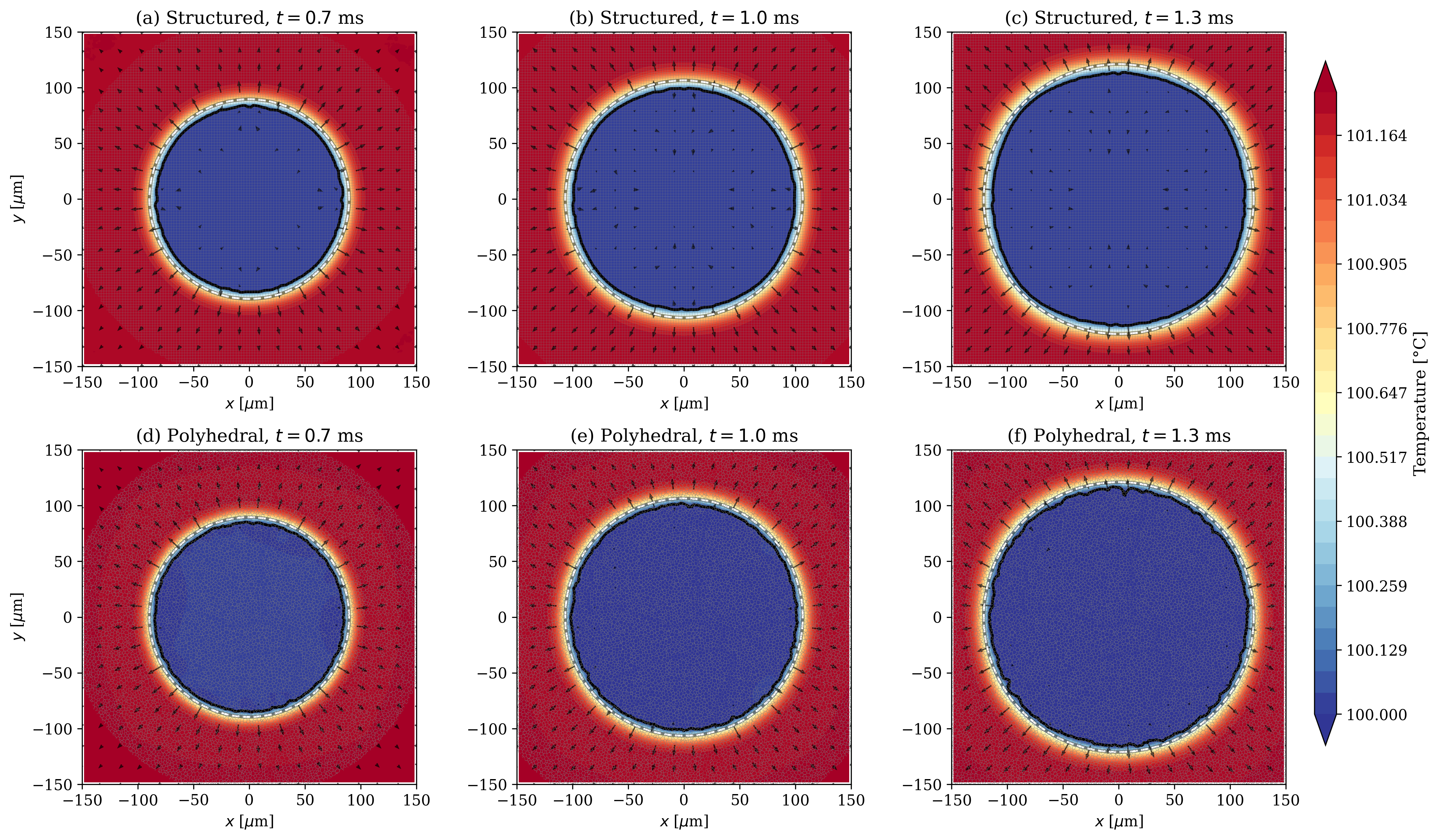}
		\caption{Scriven problem: temperature contours and velocity vectors on a cross-section through the bubble centre at $t = 0.73$, $1.03$, and $1.33$~ms, for (a--c)~the structured $125^3$ mesh and (d--f)~the polyhedral $125^3$ mesh. The black line shows the numerical interface ($\alpha$~$= 0.5$), and the white dashed line the analytical Scriven radius.}
		\label{fig:scriven_fields}
	\end{figure}

	Figure~\ref{fig:scriven_fields} presents the instantaneous temperature distribution and velocity vectors on a cross-section through the bubble centre at three time instants, comparing results on the structured hexahedral and polyhedral meshes. The thin thermal boundary layer surrounding the bubble is clearly visible on both mesh types, and the radial velocity field demonstrates the vapour expansion driving the liquid outward. As the bubble grows, the thermal boundary layer thins relative to the bubble radius, consistent with the $\sqrt{t}$-growth predicted by Scriven's solution. A close inspection of the structured mesh results (top row) reveals that the bubble assumes a slightly non-spherical shape, with the interface radius approximately $5\%$ larger along the grid diagonals than along the grid-aligned directions. This four-fold anisotropy arises from the orientation dependence of the temperature gradient computation on Cartesian meshes and is analysed in detail in Section~\ref{sec:gradient_anisotropy}. The polyhedral $125^3$ mesh (bottom row), by contrast, produces a nearly isotropic bubble shape owing to the random orientation of cell faces, which eliminates preferential alignment with any coordinate direction.

\begin{figure}[htb!]
		\centering
		\includegraphics[width=\linewidth]{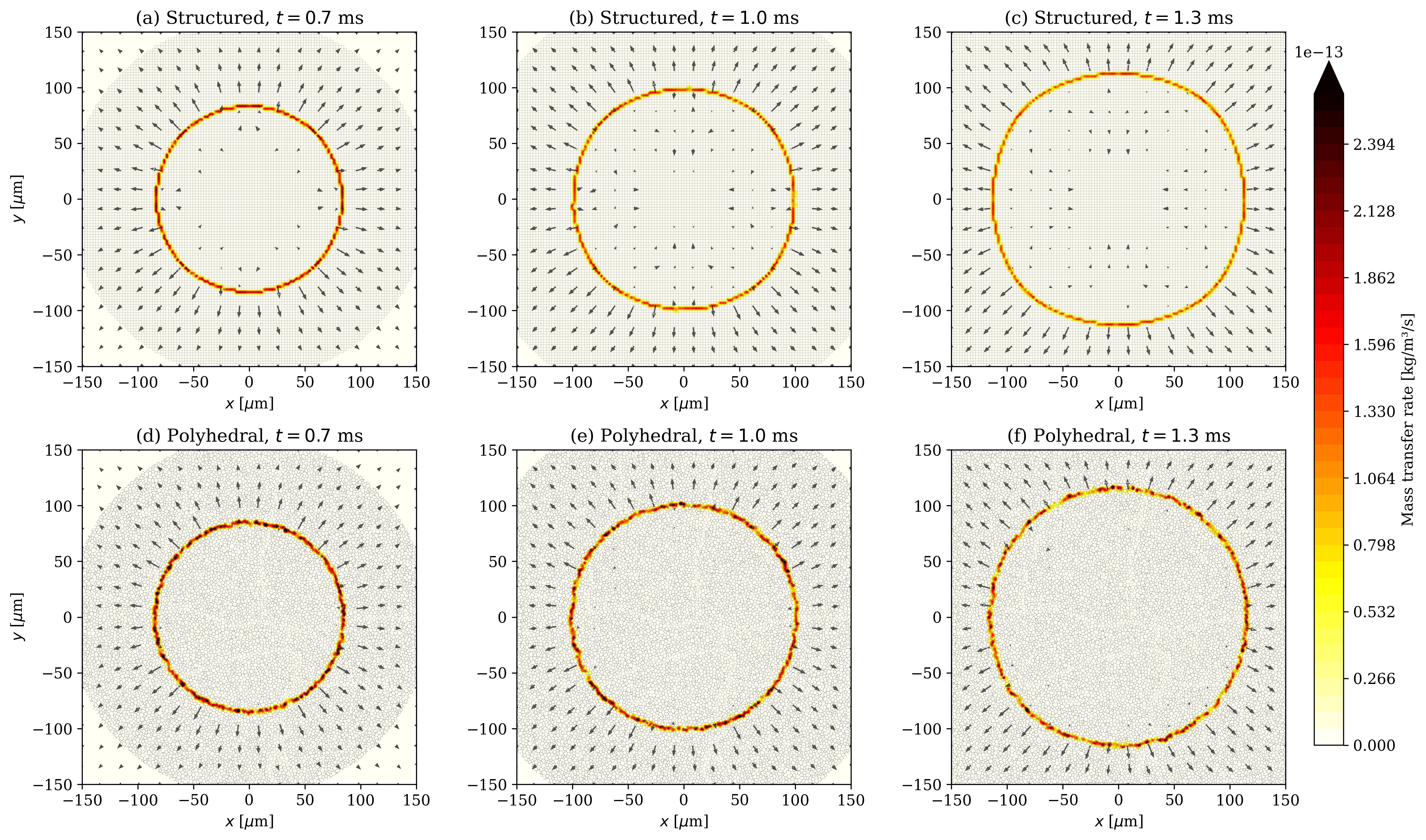}
		\caption{Scriven problem: mass transfer rate $\dot{m}$ on the $z = 0$ cross-section at three time instants, for (a--c)~the structured $125^3$ mesh and (d--f)~the polyhedral $125^3$ mesh. The mass transfer is localised at the interface and shows near-uniform angular distribution on both mesh types, despite the gradient anisotropy on the structured mesh.}
		\label{fig:scriven_mdot}
	\end{figure}

	Figure~\ref{fig:scriven_mdot} shows the computed mass transfer rate $\dot{m}$ on the $z = 0$ cross-section at three time instants. On both the structured and polyhedral meshes, the mass transfer is concentrated in a narrow band at the interface, as expected from the localised nature of the phase-change model. The mass transfer distribution appears nearly uniform around the bubble on both mesh types, with comparable magnitude. However, a close inspection of the structured mesh results reveals the same subtle four-fold anisotropy visible in the temperature field (Figure~\ref{fig:scriven_fields}). The origin and implications of this anisotropy are analysed in detail in Section~\ref{sec:gradient_anisotropy}.

	\subsubsection{Temporal sensitivity and operator splitting}
	\label{sec:temporal}

	Before examining mesh convergence, the sensitivity of the solution to the time-step size is assessed by running the polyhedral $75^3$ case at $\Delta t = 1$, $2$, and $10~\mu$s (Figure~\ref{fig:scriven_sensitivity}(b)). The difference between $\Delta t = 1~\mu$s and $\Delta t = 2~\mu$s is small, not exceeding $1.1\%$ at $t = 1.5$~ms, confirming that $\Delta t = 2~\mu$s is temporally converged for this problem. Increasing the time step to $\Delta t = 10~\mu$s introduces a systematic underprediction of $8$--$10\%$ at later times.

	The primary source of this temporal error is the sequential operator splitting in the solution algorithm (Section~\ref{sec:solution_algorithm}): the VOF equation is solved first, using the mass-transfer rate computed from the previous time step's temperature field to advect the interface. Within the subsequent outer iteration loop, the mass transfer rate is recomputed from the current temperature gradients and the energy equation is solved with the updated source terms, so the thermal field and mass transfer are iteratively coupled within each time step. The interface position, however, is fixed after the VOF solve and therefore lags by $\mathcal{O}(\Delta t)$. This error is mesh-independent: the same ten-fold time-step increase produces a nearly identical deficit on the structured $75^3$ mesh (Table~\ref{tab:scriven_dt}). The polyhedral mesh convergence study below accordingly uses $\Delta t = 2~\mu$s; the structured mesh convergence study uses $\Delta t = 10~\mu$s (the smallest time step for which all four refinement levels completed), with the understanding that this shifts all errors by a uniform $8$--$10\%$ without affecting convergence rates. No polyhedral mesh produces an overshoot above $R_\mathrm{Scriven}$ at any time step.

	\begin{table}[htb!]
		\centering
		\caption{Operator-splitting error: difference in $R/R_\mathrm{Scriven}$ between $\Delta t = 10~\mu$s and $\Delta t = 1~\mu$s on two mesh topologies. The nearly identical values confirm that the temporal error is mesh-independent.}
		\label{tab:scriven_dt}
		\begin{tabular}{ccc}
			\hline
			$t$ [ms] & $\Delta(R/R_\mathrm{Scriven})$, polyhedral $75^3$
			         & $\Delta(R/R_\mathrm{Scriven})$, structured $75^3$ \\
			\hline
			0.5 & $-5.4\%$ & $-5.9\%$ \\
			1.0 & $-8.2\%$ & $-8.9\%$ \\
			1.5 & $-10.0\%$ & $-10.5\%$ \\
			\hline
		\end{tabular}
	\end{table}

	\subsubsection{Mesh convergence}

	Figure~\ref{fig:scriven_sensitivity}(a) shows the normalised bubble radius $R/R_\mathrm{Scriven}$ on the four polyhedral meshes, all computed with $\Delta t = 2~\mu$s (temporally converged per the analysis above). The solution converges monotonically toward the analytical value with mesh refinement: at $t = 1.0$~ms, the relative radius error decreases from $7.1\%$ on the polyhedral $75^3$ mesh to $3.9\%$ on the polyhedral $150^3$ mesh (Table~\ref{tab:scriven_poly_convergence}). At no point does any polyhedral mesh produce a radius that exceeds the analytical value. All meshes exhibit a characteristic initial error growth as the thermal boundary layer thins relative to the cell size, followed by a gradual recovery as the bubble grows and the interface region becomes better resolved. The remaining discrepancy is attributed to the under-resolution of the thin thermal boundary layer ($\delta_T \sim \sqrt{\alpha_l t} \approx 10$--$13~\mu$m at $t = 0.5$--$1.0$~ms), which spans only 3--5 cells even on the finest mesh.

	\begin{table}[htb!]
		\centering
		\caption{Polyhedral mesh convergence for the Scriven problem: relative radius error at two fixed times ($\Delta t = 2~\mu$s for all meshes).}
		\label{tab:scriven_poly_convergence}
		\begin{tabular}{lccc}
			\hline
			\textbf{Mesh} & $h$ [$\mu$m] & Error at $t = 0.5$~ms [\%] & Error at $t = 1.0$~ms [\%] \\
			\hline
			Polyhedral $75^3$  & 4.0 & 8.4 & 7.1 \\
			Polyhedral $100^3$ & 3.0 & 6.8 & 6.0 \\
			Polyhedral $125^3$ & 2.4 & 5.3 & 4.3 \\
			Polyhedral $150^3$ & 2.0 & 4.5 & 3.9 \\
			\hline
		\end{tabular}
	\end{table}

\begin{figure}[htb!]
		\centering
		\begin{subfigure}[b]{0.48\textwidth}
			\centering
			\includegraphics[width=\linewidth]{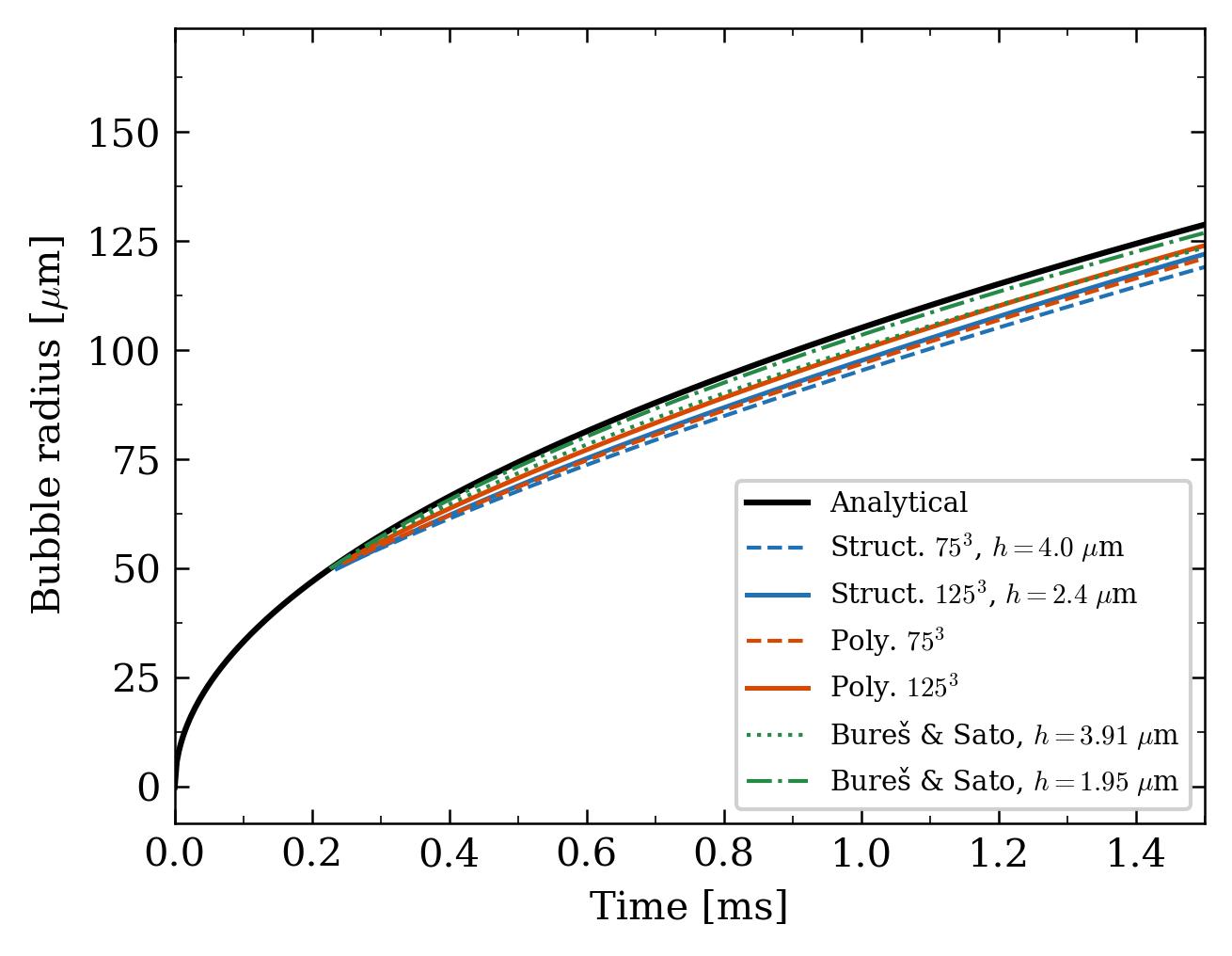}
			\caption{}
			\label{fig:scriven_mass}
		\end{subfigure}
		\hfill
		\begin{subfigure}[b]{0.48\textwidth}
			\centering
			\includegraphics[width=\linewidth]{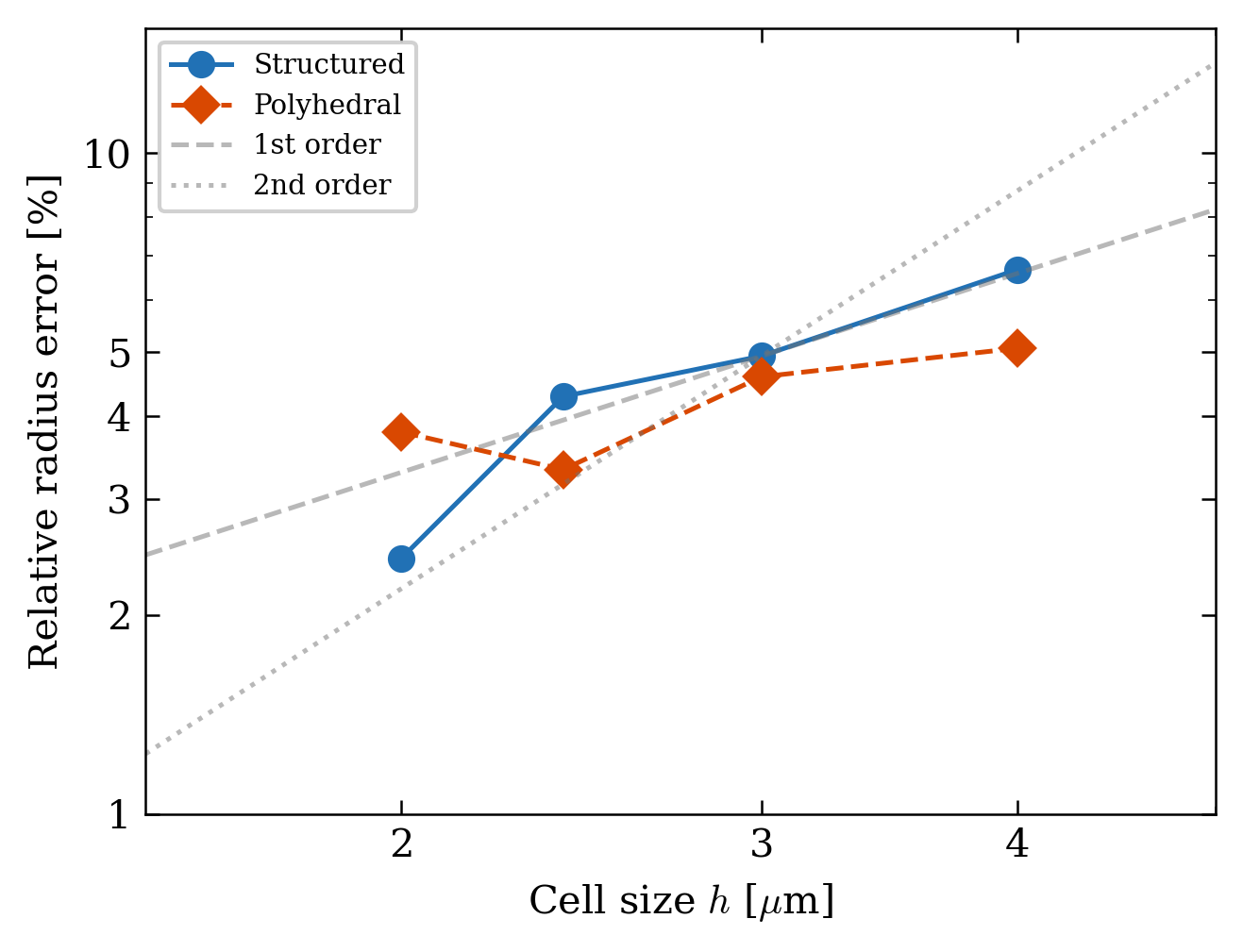}
			\caption{}
			\label{fig:scriven_convergence}
		\end{subfigure}
		\caption{Scriven problem: (a)~bubble radius growth on the $75^3$ and $125^3$ structured ($\Delta t = 10~\mu$s) and polyhedral ($\Delta t = 2~\mu$s) meshes, compared with the analytical Scriven solution and with reference data from Bure\v{s} and Sato~\cite{bures_direct_2021}; (b)~relative radius error at $t = 1.5$~ms versus cell size for structured and polyhedral meshes, with first- and second-order reference slopes.}
		\label{fig:scriven_results}
	\end{figure}

	Figure~\ref{fig:scriven_mass} compares the bubble radius on the $75^3$ and $125^3$ structured and polyhedral meshes with the analytical Scriven solution and with reference data from Bure\v{s} and Sato~\cite{bures_direct_2021} obtained with a geometric VOF method on two Cartesian grids. For the structured meshes, the bubble radius is computed from the VOF-integrated volume as $R_\mathrm{vol} = (3V/4\pi)^{1/3}$; for the polyhedral meshes, from the area-weighted mean distance of the reconstructed front element centroids to the bubble centre. All meshes underpredict the analytical radius at comparable resolution, with errors consistent with those reported by Bure\v{s} and Sato. The structured mesh results use $\Delta t = 10~\mu$s; the qualitatively different behavior observed at temporally converged time steps is analysed in Section~\ref{sec:struct_hex}. The CICSAM advection scheme solves the VOF transport equation in conservative form, ensuring that the global volume fraction is conserved to solver tolerance at each time step. The bubble volume used to compute $R_\mathrm{vol}$ is obtained by summing $(1-\alpha)\,V_c$ over all cells; any mass conservation error would manifest as a drift in $R_\mathrm{vol}$ independent of the thermal growth, which is not observed.

	Figure~\ref{fig:scriven_convergence} shows the convergence at $t = 1.5$~ms for both structured and polyhedral meshes. The error decreases monotonically with refinement, with an observed order of convergence $p \approx 1.3$. The sub-quadratic rate is expected: the CICSAM advection scheme is formally first-order at the interface, and the thermal boundary layer is marginally resolved (3--5 cells) on all four meshes. The error magnitudes are comparable to those reported by Bure\v{s} and Sato~\cite{bures_direct_2021} for a geometric VOF method on Cartesian grids at the same superheat. On polyhedral meshes, where the coherent gradient anisotropy (Section~\ref{sec:gradient_anisotropy}) is absent, the convergence rate is comparable ($p \approx 1.2$--$1.4$ between adjacent refinement levels), suggesting that the gradient anisotropy does not significantly degrade the observed order.

	\subsubsection{Effect of mesh topology}
	\label{sec:struct_hex}

\begin{figure}[htb!]
		\centering
		\includegraphics[width=\textwidth]{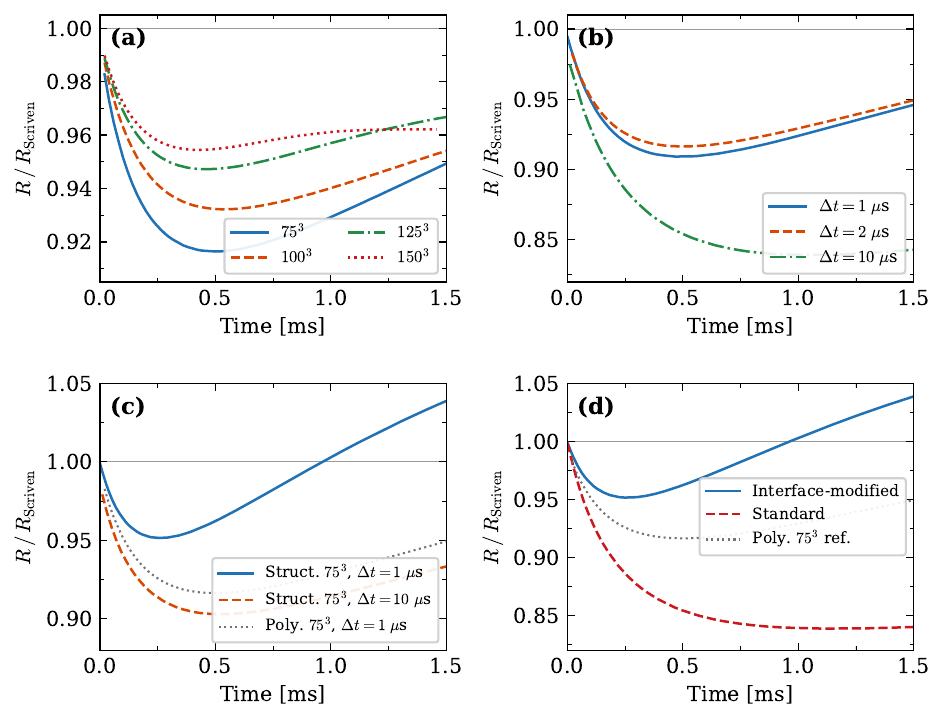}
		\caption{Scriven bubble growth: sensitivity analysis. Normalised bubble radius $R/R_\mathrm{Scriven}$ versus time for (a)~polyhedral mesh convergence at $\Delta t = 2~\mu$s; (b)~time-step sensitivity on the polyhedral $75^3$ mesh at $\Delta t = 1$, $2$, and $10~\mu$s; (c)~structured hexahedral $75^3$ mesh at $\Delta t = 1$ and $10~\mu$s showing overshoot; (d)~ablation test comparing interface-modified and standard LSQ gradient on the structured $75^3$ mesh at $\Delta t = 1~\mu$s. Grey dotted lines in panels (c) and (d) show the polyhedral $75^3$, $\Delta t = 1~\mu$s result for reference.}
		\label{fig:scriven_sensitivity}
	\end{figure}

	At temporally converged time steps, the structured hexahedral mesh reveals a qualitatively different behavior from the polyhedral mesh. Figure~\ref{fig:scriven_sensitivity}(c) shows the bubble radius on the structured $75^3$ mesh at $\Delta t = 1$ and $10~\mu$s. At $\Delta t = 1~\mu$s, the solution initially underpredicts the analytical radius, then reverses and overshoots $R/R_\mathrm{Scriven} = 1$ at $t \approx 0.96$~ms, reaching $+3.9\%$ by $t = 1.5$~ms. At $\Delta t = 10~\mu$s, the operator-splitting bias (Section~\ref{sec:temporal}) masks this overshoot, producing a curve that remains below the analytical value---which is why the convergence comparison in Figure~\ref{fig:scriven_mass} shows no overshoot. No polyhedral mesh produces $R > R_\mathrm{Scriven}$ at any resolution or time step.

	Despite the overshoot, the interface-modified gradient remains essential. An ablation test using the standard (unmodified) LSQ gradient on the same mesh (Figure~\ref{fig:scriven_sensitivity}(d)) produces a $16\%$ underprediction ($R/R_\mathrm{Scriven} \approx 0.84$), far worse than the ${\sim}4\%$ overshoot with the interface-modified gradient. The standard gradient cannot resolve the steep temperature variation across the thermal boundary layer without knowledge of the interface location; the interface modification corrects this by imposing $T_\mathrm{sat}$ at the known intersection point. The overshoot on the hexahedral mesh arises from two topology-dependent mechanisms, both originating from the interaction between the interface-modified stencil and the regular cell arrangement. First, the interface modification systematically amplifies the mean gradient magnitude: the mass transfer heat sink maintains the cell-centre temperature close to $T_\mathrm{sat}$, which converts the symmetric central difference into an effectively one-sided estimate and amplifies the computed gradient by a factor of approximately $1.4$ (Eq.~\eqref{eq:boost_limit} in~\ref{app:gradient_magnitude}, evaluated at $d/h \approx 0.5$). Because the unmodified gradient already underestimates the true interface gradient by ${\sim}30\%$ on these meshes, this amplification overcompensates the resolution deficit, driving $R > R_\mathrm{Scriven}$. Second, the diagonal structure of $\mathbf{G}_P$ on a Cartesian mesh allows the interface modification to perturb each gradient component independently, producing a coherent four-fold anisotropy that distorts the bubble shape and is analysed in detail below. On polyhedral meshes, the off-diagonal coupling in $\mathbf{G}_P$ and the richer stencil (${\sim}14$ faces vs.\ $6$) damp both effects: the gradient amplification is reduced to a factor of $1.15$--$1.2$ (insufficient to overcome the baseline underestimate), and the random face orientations prevent coherent accumulation of anisotropy. This is consistent with the analysis of Diskin and Thomas~\cite{Diskin2012} and Syrakos~et~al.~\cite{Syrakos2017}, who showed that LSQ gradient accuracy on structured meshes depends critically on stencil symmetry---a symmetry that the interface modification necessarily breaks.

	The improved isotropy of polyhedral meshes comes at a cost. \ref{app:computational_cost} reports that polyhedral cells are approximately $1.8\times$ more expensive per cell due to the larger stencil (${\sim}14$ faces vs.\ $6$) and the associated gradient and interpolation overhead. Moreover, the ``$75^3$ polyhedral'' mesh contains $319$k cells compared to $422$k for the structured $75^3$, because the tetrahedral-to-dual conversion does not preserve the cell count. The effective resolution per cell may also differ: the richer stencil provides more gradient information per cell but each cell samples a larger volume. A cost-equivalent comparison---matching total wall-clock time rather than nominal resolution---would require a coarser polyhedral mesh, and the relative advantage might narrow. Nevertheless, the elimination of coherent anisotropy is a qualitative benefit that cannot be achieved on structured meshes regardless of resolution.

	\subsubsection{Temperature gradient anisotropy}
	\label{sec:gradient_anisotropy}

	The structured mesh produces two distinct effects on the temperature gradient at the interface: (i)~a coherent four-fold angular anisotropy that distorts the bubble shape (analysed in this section and Section~\ref{sec:struct_hex_origin}), and (ii)~a systematic overestimate of the mean gradient magnitude that drives the radius overshoot above $R_\mathrm{Scriven}$ (derived in~\ref{app:gradient_magnitude}). Both originate from the interaction between the interface-modified stencil and the regular Cartesian topology, but they are independent mechanisms with different consequences. The mechanism is verified independently using a synthetic two-dimensional model (Figure~\ref{fig:stencil_polar}).

	To quantify the anisotropic bubble shape observed on the structured mesh, we examine the temperature gradient field in detail. The temperature gradient at the interface directly determines the mass transfer rate through Eq.~\eqref{eq:mass_transfer}, and any systematic angular dependence in the gradient magnitude translates to non-uniform phase change and deformation of the bubble from its theoretically spherical shape.

\begin{figure}[htb!]
		\centering
		\includegraphics[width=\linewidth]{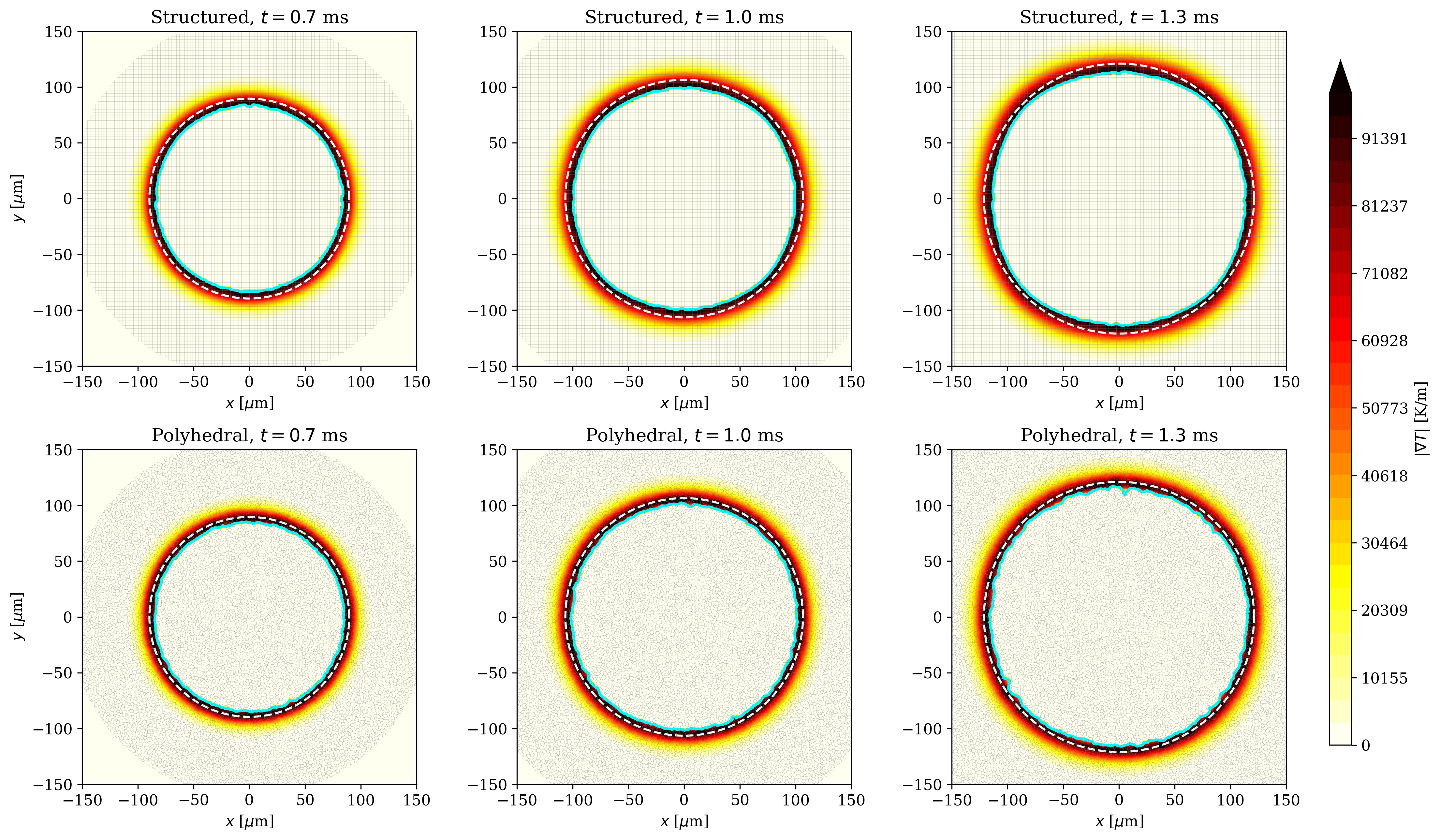}
		\caption{Scriven problem: temperature gradient magnitude $|\nabla T|$ on the $z = 0$ cross-section at three time instants, for the structured $125^3$ mesh (top) and the polyhedral $125^3$ mesh (bottom). The cyan line shows the $\alpha = 0.5$ interface and the white dashed line the analytical Scriven radius. Note the four-fold pattern on the structured mesh, absent on the polyhedral mesh.}
		\label{fig:scriven_grad_T}
	\end{figure}

	Figure~\ref{fig:scriven_grad_T} shows the magnitude of the temperature gradient $|\nabla T|$ on the $z = 0$ cross-section for both mesh types at three time instants. On the structured mesh (top row), the gradient magnitude exhibits a clear four-fold symmetry around the interface: the gradient is strongest along the mesh diagonals ($45^\circ$, $135^\circ$, $225^\circ$, $315^\circ$) and weakest along the coordinate axes ($0^\circ$, $90^\circ$, $180^\circ$, $270^\circ$). This pattern persists and intensifies as the bubble grows, because the thermal boundary layer thins and the gradient magnitude increases, amplifying the absolute difference between the diagonal and axial directions. On the polyhedral mesh (bottom row), the gradient magnitude is distributed much more uniformly around the bubble circumference, with no discernible preferential directions.

\begin{figure}[htb!]
		\centering
		\includegraphics[width=\linewidth]{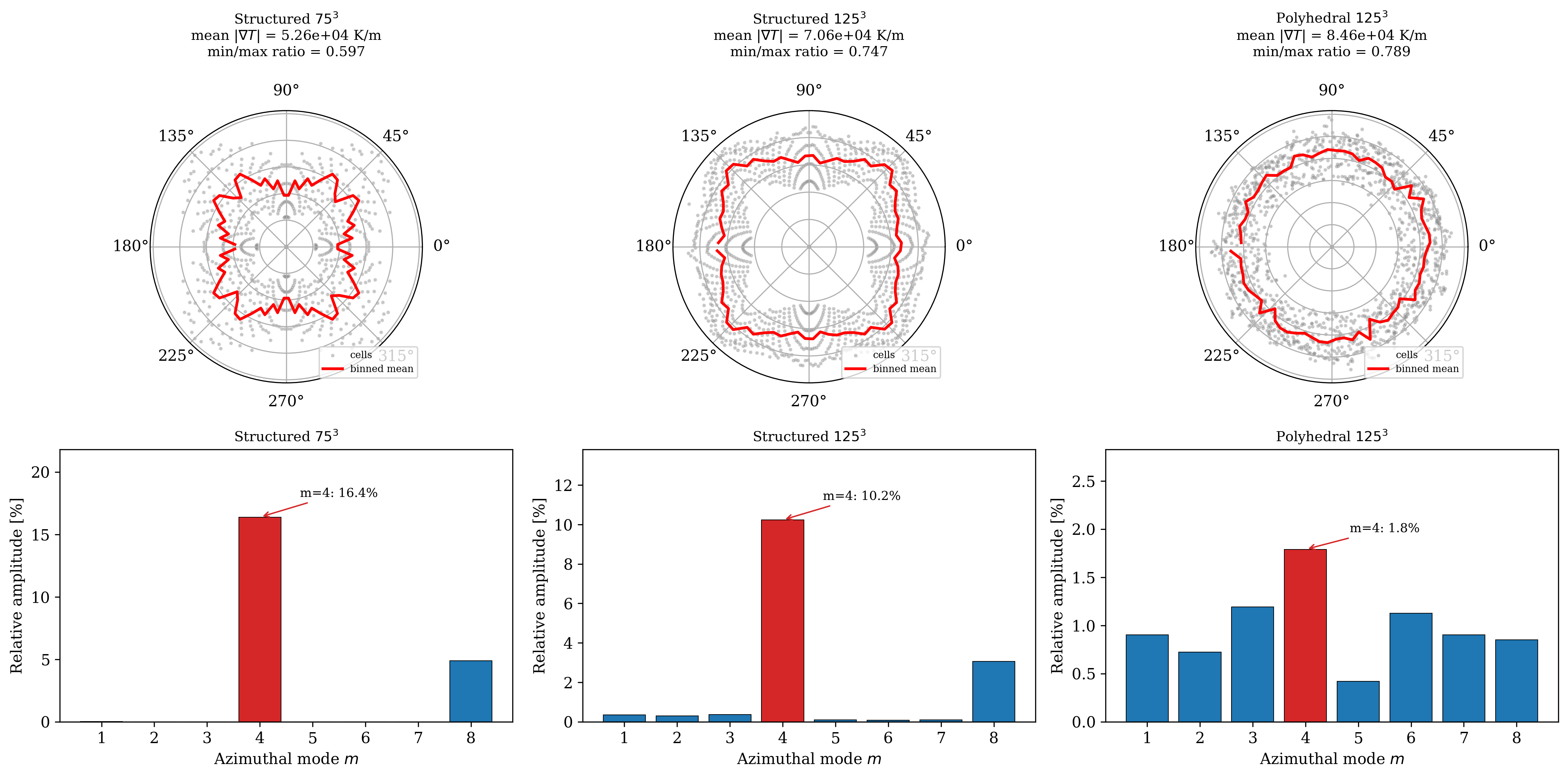}
		\caption{Polar distribution of temperature gradient magnitude $|\nabla T|$ around the bubble interface (top row) and azimuthal Fourier decomposition (bottom row) at $t = 1.23$~ms for (a,\,d)~structured $75^3$ mesh, (b,\,e)~structured $125^3$ mesh, and (c,\,f)~polyhedral $125^3$ mesh. Grey dots represent individual cell values; red lines show binned angular means. The Fourier spectra show the amplitude of each azimuthal mode $m$ as a percentage of the mean gradient; the $m = 4$ bar is highlighted in red. The structured meshes concentrate $16\%$ ($75^3$) and $10\%$ ($125^3$) of the mean in the coherent $m = 4$ mode, showing clear convergence with refinement, whereas the polyhedral mesh distributes the scatter quasi-randomly across modes. Note that the Fourier amplitude scales differ between panels (d)--(f) to resolve the smaller amplitudes on finer and polyhedral meshes.}
		\label{fig:scriven_grad_polar}
	\end{figure}

	The angular dependence of the gradient is quantified in Figure~\ref{fig:scriven_grad_polar}, which shows polar plots of $|\nabla T|$ for interface-adjacent cells at $t = 1.23$~ms, together with the azimuthal Fourier decomposition of the binned gradient magnitude. Three cases are compared: the structured $75^3$ and $125^3$ meshes, and the polyhedral $125^3$ mesh. A conventional measure of anisotropy---the minimum-to-maximum ratio of the binned gradient---gives $0.60$ (structured $75^3$), $0.75$ (structured $125^3$), and $0.79$ (polyhedral $125^3$).

	However, the Fourier decomposition reveals an important qualitative difference. On the structured $75^3$ mesh, the azimuthal mode $m = 4$ dominates the spectrum with an amplitude of approximately $16\%$ of the mean gradient, while all other modes remain below $1\%$. On the finer structured $125^3$ mesh, the $m = 4$ amplitude decreases to approximately $10\%$, which indicates convergence with mesh refinement, but the four-fold pattern remains the dominant mode. This means nearly all angular variation is concentrated in a single coherent pattern aligned with the Cartesian mesh topology. On the polyhedral mesh, the $m = 4$ amplitude drops to approximately $2\%$, and the remaining scatter is distributed across modes $m = 1$--$8$ at comparable amplitudes ($0.4$--$1.2\%$ each); the angular variation is quasi-random with no dominant preferential direction. This distinction is important: a coherent mode accumulates systematically over time and produces the persistent bubble deformation observed on structured meshes, whereas random angular scatter tends to cancel, which preserves the spherical shape on polyhedral meshes.

	\subsubsection{Origin of the gradient anisotropy}
	\label{sec:struct_hex_origin}

\begin{figure}[htb!]
		\centering
		\includegraphics[width=\linewidth]{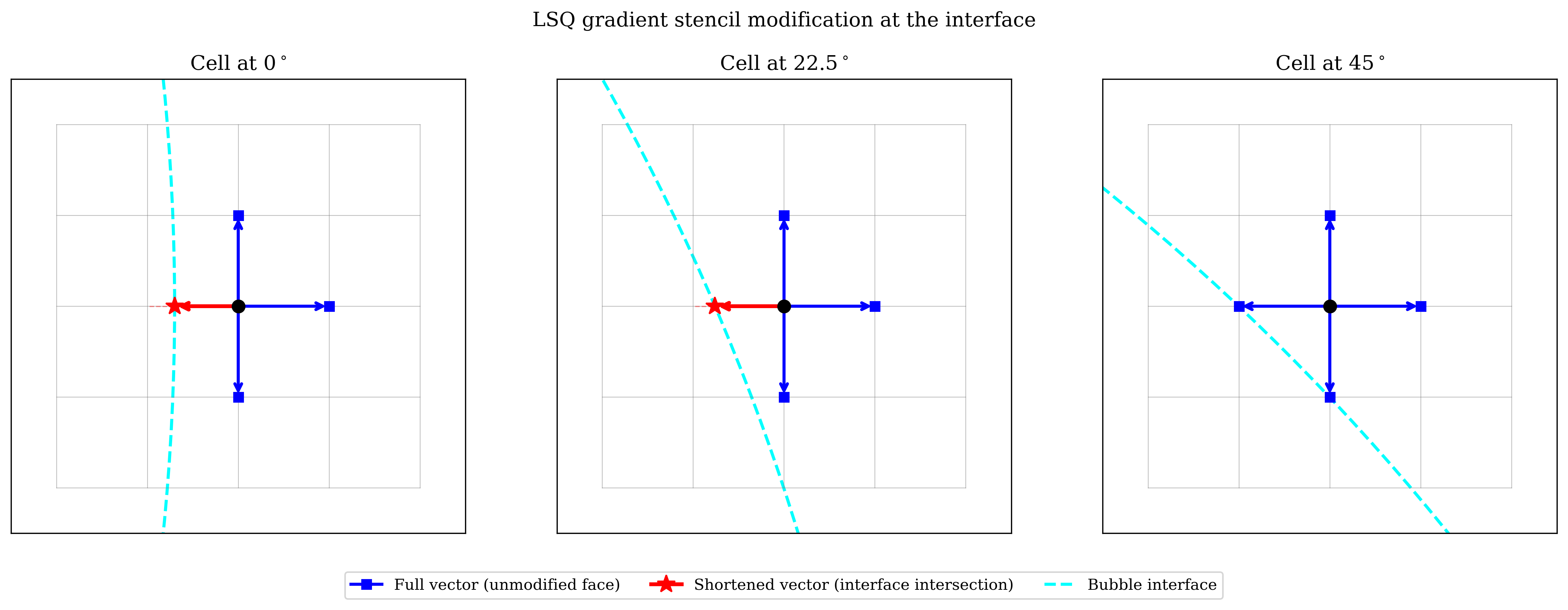}
		\caption{Illustration of the interface-modified gradient stencil on a structured mesh at three angular positions relative to the spherical interface. Blue arrows: unmodified displacement vectors to neighbour cells. Red arrows: displacement vectors shortened to the interface intersection points (red stars). At $0^\circ$, only one face is intersected and the stencil remains nearly symmetric; at $45^\circ$, two faces are intersected with different shortening ratios, which creates an asymmetric gradient matrix.}
		\label{fig:stencil_geometry}
	\end{figure}

	The four-fold anisotropy originates from the interaction between the interface-modified least-squares gradient stencil (\ref{app:LeastSQGR}) and the regular topology of a Cartesian mesh. Figure~\ref{fig:stencil_geometry} illustrates this mechanism for three representative cell orientations. On a structured hexahedral mesh, each cell has six face-connected neighbours. When a cell face is intersected by the vapour--liquid interface, the displacement vector in the gradient matrix $\mathbf{G}_P$ (defined in \ref{app:LeastSQGR}) is shortened from the cell centre to the interface intersection point (rather than the neighbour cell centre), and the corresponding scalar value is replaced by $T_\mathrm{sat}$.

	We observe that the number of intersected faces and the degree of shortening depend on the cell's angular position relative to the bubble. At $\theta = 0^\circ$ (along a coordinate axis), typically only the face in the radial direction is intersected, and the shortened displacement vector remains aligned with the radial direction, preserving the symmetry of the $3 \times 3$ gradient matrix. At $\theta = 45^\circ$ (along the mesh diagonal), two faces are intersected with potentially different shortening ratios. This creates a highly asymmetric gradient matrix $\mathbf{G}_P$ whose eigenvalue ratio (condition number) differs substantially from the axis-aligned case. When the matrix is inverted to obtain the gradient, the direction-dependent conditioning produces a gradient magnitude that varies with orientation even for a perfectly spherical temperature field.

\begin{figure}[htb!]
		\centering
		\includegraphics[width=\linewidth]{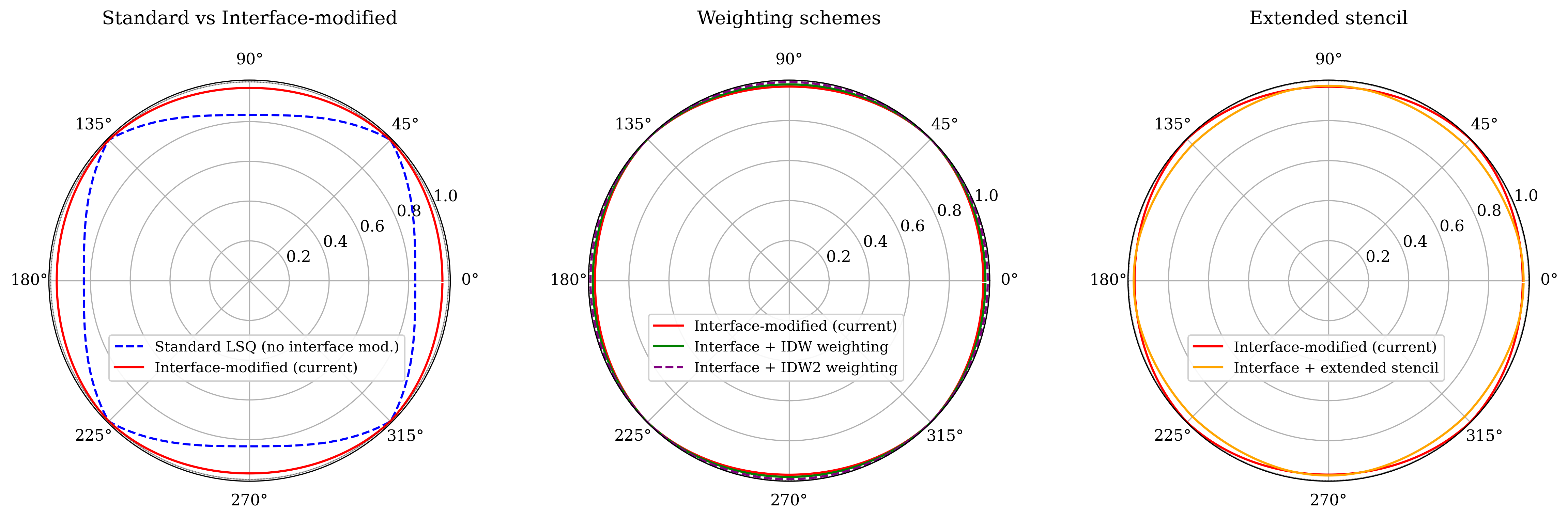}
		\caption{Synthetic analysis of gradient stencil anisotropy: polar plots of the ratio $|\nabla T|_\mathrm{computed}/|\nabla T|_\mathrm{analytical}$ for (a)~standard vs interface-modified gradient, (b)~effect of inverse-distance weighting, and (c)~extended stencil. A circle with radius 1.0 indicates the ideal solution. Both standard and interface-modified gradients exhibit four-fold anisotropy, but with different orientations and magnitudes. Results are computed from an analytical temperature field on a 2D Cartesian mesh.}
		\label{fig:stencil_polar}
	\end{figure}

	To verify this mechanism independently of the CFD simulation, we constructed a synthetic two-dimensional model that applies the interface-modified gradient stencil to an analytical temperature field around a circle. Figure~\ref{fig:stencil_polar} presents the resulting gradient magnitude as a function of angle for different stencil configurations. The standard (unmodified) gradient (blue dashed) already exhibits a four-fold pattern because the stencil includes neighbours on both sides of the interface: at some angles, one or two neighbours lie inside the bubble (where $T = T_\mathrm{sat}$), and the gradient matrix samples across the phase discontinuity without accounting for the interface location. The interface-modified gradient (red) changes the character of this anisotropy by replacing cross-interface neighbours with the intersection point at $T_\mathrm{sat}$, which shortens the displacement vectors in the stencil. The resulting pattern differs from the standard case in both orientation and magnitude, as the degree of shortening depends on the cell's angular position. Panel~(b) shows the effect of inverse-distance weighting (IDW): IDW amplifies the contribution of the shortened interface vectors, which worsens the anisotropy. Panel~(c) shows that extending the stencil to include vertex neighbours partially restores isotropy by reducing the relative weight of the modified faces. The IDW prediction was confirmed by implementing IDW weighting in the full three-dimensional solver: the resulting gradient anisotropy was greater than with the unweighted scheme, which validates the mechanism identified by the two-dimensional model.

	The condition number of the gradient matrix $\mathbf{G}_P$ confirms this mechanism: whereas the standard gradient produces a uniform condition number (the matrix depends only on mesh geometry, which is translationally invariant on a Cartesian mesh), the interface-modified matrix shows pronounced peaks at the cardinal directions where a single face is shortened significantly, varying by approximately a factor of~3 around the interface and directly correlating with the gradient magnitude variation in Figure~\ref{fig:stencil_polar}.

	The anisotropy primarily affects the local gradient magnitude rather than the global mass transfer rate, because the subsequent extrapolation procedure (Eq.~\eqref{eq:extrapolation}) partially compensates by averaging gradient information along the interface normal. Analysis of the computed mass transfer rate $\dot{m}$ shows that the angular variation in $\dot{m}$ is approximately $4\%$, substantially less than the $10$--$16\%$ coherent $m = 4$ variation in gradient magnitude; the extrapolation step thus acts as an effective regulariser. Nevertheless, the residual coherent anisotropy in $\dot{m}$ accumulates over time and produces the observed diagonal elongation of the bubble on structured meshes.

	\FloatBarrier
	\section{Upward co-current annular boiling flow}
	\label{sec:annular}
	Annular flow is one of the most commonly encountered two-phase flow regimes in industrial applications, particularly in steam generators, boiling water reactors (BWRs), and heat exchangers. In upward co-current annular flow, the vapour phase flows through the central core of the channel while a thin liquid film adheres to the wall, with both phases moving in the same direction. This configuration arises naturally during flow boiling as the void fraction increases: the flow pattern transitions from bubbly to slug, churn, and finally annular flow as evaporation progresses along the heated channel \cite{kharangate_review_2017}.
	
	The annular flow regime is of particular importance for nuclear safety analysis because it precedes the critical phenomenon of dryout, where the liquid film becomes depleted and the heated wall loses its cooling capability \cite{nishida_front_2023}. Accurate prediction of the liquid film thickness, entrainment and deposition rates, and the onset of dryout is therefore essential for determining safety margins in reactor design. From a computational perspective, simulating annular flow with phase change presents significant challenges: the thin liquid film must be adequately resolved, and the complex interactions between the film and the droplet-laden vapour core---including interfacial shear, wave dynamics, and mass transfer---must be captured accurately.
	
	We apply the framework to turbulent upward co-current annular flow with evaporation. The simulation uses Large Eddy Simulation (LES) with the Wall-Adapting Local Eddy-viscosity (WALE) subgrid-scale model \cite{nicoud_subgrid-scale_1999} to capture the turbulent transport in both the liquid film and the vapour core. The WALE model is particularly well suited for VOF-coupled simulations because its eddy viscosity, constructed from the traceless symmetric part of the square of the velocity gradient tensor, naturally vanishes in laminar shear flows and near walls without requiring explicit damping functions. Near the liquid--vapour interface, where surface tension forces dominate and the flow is locally laminar, the WALE eddy viscosity therefore tends to zero and does not artificially diffuse the interface or the temperature field. No additional turbulence damping or suppression at the interface is applied. The simulation captures the characteristic features of annular flow, including the wavy liquid film structure and the interaction between the phases. This case serves as a demonstration of capability rather than a full quantitative validation: the simulated physical time (4.9~ms) is substantially shorter than the ${\sim}100$~ms required to reach statistical stationarity \cite{sato_large_2017}, and the results should therefore be interpreted as representing the early transient development of the flow. Qualitative comparison is made with the experimental observations of Barbosa Jr.\ \cite{barbosa_high-speed_2003} and the LES study of Sato and Ni\v{c}eno \cite{sato_large_2017}, who simulated the same configuration for 300~ms.
	

\subsection{Case description and simulation setup}
\label{sec:annular_setup}

The upward co-current annular flow simulation is performed in a cylindrical pipe geometry with an inner radius of $R = 9.5$~mm, an outer radius of $16$~mm, and an axial length of $L = 52$~mm. The liquid film is initialised on the inner wall with a uniform thickness of $\delta_0 = 0.25$~mm, while the vapour phase occupies the central core region. Both phases flow upward in the positive axial direction $y$, with gravity ($g = 9.81$~m/s$^2$) acting downward opposing the flow. To reduce computational cost, only a 45$^\circ$ azimuthal sector (one-eighth of the full domain) is simulated, with symmetry boundary conditions applied at the circumferential boundaries. This follows the approach of the previous LES study by Sato and Ni\v{c}eno~\cite{sato_large_2017}, who simulated a comparable azimuthal sector (approximated as a rectangular Cartesian domain) for their LES of annular boiling flow at identical conditions. Using a sector instead of the full pipe is not a concern because the waves that matter in annular flow are axisymmetric --- they form as rings around the pipe driven by the shear between the fast vapour core and the slow liquid film \cite{sato_large_2017} --- and the radial heat transfer that drives evaporation is fully captured by the sector geometry. This symmetry assumption suppresses non-axisymmetric modes ($m \neq 0, 8, 16, \ldots$), which may play a role in three-dimensional wave breaking and droplet entrainment; however, these effects are secondary to the axisymmetric disturbance waves that dominate mass transfer in the thin-film regime considered here.

At the inlet ($y = 0$), parabolic velocity profiles are prescribed independently in each phase with the bulk velocities listed in Table~\ref{tab:annular_params}. Turbulent velocity fluctuations generated by the synthetic eddy method are superimposed on the mean profiles to promote the development of turbulence within the domain. The inlet temperature is set to $T_\mathrm{sat}$ for the vapour phase and $T_\mathrm{sat} + \Delta T$ for the liquid phase. The volume fraction is $\alpha = 1$ in the film region and $\alpha = 0$ in the vapour core.

The inner wall ($r = R$) is a no-slip boundary held at a constant temperature $T_\mathrm{wall} = T_\mathrm{sat} + \Delta T$; the resulting superheat drives conduction through the liquid film and evaporation at the interface. The outer wall ($r = 16$~mm) is adiabatic with a no-slip condition. Symmetry conditions are imposed on the two azimuthal boundaries of the 45$^\circ$ sector.

At the outlet ($y = L$), a convective boundary condition proposed by Bottaro \cite{bottaro_note_1990} is applied:
\begin{equation}
    \frac{\partial \phi}{\partial t} + U \frac{\partial \phi}{\partial x} = 0,
    \label{eq:convective_bc}
\end{equation}
where $\phi$ represents any transported quantity and $U$ is the mean streamwise velocity. This Sommerfeld-type radiation condition allows disturbances to exit the domain without reflection and improves numerical stability compared to standard Neumann (zero-gradient) conditions when local backflow occurs at the outlet.

The flow field is initialised uniformly with the inlet profiles: the velocity, volume fraction, and temperature fields are set to their respective inlet values throughout the domain. No perturbation is applied to the initial interface position; disturbances develop from the inlet turbulence and the shear-driven instability of the liquid film.

The thermophysical properties of water and steam at 1.9~bar are listed in Table~\ref{tab:annular_params}, together with the simulation parameters and resulting flow conditions. The large density ratio ($\rho_l/\rho_v \approx 877$) is characteristic of boiling systems and poses significant numerical challenges for interface-capturing methods.

\begin{figure}[htbp]
    \centering
    \includegraphics[width=\textwidth]{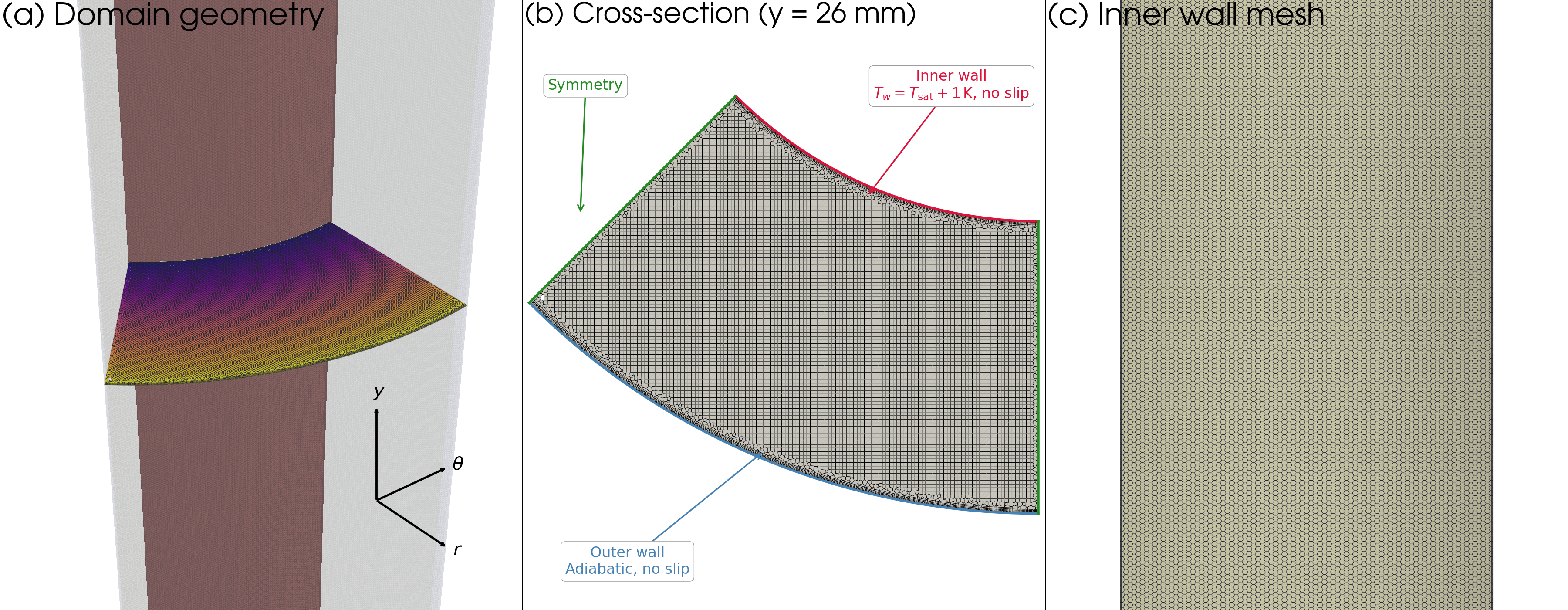}
    \caption{Computational mesh for the annular flow simulation: (a)~perspective view of the 45$^\circ$ azimuthal sector domain showing the inner wall, outer wall, azimuthal symmetry planes, and a cross-sectional slice coloured by radial coordinate; (b)~face-on cross-section at $y = 26$~mm illustrating the polyhedral mesh structure across the annular gap; (c)~detail of the inner wall boundary mesh showing the polyhedral face topology at the heated surface.}
    \label{fig:annular_mesh}
\end{figure}

The computational mesh consists of approximately 7.2 million cells and is illustrated in Figure~\ref{fig:annular_mesh}. The bulk of the annular gap is discretised with hexahedral cells (78\% of the total), while polyhedral cells
(22\%) fill the transition regions between structured and unstructured zones. The mesh quality is characterised by a minimum orthogonal quality of 0.25. Near-wall refinement is applied at the heated inner surface to resolve the thin liquid film and thermal boundary layer. The first cell centre lies at a wall-normal distance of approximately 4~$\mu$m (wall-normal cell width $\sim$8~$\mu$m), and successive layers expand with a geometric ratio of 1.1--1.3, which reaches $\sim$33~$\mu$m at the eighth layer. Beyond this point the mesh transitions from structured hexahedra to polyhedral cells before returning to a uniform hexahedral spacing of $\sim$80~$\mu$m in the vapour core. This grading places approximately 12 radial cell layers across the mean liquid film thickness ($\bar{\delta} \approx 0.19$~mm) and 16 layers across the initial film ($\delta_0 = 0.25$~mm), which is sufficient  
to resolve the wall-normal temperature gradient that governs conduction-limited evaporation. Subgrid-scale stresses are modelled with the Wall-Adapting Local Eddy-viscosity (WALE) model \cite{nicoud_subgrid-scale_1999}. The simulation is advanced with a time step of $\Delta t = 0.1$~$\mu$s, and data are extracted every 10 time steps, which yields a temporal resolution of 1~$\mu$s for the interfacial wave analysis. The total simulation time spans approximately 4.9~ms of physical time, comprising 4904 saved time steps.

\begin{table}[htbp]
\centering
\caption{Simulation parameters and thermophysical properties for the annular flow case.}
\label{tab:annular_params}
\small
\begin{tabular}{lccr}
\toprule
\textbf{Parameter} & \textbf{Liquid} & \textbf{Vapour} & \textbf{Unit} \\
\midrule
\multicolumn{4}{l}{\textit{Geometry \& conditions}} \\
Inner radius, $R$ & \multicolumn{2}{c}{9.5} & mm \\
Outer radius & \multicolumn{2}{c}{16} & mm \\
Axial length, $L$ & \multicolumn{2}{c}{52} & mm \\
Initial film thickness, $\delta_0$ & \multicolumn{2}{c}{0.25} & mm \\
Wall superheat, $\Delta T$ & \multicolumn{2}{c}{1} & K \\
Mean velocity, $U_y$ & 0.6 & 18.5 & m/s \\
\midrule
\multicolumn{4}{l}{\textit{Thermophysical properties (water--steam, 1.9~bar, $T_\mathrm{sat} = 119.6$\,°C)}} \\
Density, $\rho$ & 943.9 & 1.076 & kg/m$^3$ \\
Dynamic viscosity, $\mu$ & $2.35 \times 10^{-4}$ & $1.29 \times 10^{-5}$ & Pa$\cdot$s \\
Thermal conductivity, $k$ & 0.683 & 0.027 & W/(m$\cdot$K) \\
Specific heat, $c_p$ & 4242 & 2112 & J/(kg$\cdot$K) \\
Surface tension, $\sigma$ & \multicolumn{2}{c}{0.0553} & N/m \\
Latent heat, $h_{lv}$ & \multicolumn{2}{c}{$2.21 \times 10^{6}$} & J/kg \\
\midrule
\multicolumn{4}{l}{\textit{Dimensionless numbers (based on $\delta_0$)}} \\
$Re_l = \rho_l U_{y,l} \delta_0 / \mu_l = 602$ & \multicolumn{2}{c}{} & $We_v = \rho_v \Delta U^2 \delta_0 / \sigma = 1.56$ \\
$Fr = U_{y,l}/\sqrt{g\delta_0} = 12.1$ & \multicolumn{2}{c}{} & $Oh = \mu_l / \sqrt{\rho_l \sigma \delta_0} = 0.0021$ \\
\bottomrule
\end{tabular}
\smallskip

\noindent\textit{Note:} Dimensionless numbers are based on the initial film thickness $\delta_0 = 0.25$~mm. By the end of the simulation, the mean film thickness decreases to $\bar{\delta} \approx 0.16$~mm, which would increase $Re_l$ by a factor of $\delta_0/\bar{\delta} \approx 1.6$ and decrease $We_v$ proportionally.
\end{table}
\FloatBarrier
All annular flow simulations were performed on an AMD EPYC-based in-house Paul Scherrer Institute Merlin7 cluster, whereas the benchmark simulations were performed on a local HPC machine. The annular flow simulation, with 7.2~million cells and a time step ($\Delta t = 0.1~\mu$s, CFL~$\approx 0.06$), ran on 256 MPI processes for approximately 1000~hours of wall-clock time (49\,000 time steps). The dominant cost per time step arises from the iterative pressure--velocity coupling (PISO algorithm with PETSc linear solvers) and the VOF advection with interface reconstruction; the pseudo-time temperature extrapolation and mass transfer computation together account for less than 10\% of the total time per step.

\subsection{Spatial structure of the annular flow}
\label{sec:annular_spatial}

\begin{figure}[htbp]
    \centering
    \includegraphics[width=0.85\textwidth]{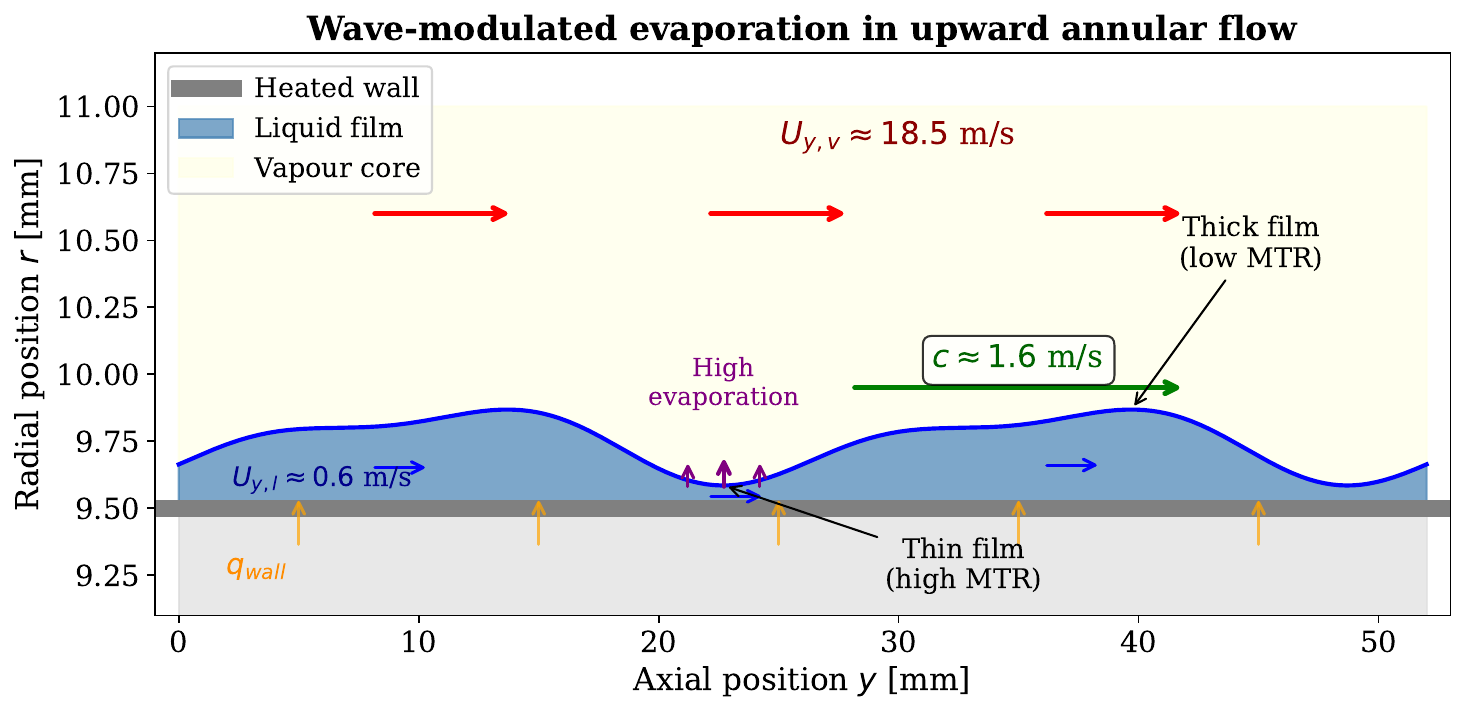}
    \caption{Schematic of wave-modulated evaporation in upward co-current annular flow. The liquid film (blue) coats the heated wall while vapour flows through the core. Interfacial waves propagate along the film surface. At wave troughs where the film is thin, thermal resistance is low and evaporation is enhanced; at wave crests where the film is thick, evaporation is suppressed. In our simulation, the wall temperature is set to be constant at $T_{sat} + 1K$, instead of applying heat flux.}
    \label{fig:annular_schematic}
\end{figure}

Figure~\ref{fig:annular_schematic} illustrates the wave-modulated evaporation mechanism. Interfacial waves create alternating regions of thin and thick film; heat conducted through the film drives evaporation at the interface, with the local mass transfer rate approximately proportional to the inverse of the liquid film thickness ($\dot{m} \propto 1/\delta$).

\begin{figure}[htbp]
    \centering
    \includegraphics[width=0.95\textwidth]{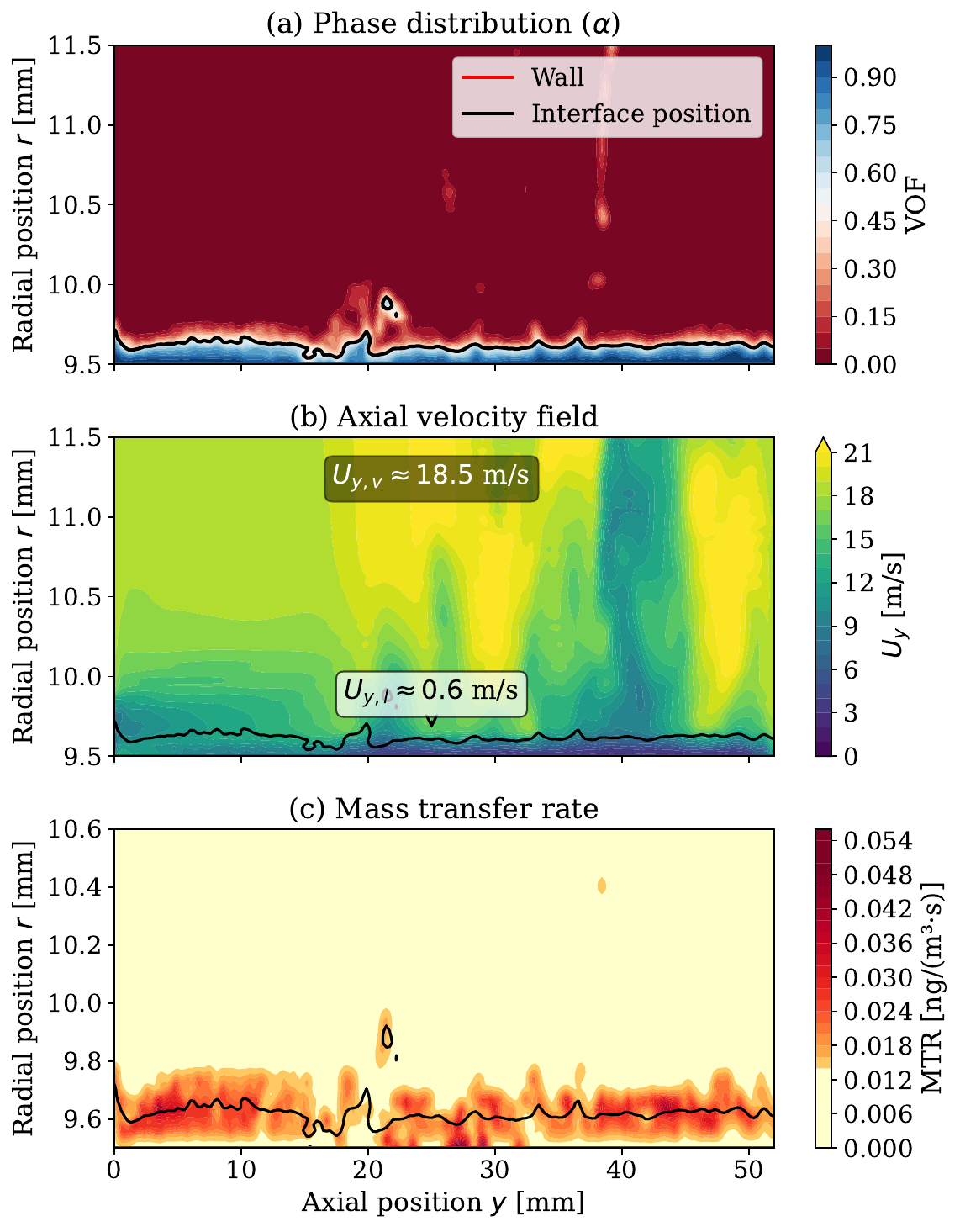}
    \caption{Instantaneous snapshot on a sliced plane of the annular flow simulation: (a)~phase distribution showing the liquid film (blue) on the wall and gas core (light blue); (b)~axial velocity field showing the fast gas core and slower liquid film; (c)~mass transfer rate field showing enhanced evaporation at wave troughs where the film is thin.}
    \label{fig:annular_snapshot}
\end{figure}

Figure~\ref{fig:annular_snapshot} shows the spatial structure at a representative instant. The wavy interface is clearly visible in the phase distribution (panel~a), with wave crests protruding into the gas phase and troughs approaching the wall. The steep velocity gradient at the interface (panel~b) drives the shear stress that excites the interfacial waves. The mass transfer rate field (panel~c) confirms that evaporation is concentrated at wave troughs where the film is thin.

\subsection{Temporal evolution and wave dynamics}
\label{sec:annular_dynamics}

\begin{figure}[htbp]
    \centering
    \includegraphics[width=0.95\textwidth]{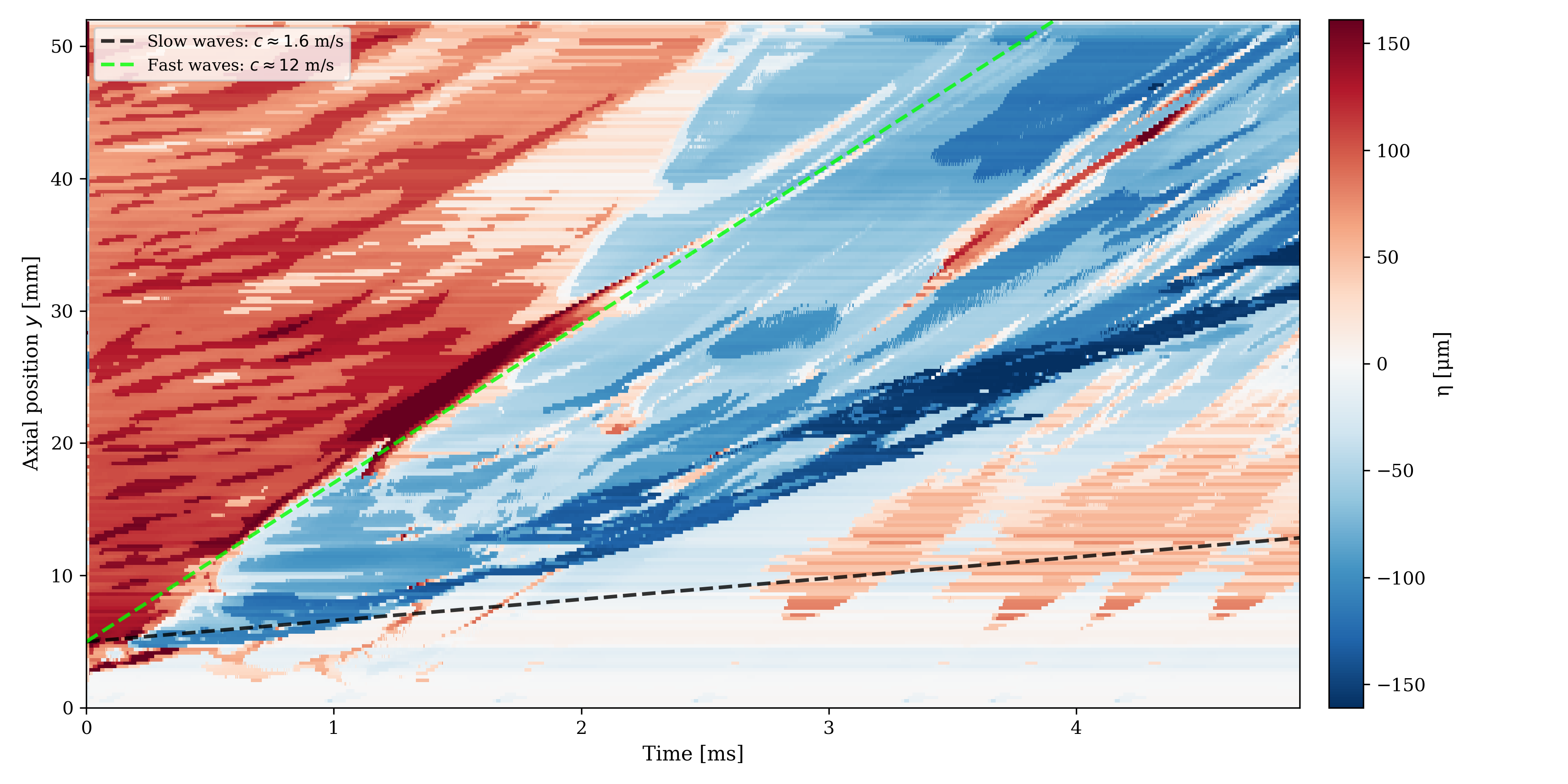}
    \caption{Hovmöller diagram of the interface perturbation $\eta(y,t)$ showing two wave populations: fast-moving structures advected by the gas phase (steep diagonals, $c \approx 12$~m/s) and slower interfacial disturbances (shallow diagonals, $c \approx 1.6$~m/s). The dashed lines indicate reference propagation velocities.}
    \label{fig:annular_hovmoller}
\end{figure}

The following wave analysis is based on 4.9~ms of data from the early transient phase; the statistics are therefore not converged and should be interpreted as preliminary characterisation of the developing flow. Figure~\ref{fig:annular_hovmoller} presents the spatio-temporal evolution of the interface through a Hovmöller diagram of the interface perturbation $\eta(y,t)$ relative to the global mean position. The diagonal patterns reveal wave propagation in the streamwise direction, with two distinct populations visible: steep diagonal features corresponding to fast structures advected by the gas phase ($c \approx 12$~m/s), and shallower features suggesting slower interfacial disturbances ($c \approx 1.6$~m/s, annotated dashed line).

The interface position $\eta(y,t)$ was extracted by identifying the radial location of the $\alpha = 0.5$ isosurface at each axial station, to produce a two-dimensional field over 4904 time steps (4.9~ms) and 202 axial locations. The local film thickness is then $\delta(y,t) = R_{\mathrm{interface}}(y,t) - R_{\mathrm{wall}}$, where $R_{\mathrm{wall}}$ is the inner pipe radius.

Phase velocity was determined via cross-correlation of $\eta(y,t)$ between streamwise stations separated by $\Delta y = 5$~mm. The analysis reveals two distinct wave populations:
\begin{enumerate}
    \item \textbf{Fast waves} ($c \approx 12$~m/s): gas-driven surface disturbances propagating at approximately 65\% of the mean gas velocity. These appear as steep diagonal features in the Hovmöller diagram, carry approximately half of the total interfacial wave energy, and are robustly detected by cross-correlation with spatial coherence of 0.86.
    \item \textbf{Slow component} ($c \approx 1.6$~m/s): a lower-velocity component is visible after two-dimensional Fourier bandpass filtering ($c < 3$~m/s), which retains approximately 3\% of the total wave energy but with lower spatial coherence (${\sim}0.6$). However, the observation window of 4.9~ms is insufficient to resolve complete wave periods at these velocities---a 1~m/s wave with $\lambda = 10$~mm completes only 0.49 periods---and this velocity should be considered an estimate.
\end{enumerate}
The limited axial extent of the domain ($L = 52$~mm) precludes a reliable measurement of the dominant wavelength from spatial Fourier analysis, as the longest resolvable mode coincides with the domain scale. The cross-correlation velocity measurements, however, are based on local phase shifts between neighbouring stations and are not affected by this limitation.

\subsection{Wave characteristics and comparison with Kelvin--Helmholtz theory}
\label{sec:annular_KH}

The measured wave characteristics can be compared with predictions from classical Kelvin--Helmholtz (K--H) instability theory. Classical inviscid K--H theory predicts a phase velocity of
\begin{equation}
    c_{\mathrm{KH}} = \frac{\rho_v U_{y,v} + \rho_l U_{y,l}}{\rho_v + \rho_l} \approx 0.62~\mathrm{m/s},
    \label{eq:c_KH}
\end{equation}
which lies below the slow-component velocity of $c_{\mathrm{slow}} \approx 1.6$~m/s estimated from bandpass filtering. The difference is expected, as viscous and thin-film effects are neglected in the inviscid K--H model.

The fast-wave population ($c_{\mathrm{fast}} \approx 12$~m/s) represents gas-phase disturbances imprinted on the interface by the turbulent vapour core. The slow component ($c_{\mathrm{slow}} \approx 1.6$~m/s) exceeds the K--H prediction and falls within the experimental range of 0.5--1.7~m/s \cite{barbosa_high-speed_2003}, supporting its identification as interfacial waves modified by viscous and thin-film effects.

The coexistence of these two populations is consistent with the wave generation mechanism described by Sato and Ni\v{c}eno~\cite{sato_large_2017}: waves originate from an imbalance between the upward interfacial shear stress and the downward gravitational body force, amplified by flow separation in the vapour phase behind wave crests. This mechanism---driven by shear stress inhomogeneity rather than the density-weighted inertial instability of classical K--H theory---explains why the observed phase velocities deviate from inviscid predictions.

\subsection{Wave-modulated mass transfer}
\label{sec:annular_mtr}

\begin{figure}[htbp]
    \centering
    \includegraphics[width=0.95\textwidth]{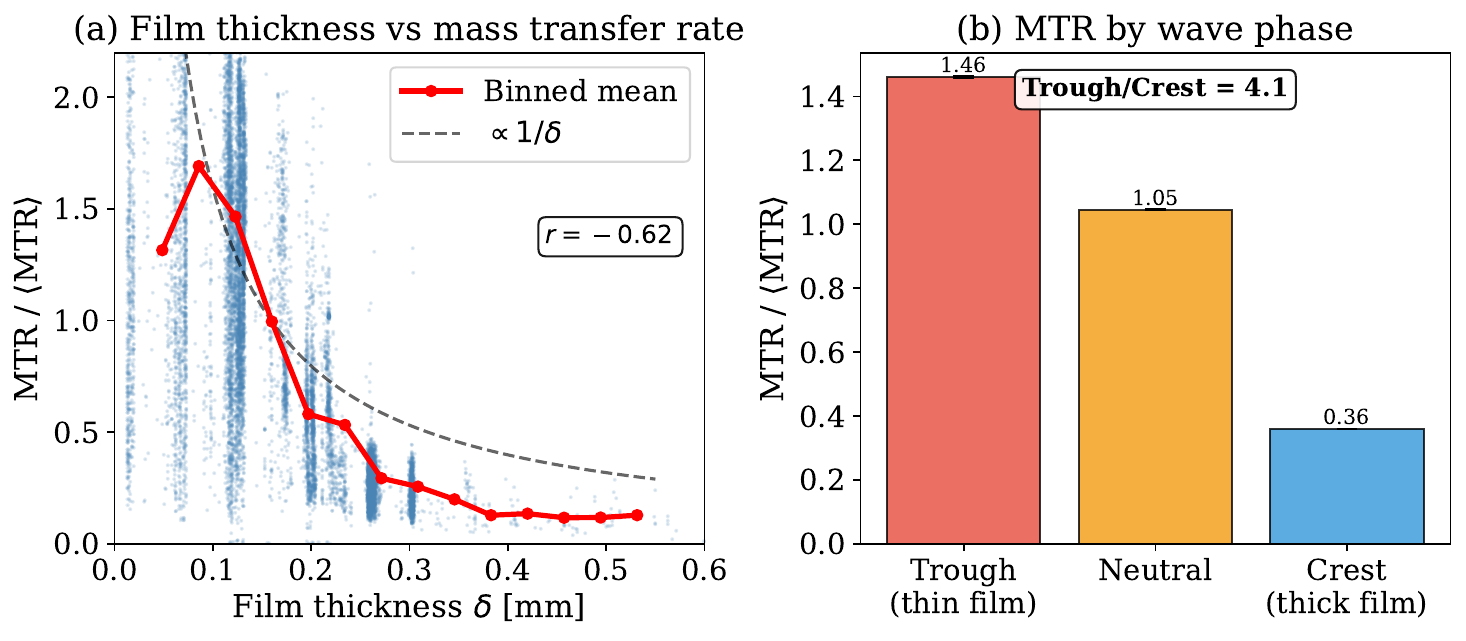}
    \caption{Quantitative analysis of wave-modulated mass transfer: (a)~scatter plot of film thickness versus normalised mass transfer rate, with binned means (red) and the theoretical $1/\delta$ scaling (dashed); the Pearson correlation is $r = -0.62$; (b)~conditional MTR by wave phase ($\eta$ relative to one standard deviation $s_\eta$), which shows that troughs (thin film) evaporate approximately four times faster than crests (thick film).}
    \label{fig:annular_quantitative}
\end{figure}

Figure~\ref{fig:annular_quantitative} quantifies the coupling between wave dynamics and evaporative mass transfer. Panel~(a) shows the correlation between local film thickness and mass transfer rate (Pearson $r = -0.62$ over the 4.9~ms observation window); the binned mean closely follows the $\dot{m} \propto 1/\delta$ scaling expected for conduction-limited evaporation ($q \sim k_l \Delta T / \delta$). Panel~(b) presents conditional statistics by wave phase, where each $(y,t)$ location is classified as trough ($\eta < -s_\eta$), neutral ($|\eta| \leq s_\eta$), or crest ($\eta > s_\eta$). The mass transfer rate in wave troughs is approximately 4 times higher than in wave crests in this early transient; these ratios may evolve as the flow approaches statistical stationarity.

The progressive thinning of the liquid film by evaporation is also apparent from the interface data. The spatially-averaged film thickness decreases from approximately 0.25~mm early in the simulation to 0.16~mm by $t = 4.9$~ms---a drift of roughly 90~$\mu$m over the observation window. This drift creates a systematic bias when using a global mean as the reference for $\eta$: late-time regions appear predominantly as troughs. For the wave analysis, this drift is slow compared to the wave periods and does not affect the cross-correlation velocity measurements. The conservative formulation of the CICSAM scheme ensures that the global liquid volume fraction is preserved to solver tolerance at each time step; the observed film thinning is driven entirely by phase change, not by numerical mass loss. This positive feedback between film thinning and enhanced evaporation is characteristic of annular flow boiling: as the film thins, thermal resistance decreases, further accelerating evaporation---a process that, in the limiting case, leads to complete film depletion (dryout). All quantitative statistics reported in this section---correlation coefficients, conditional ratios, and phase velocities---are computed from the early transient and should be considered indicative rather than statistically converged values.

\subsection{Qualitative comparison with previous LES studies}
\label{sec:annular_comparison}

Although the present simulation covers only 4.9~ms of physical time---well short of the ${\sim}100$~ms required for statistical stationarity---a qualitative comparison with the LES of Sato and Ni\v{c}eno~\cite{sato_large_2017}, who simulated the same configuration for 300~ms using an interface tracking method on a structured Cartesian grid, is informative for assessing whether the framework captures the essential physics. Table~\ref{tab:annular_comparison} summarises the key numerical parameters of both studies.

\begin{table}[htbp]
\centering
\caption{Comparison of numerical parameters between the present study and \cite{sato_large_2017}.}
\label{tab:annular_comparison}
\begin{tabular}{lcc}
\toprule
\textbf{Parameter} & \textbf{Present study} & \textbf{Sato \& Ni\v{c}eno (2017)} \\
\midrule
Mesh type & Unstructured (hex + poly) & Structured Cartesian \\
Domain geometry & 45$^\circ$ cylindrical sector & Rectangular (Cartesian approx.) \\
Axial length & 52~mm & 103.6~mm \\
Number of cells & 7.2~M & 8.4~M \\
Time step, $\Delta t$ & 0.1~$\mu$s & 0.2~$\mu$s \\
Simulated time & 4.9~ms & 300~ms \\
SGS model & WALE & Smagorinsky \\
Interface detection & Algebraic VOF + isoap & Colour function (CIP-CSL2) \\
Conjugate heat transfer & No & Yes \\
\bottomrule
\end{tabular}
\end{table}

Despite differences in methodology and simulation duration, both studies agree on the key physics of annular flow boiling:

\begin{itemize}
    \item \textbf{Wave-modulated mass transfer:} The reference study reported enhanced mass transfer in thin-film regions~\cite{sato_large_2017}, consistent with the $\dot{m} \propto 1/\delta$ scaling and fourfold trough-to-crest enhancement quantified in Section~\ref{sec:annular_mtr}.

    \item \textbf{Wave dynamics:} Wave generation was attributed to inhomogeneous shear stress and vapour-phase flow separation (Section~\ref{sec:annular_KH}), consistent with the phase velocities measured here.

    \item \textbf{Film thickness:} Wave elevations of 0.15--0.35~mm (Fig.~8 of \cite{sato_large_2017}) compare with the 0.10--0.45~mm range observed here.

    \item \textbf{Wave velocity:} Quantitative wave velocities were not reported in \cite{sato_large_2017}; the cross-correlation analysis presented here provides a first estimate.
\end{itemize}

The present study complements \cite{sato_large_2017} by adding quantitative wave velocity measurements and mass transfer statistics on an unstructured polyhedral mesh, while their inclusion of conjugate heat transfer captured elevated wall temperatures beneath wave crests---relevant to bubble nucleation, which is not modelled here.

	\section{Conclusions}
	\label{sec:conclusions}
	\vspace{1pt}
	This paper presented a framework for simulating boiling on unstructured meshes using an algebraic Volume-of-Fluid method with geometric interface reconstruction. The main contributions are: (i)~a direct heat-flux-driven mass transfer model that computes phase-change rates from local temperature gradients at the geometrically reconstructed interface, combining CICSAM algebraic VOF transport with isoap piecewise-linear interface reconstruction on arbitrary polyhedral cells; (ii)~a systematic analysis of the interaction between the interface-modified least-squares gradient stencil and mesh topology, revealing two distinct mechanisms on structured hexahedral meshes---a coherent four-fold anisotropy that distorts the bubble shape, and a systematic overestimate of the mean gradient magnitude driven by the mass transfer heat sink---both of which are naturally damped on polyhedral meshes; and (iii)~demonstration of the framework on turbulent annular boiling flow requiring unstructured meshing.

	The coupling of CICSAM algebraic VOF transport with isoap geometric interface reconstruction enables direct computation of phase-change rates from local temperature gradients on arbitrary polyhedral cells, without empirical closure models. The interface-modified least-squares gradient provides the interfacial heat flux, while an implicit interface treatment in the energy equation maintains cell temperatures close to $T_\mathrm{sat}$ near the phase boundary. Because the method relies on standard finite-volume building blocks --- least-squares gradients, face-based fluxes, and cell-centred storage --- it can be integrated into other unstructured finite-volume codes. Validation against three canonical benchmarks confirms the accuracy: the Stefan and Sucking problems give interface position errors below $1\%$, and the Scriven bubble growth converges monotonically with mesh refinement on polyhedral meshes. The geometric interface reconstruction and VOF advection together account for less than $10\%$ of the total solver time.

	Analysis of the temperature gradient field revealed a four-fold anisotropy on structured hexahedral meshes that originates from the interaction between the interface-modified least-squares gradient stencil and the regular Cartesian cell topology. Fourier decomposition showed that $16\%$ of the mean gradient magnitude is concentrated in the coherent $m = 4$ azimuthal mode on the coarse structured mesh, decreasing to $9\%$ on the fine structured mesh; the polyhedral mesh reduces this to ${\sim}1\%$ with the remaining scatter distributed quasi-randomly. Despite the $9$--$16\%$ gradient anisotropy on structured meshes, the pseudo-time extrapolation procedure acts as a regulariser, which limits the coherent variation in the actual mass transfer rate to approximately $4\%$. In addition to this angular variation, a systematic overestimate of the mean gradient magnitude was identified on structured meshes: the mass transfer heat sink maintains interface-adjacent cell temperatures close to~$T_\mathrm{sat}$, which converts the symmetric central difference into an effectively one-sided estimate and amplifies the computed gradient by a factor of approximately $1.4$ relative to the central difference; because the unmodified gradient already underestimates the true interface gradient by~${\sim}30\%$, this amplification overcompensates the resolution deficit and accounts for the radius overshoot on structured grids. Moreover, the gradient overestimate does not diminish with mesh refinement: the amplification factor depends only on the ratio $d/h$, which is uniformly distributed regardless of mesh size, while the physical cell-centre temperature approaches $T_\mathrm{sat}$ on finer meshes, which makes the heat-sink limit increasingly representative of the actual simulation state (\ref{app:gradient_magnitude}). Polyhedral meshes, with their irregular face orientations and richer stencils, naturally eliminate both the preferential alignment and the magnitude bias, which produces near-isotropic gradients, spherical bubble shapes, and monotonic convergence without overshoot.

	The proposed simulation method was also applied to turbulent upward co-current annular flow with evaporation. The simulated physical time (4.9~ms) is well short of the ${\sim}100$~ms required for statistical stationarity, but the early transient results are qualitatively consistent with the LES of Sato and Ni\v{c}eno \cite{sato_large_2017} and the experiments of Barbosa Jr. \cite{barbosa_high-speed_2003}. The simulation reproduces wave-modulated evaporation with enhanced mass transfer at film troughs, progressive film thinning, and the coexistence of fast gas-driven ($c \approx 12$~m/s) and slower interfacial ($c \approx 1.6$~m/s) wave populations. Film thickness variations produce approximately fourfold enhancement of mass transfer rate in troughs compared to crests ($r = -0.62$), consistent with conduction-limited evaporation where $\dot{m} \sim 1/\delta$. The slow wave component falls within the experimental range of 0.5--1.7~m/s, though longer runs are needed for a definitive comparison.

	Future work includes extending the annular flow simulation to longer physical times for quantitative comparison with experiments, adding conjugate heat transfer for coupled fluid--solid simulations, and replacing the algebraic VOF with a full geometric method to eliminate numerical interface diffusion.
	
	\textbf{CRediT authorship contribution statement}

	\textbf{Jan Kren}: Conceptualization, Methodology, Software, Investigation, Validation, Visualization, Writing -- original draft. \textbf{Bojan Ni\v{c}eno}: Conceptualization, Methodology, Software, Writing -- review \& editing. \textbf{Yohei Sato}: Conceptualization, Methodology, Software, Investigation, Supervision, Writing -- review \& editing.

	\textbf{Acknowledgments}
	
	The authors would like to thank Swissnuclear for providing funding for this research under the project BRAVA LSM\_22\_02.
	
	\textbf{Data availability}
	
	The base version of the T-Flows code is publicly available at \url{https://github.com/DelNov/T-Flows}. The latest implementation, including the phase-change extensions presented in this work, is available from the authors upon request. The post-processing scripts and data used to generate the figures in this paper are available at \url{https://github.com/JanKren/Unstructured-Boiling-Paper-Repo}.
	
	\bibliographystyle{elsarticle-num}   
	\bibliography{references}

\begin{thebibliography}{10}
\expandafter\ifx\csname url\endcsname\relax
  \def\url#1{\texttt{#1}}\fi
\expandafter\ifx\csname urlprefix\endcsname\relax\def\urlprefix{URL }\fi
\expandafter\ifx\csname href\endcsname\relax
  \def\href#1#2{#2} \def\path#1{#1}\fi

\bibitem{nishida_front_2023}
K.~Nishida, S.~Morooka, M.~Mori, Y.~Koizumi (Eds.), Boiling Water Reactors,
  Vol.~4 of {JSME} {Series} in {Thermal} and {Nuclear} {Power} {Generation},
  Elsevier, 2023.
\newblock \href {https://doi.org/10.1016/B978-0-12-821361-2.09988-8}
  {\path{doi:10.1016/B978-0-12-821361-2.09988-8}}.

\bibitem{lim_exploring_2022}
J.~H. Lim, M.~Park, S.~M. Shin, S.~S. Chung, Exploring the onset of nucleate
  boiling with hypervapotron channel for tokamak cooling system application,
  Applied Thermal Engineering 209 (2022) 118334.
\newblock \href {https://doi.org/10.1016/j.applthermaleng.2022.118334}
  {\path{doi:10.1016/j.applthermaleng.2022.118334}}.

\bibitem{sohel_murshed_critical_2017}
S.~M. Sohel~Murshed, C.~A. Nieto~de Castro, A critical review of traditional
  and emerging techniques and fluids for electronics cooling, Renewable and
  Sustainable Energy Reviews 78 (2017) 821--833.
\newblock \href {https://doi.org/10.1016/j.rser.2017.04.112}
  {\path{doi:10.1016/j.rser.2017.04.112}}.

\bibitem{kharangate_review_2017}
C.~R. Kharangate, I.~Mudawar, Review of computational studies on boiling and
  condensation, International Journal of Heat and Mass Transfer 108 (2017)
  1164--1196.
\newblock \href {https://doi.org/10.1016/j.ijheatmasstransfer.2016.12.065}
  {\path{doi:10.1016/j.ijheatmasstransfer.2016.12.065}}.

\bibitem{hu_comprehensive_2025}
Y.~Hu, H.~Gao, Y.~Yan, A comprehensive review of boiling heat transfer on
  multi-scale hybrid surfaces and applications, Journal of Industrial and
  Engineering Chemistry 143 (2025) 123--143.
\newblock \href {https://doi.org/10.1016/j.jiec.2024.09.002}
  {\path{doi:10.1016/j.jiec.2024.09.002}}.

\bibitem{mb_numerical_2024}
{Darshan M. B.}, M.~Magnini, O.~K. Matar, Numerical modelling of flow boiling
  inside microchannels: {A} critical review of methods and applications,
  Applied Thermal Engineering 257 (2024) 124464.
\newblock \href {https://doi.org/10.1016/j.applthermaleng.2024.124464}
  {\path{doi:10.1016/j.applthermaleng.2024.124464}}.

\bibitem{chen_review_2024}
Y.~Chen, B.~Yu, W.~Lu, B.~Wang, D.~Sun, K.~Jiao, W.~Zhang, W.~Tao, Review on
  numerical simulation of boiling heat transfer from atomistic to mesoscopic
  and macroscopic scales, International Journal of Heat and Mass Transfer 225
  (2024) 125396.
\newblock \href {https://doi.org/10.1016/j.ijheatmasstransfer.2024.125396}
  {\path{doi:10.1016/j.ijheatmasstransfer.2024.125396}}.

\bibitem{juric_computations_1998}
D.~Juric, G.~Tryggvason, Computations of boiling flows, International Journal
  of Multiphase Flow 24~(3) (1998) 387--410.
\newblock \href {https://doi.org/10.1016/S0301-9322(97)00050-5}
  {\path{doi:10.1016/S0301-9322(97)00050-5}}.

\bibitem{esmaeeli_front_2004}
A.~Esmaeeli, G.~Tryggvason, A front tracking method for computations of boiling
  in complex geometries, International Journal of Multiphase Flow 30~(7) (2004)
  1037--1050.
\newblock \href {https://doi.org/10.1016/j.ijmultiphaseflow.2004.04.008}
  {\path{doi:10.1016/j.ijmultiphaseflow.2004.04.008}}.

\bibitem{anderson-diffuse1998}
D.~M. Anderson, G.~B. McFadden, A.~A. Wheeler, Diffuse-interface methods in
  fluid mechanics, Annual Review of Fluid Mechanics 30 (1998) 139--165.
\newblock \href {https://doi.org/10.1146/annurev.fluid.30.1.139}
  {\path{doi:10.1146/annurev.fluid.30.1.139}}.

\bibitem{roccon_boiling_2025}
A.~Roccon, Boiling heat transfer by phase-field method, Acta Mechanica 236
  (2025) 5623--5638.
\newblock \href {https://doi.org/10.1007/s00707-024-04122-7}
  {\path{doi:10.1007/s00707-024-04122-7}}.

\bibitem{sussman_level_1994}
M.~Sussman, P.~Smereka, S.~Osher, A level set approach for computing solutions
  to incompressible two-phase flow, Journal of Computational Physics 114~(1)
  (1994) 146--159.
\newblock \href {https://doi.org/10.1006/jcph.1994.1155}
  {\path{doi:10.1006/jcph.1994.1155}}.

\bibitem{tanguy_level_2007}
S.~Tanguy, T.~M\'{e}nard, A.~Berlemont, A level set method for vaporizing
  two-phase flows, Journal of Computational Physics 221~(2) (2007) 837--853.
\newblock \href {https://doi.org/10.1016/j.jcp.2006.07.003}
  {\path{doi:10.1016/j.jcp.2006.07.003}}.

\bibitem{sahut_numerical_2021}
G.~Sahut, G.~Ghigliotti, G.~Balarac, M.~Bernard, V.~Moureau, P.~Marty,
  Numerical simulation of boiling on unstructured grids, Journal of
  Computational Physics 432 (2021) 110161.
\newblock \href {https://doi.org/10.1016/j.jcp.2021.110161}
  {\path{doi:10.1016/j.jcp.2021.110161}}.

\bibitem{fan_high-fidelity_2022}
Y.~Fan, M.~Li, W.~D. Pointer, I.~A. Bolotnov, High-fidelity pool boiling
  simulations on multiple nucleation sites using interface capturing method,
  Nuclear Engineering and Design 399 (2022) 112004.
\newblock \href {https://doi.org/10.1016/j.nucengdes.2022.112004}
  {\path{doi:10.1016/j.nucengdes.2022.112004}}.

\bibitem{iskhakova_interface_2023}
A.~Iskhakova, Y.~Kondo, K.~Tanimoto, N.~T. Dinh, I.~A. Bolotnov, Interface
  capturing flow boiling simulations in a compact heat exchanger, ASME Journal
  of Heat and Mass Transfer 145 (2023) 041605.
\newblock \href {https://doi.org/10.1115/1.4056688}
  {\path{doi:10.1115/1.4056688}}.

\bibitem{hirt_volume_1981}
C.~W. Hirt, B.~D. Nichols, Volume of fluid ({VOF}) method for the dynamics of
  free boundaries, Journal of Computational Physics 39~(1) (1981) 201--225.
\newblock \href {https://doi.org/10.1016/0021-9991(81)90145-5}
  {\path{doi:10.1016/0021-9991(81)90145-5}}.

\bibitem{zhang_refined_2014}
D.~Zhang, C.~Jiang, D.~Liang, Z.~Chen, Y.~Yang, Y.~Shi, A refined
  volume-of-fluid algorithm for capturing sharp fluid interfaces on arbitrary
  meshes, Journal of Computational Physics 274 (2014) 709--736.
\newblock \href {https://doi.org/10.1016/j.jcp.2014.06.043}
  {\path{doi:10.1016/j.jcp.2014.06.043}}.

\bibitem{saincher_efficient_2022}
S.~Saincher, V.~Sriram, An efficient operator-split {CICSAM} scheme for
  three-dimensional multiphase-flow problems on {Cartesian} grids, Computers \&
  Fluids 240 (2022) 105440.
\newblock \href {https://doi.org/10.1016/j.compfluid.2022.105440}
  {\path{doi:10.1016/j.compfluid.2022.105440}}.

\bibitem{bures_piecewise_2021}
L.~Bureš, Y.~Sato, A.~Pautz, Piecewise linear interface-capturing
  volume-of-fluid method in axisymmetric cylindrical coordinates, Journal of
  Computational Physics 436 (2021) 110291.
\newblock \href {https://doi.org/10.1016/j.jcp.2021.110291}
  {\path{doi:10.1016/j.jcp.2021.110291}}.

\bibitem{long_edge-based_2024}
T.~Long, J.~Pan, S.~Zaleski, An edge-based interface tracking ({EBIT}) method
  for multiphase flows with phase change, Journal of Computational Physics 513
  (2024) 113159.
\newblock \href {https://doi.org/10.1016/j.jcp.2024.113159}
  {\path{doi:10.1016/j.jcp.2024.113159}}.

\bibitem{sato_sharp-interface_2013}
Y.~Sato, B.~Ničeno, A sharp-interface phase change model for a
  mass-conservative interface tracking method, Journal of Computational Physics
  249 (2013) 127--161.
\newblock \href {https://doi.org/10.1016/j.jcp.2013.04.035}
  {\path{doi:10.1016/j.jcp.2013.04.035}}.

\bibitem{bures_direct_2021}
L.~Bureš, Y.~Sato, Direct numerical simulation of evaporation and condensation
  with the geometric {VOF} method and a sharp-interface phase-change model,
  International Journal of Heat and Mass Transfer 173 (2021) 121233.
\newblock \href {https://doi.org/10.1016/j.ijheatmasstransfer.2021.121233}
  {\path{doi:10.1016/j.ijheatmasstransfer.2021.121233}}.

\bibitem{poblador-ibanez_momentum_2025}
J.~Poblador-Ibanez, N.~Valle, B.~J. Boersma, A momentum balance correction to
  the non-conservative one-fluid formulation in boiling flows using
  volume-of-fluid, Journal of Computational Physics 524 (2025) 113704.
\newblock \href {https://doi.org/10.1016/j.jcp.2024.113704}
  {\path{doi:10.1016/j.jcp.2024.113704}}.

\bibitem{fedkiw_non-oscillatory_1999}
R.~P. Fedkiw, T.~Aslam, B.~Merriman, S.~Osher, A non-oscillatory {Eulerian}
  approach to interfaces in multimaterial flows (the ghost fluid method),
  Journal of Computational Physics 152~(2) (1999) 457--492.
\newblock \href {https://doi.org/10.1006/jcph.1999.6236}
  {\path{doi:10.1006/jcph.1999.6236}}.

\bibitem{gibou_level_2007}
F.~Gibou, L.~Chen, D.~Nguyen, S.~Banerjee, A level set based sharp interface
  method for the multiphase incompressible {Navier}–{Stokes} equations with
  phase change, Journal of Computational Physics 222~(2) (2007) 536--555.
\newblock \href {https://doi.org/10.1016/j.jcp.2006.07.035}
  {\path{doi:10.1016/j.jcp.2006.07.035}}.

\bibitem{rueda_villegas_ghost_2016}
L.~Rueda~Villegas, R.~Alis, M.~Lepilliez, S.~Tanguy, A ghost fluid/level set
  method for boiling flows and liquid evaporation: Application to the
  {Leidenfrost} effect, Journal of Computational Physics 316 (2016) 789--813.
\newblock \href {https://doi.org/10.1016/j.jcp.2016.04.031}
  {\path{doi:10.1016/j.jcp.2016.04.031}}.

\bibitem{Maric2020}
T.~Mari\'{c}, D.~B. Kothe, D.~Bothe, Unstructured un-split geometrical
  {Volume-of-Fluid} methods -- {A} review, Journal of Computational Physics 420
  (2020) 109695.
\newblock \href {https://doi.org/10.1016/j.jcp.2020.109695}
  {\path{doi:10.1016/j.jcp.2020.109695}}.

\bibitem{Herrmann2008}
M.~Herrmann, A balanced force refined level set grid method for two-phase flows
  on unstructured flow solver grids, Journal of Computational Physics 227
  (2008) 2674--2706.
\newblock \href {https://doi.org/10.1016/j.jcp.2007.11.002}
  {\path{doi:10.1016/j.jcp.2007.11.002}}.

\bibitem{Nishikawa2018}
H.~Nishikawa, From hyperbolic diffusion scheme to gradient method: implicit
  {Green--Gauss} gradients for unstructured grids, Journal of Computational
  Physics 372 (2018) 126--160.
\newblock \href {https://doi.org/10.1016/j.jcp.2018.06.019}
  {\path{doi:10.1016/j.jcp.2018.06.019}}.

\bibitem{Syrakos2017}
A.~Syrakos, S.~Varchanis, Y.~Dimakopoulos, A.~Goulas, J.~Tsamopoulos, A
  critical analysis of some popular methods for the discretisation of the
  gradient operator in finite volume methods, Physics of Fluids 29 (2017)
  127103.
\newblock \href {https://doi.org/10.1063/1.4997985}
  {\path{doi:10.1063/1.4997985}}.

\bibitem{giustini_method_2021}
G.~Giustini, R.~I. Issa, A method for simulating interfacial mass transfer on
  arbitrary meshes, Physics of Fluids 33~(8) (2021) 087102.
\newblock \href {https://doi.org/10.1063/5.0058987}
  {\path{doi:10.1063/5.0058987}}.

\bibitem{lee1980pressure}
W.~H. Lee, A pressure iteration scheme for two-phase flow modeling, in: T.~N.
  Veziroglu (Ed.), Multiphase Transport: Fundamentals, Reactor Safety,
  Applications, Vol.~1, Hemisphere Publishing, Washington, DC, 1980, pp.
  407--431.

\bibitem{Scheufler2023}
H.~Scheufler, J.~Roenby, {TwoPhaseFlow}: {A} framework for developing two phase
  flow solvers in {OpenFOAM}, OpenFOAM Journal 3 (2023) 200--224.
\newblock \href {https://doi.org/10.51560/ofj.v3.80}
  {\path{doi:10.51560/ofj.v3.80}}.

\bibitem{lopez_isoap_2021}
J.~López, J.~Hernández, isoap: {A} software for isosurface extraction on
  arbitrary polyhedra (Jul. 2021).
\newblock \href {https://doi.org/10.17632/4rcf98s74c.1}
  {\path{doi:10.17632/4rcf98s74c.1}}.

\bibitem{DelNovTFlows}
B.~Niceno, J.~Kren, Y.~Sato,
  \href{https://github.com/DelNov/T-Flows}{{T‑Flows}} (2025).
\newline\urlprefix\url{https://github.com/DelNov/T-Flows}

\bibitem{petsc-user-ref}
S.~Balay, S.~Abhyankar, M.~F. Adams, S.~Benson, J.~Brown, P.~Brune,
  K.~Buschelman, E.~Constantinescu, L.~Dalcin, A.~Dener, V.~Eijkhout,
  J.~Faibussowitsch, W.~D. Gropp, V.~Hapla, T.~Isaac, P.~Jolivet, D.~Karpeev,
  D.~Kaushik, M.~G. Knepley, F.~Kong, S.~Kruger, D.~A. May, L.~C. McInnes,
  R.~T. Mills, L.~Mitchell, T.~Munson, J.~E. Roman, K.~Rupp, P.~Sanan,
  J.~Sarich, B.~F. Smith, H.~Suh, S.~Zampini, H.~Zhang, H.~Zhang, J.~Zhang,
  Petsc/tao users manual, Tech. Rep. ANL-21/39 - Revision 3.22, Argonne
  National Laboratory (2024).
\newblock \href {https://doi.org/10.2172/2205494} {\path{doi:10.2172/2205494}}.

\bibitem{tryggvason_direct_2011}
G.~Tryggvason, R.~Scardovelli, S.~Zaleski, Direct numerical simulations of
  gas-liquid multiphase flows, Cambridge University Press, Cambridge ; New
  York, 2011.
\newblock \href {https://doi.org/10.1017/CBO9780511975264}
  {\path{doi:10.1017/CBO9780511975264}}.

\bibitem{ubbink_method_1999}
O.~Ubbink, R.~I. Issa, A method for capturing sharp fluid interfaces on
  arbitrary meshes, Journal of Computational Physics 153~(1) (1999) 26--50.
\newblock \href {https://doi.org/10.1006/jcph.1999.6276}
  {\path{doi:10.1006/jcph.1999.6276}}.

\bibitem{jasak_high_1999}
H.~Jasak, H.~G. Weller, A.~D. Gosman, High resolution {NVD} differencing scheme
  for arbitrarily unstructured meshes, International Journal for Numerical
  Methods in Fluids 31~(2) (1999) 431--449.
\newblock \href
  {https://doi.org/10.1002/(SICI)1097-0363(19990930)31:2<431::AID-FLD884>3.0.CO;2-T}
  {\path{doi:10.1002/(SICI)1097-0363(19990930)31:2<431::AID-FLD884>3.0.CO;2-T}}.

\bibitem{denner_balanced-force_2013}
F.~Denner,
  \href{https://spiral.imperial.ac.uk/handle/10044/1/28101}{Balanced-force
  two-phase flow modelling on unstructured and adaptive meshes}, Ph.D. thesis,
  Imperial College London (2013).
\newline\urlprefix\url{https://spiral.imperial.ac.uk/handle/10044/1/28101}

\bibitem{mathur_pressure-based_1997}
S.~R. Mathur, J.~Y. Murthy, A pressure-based method for unstructured meshes,
  Numerical Heat Transfer, Part B: Fundamentals 31~(2) (1997) 195--215.
\newblock \href {https://doi.org/10.1080/10407799708915105}
  {\path{doi:10.1080/10407799708915105}}.

\bibitem{brackbill_continuum_1992}
J.~U. Brackbill, D.~B. Kothe, C.~Zemach, A continuum method for modeling
  surface tension, Journal of Computational Physics 100~(2) (1992) 335--354.
\newblock \href {https://doi.org/10.1016/0021-9991(92)90240-Y}
  {\path{doi:10.1016/0021-9991(92)90240-Y}}.

\bibitem{rhie_numerical_1983}
C.~M. Rhie, W.~L. Chow, Numerical study of the turbulent flow past an airfoil
  with trailing edge separation, AIAA Journal 21~(11) (1983) 1526--1532.
\newblock \href {https://doi.org/10.2514/3.8284} {\path{doi:10.2514/3.8284}}.

\bibitem{choi1999}
S.-K. Choi, Note on the use of momentum interpolation method for unsteady
  flows, Numerical Heat Transfer, Part A: Applications 36 (1999) 545--550.
\newblock \href {https://doi.org/10.1080/104077899274679}
  {\path{doi:10.1080/104077899274679}}.

\bibitem{choi_use_2003}
S.-K. Choi, S.-O. Kim, C.-H. Lee, H.-K. Choi, Use of the momentum interpolation
  method for flows with a large body force, Numerical Heat Transfer, Part B:
  Fundamentals 43~(3) (2003) 267--287.
\newblock \href {https://doi.org/10.1080/713836204}
  {\path{doi:10.1080/713836204}}.

\bibitem{mencinger_finite_2007}
J.~Mencinger, I.~{\v{Z}}un, On the finite volume discretization of
  discontinuous body force field on collocated grid: Application to {VOF}
  method, Journal of Computational Physics 221~(2) (2007) 524--538.
\newblock \href {https://doi.org/10.1016/j.jcp.2006.06.021}
  {\path{doi:10.1016/j.jcp.2006.06.021}}.

\bibitem{yu_discussion_2002}
B.~Yu, W.~Q. Tao, J.~J. Wei, Y.~Kawaguchi, T.~Tagawa, H.~Ozoe, Discussion on
  momentum interpolation method for collocated grids of incompressible flow,
  Numerical Heat Transfer, Part B: Fundamentals 42~(2) (2002) 141--166.
\newblock \href {https://doi.org/10.1080/10407790190053824}
  {\path{doi:10.1080/10407790190053824}}.

\bibitem{majumdar_role_1988}
S.~Majumdar, Role of underrelaxation in momentum interpolation for calculation
  of flow with nonstaggered grids, Numerical Heat Transfer, Part B:
  Fundamentals 13~(1) (1988) 125--132.
\newblock \href {https://doi.org/10.1080/10407788808913607}
  {\path{doi:10.1080/10407788808913607}}.

\bibitem{papageorgakopoulos_enhancement_2000}
J.~Papageorgakopoulos, G.~Arampatzis, D.~Assimacopoulos, N.~C. Markatos,
  Enhancement of the momentum interpolation method on non-staggered grids,
  International Journal for Numerical Methods in Fluids 33~(1) (2000) 1--22.
\newblock \href
  {https://doi.org/10.1002/(SICI)1097-0363(20000515)33:1<1::AID-FLD978>3.0.CO;2-0}
  {\path{doi:10.1002/(SICI)1097-0363(20000515)33:1<1::AID-FLD978>3.0.CO;2-0}}.

\bibitem{ferziger_computational_2020}
J.~H. Ferziger, M.~Perić, R.~L. Street, Computational Methods for Fluid
  Dynamics, 4th Edition, Springer, 2020.
\newblock \href {https://doi.org/10.1007/978-3-319-99693-6}
  {\path{doi:10.1007/978-3-319-99693-6}}.

\bibitem{tanguy_benchmarks_2014}
S.~Tanguy, M.~Sagan, B.~Lalanne, F.~Couderc, C.~Colin, Benchmarks and numerical
  methods for the simulation of boiling flows, Journal of Computational Physics
  264 (2014) 1--22.
\newblock \href {https://doi.org/10.1016/j.jcp.2014.01.014}
  {\path{doi:10.1016/j.jcp.2014.01.014}}.

\bibitem{scriven_dynamics_1959}
L.~E. Scriven, On the dynamics of phase growth, Chemical Engineering Science
  10~(1-2) (1959) 1--13.
\newblock \href {https://doi.org/10.1016/0009-2509(59)80019-1}
  {\path{doi:10.1016/0009-2509(59)80019-1}}.

\bibitem{geuzaine_gmsh_2009}
C.~Geuzaine, J.-F. Remacle, Gmsh: {A} 3-{D} finite element mesh generator with
  built-in pre- and post-processing facilities, International Journal for
  Numerical Methods in Engineering 79 (2009) 1309--1331.
\newblock \href {https://doi.org/10.1002/nme.2579}
  {\path{doi:10.1002/nme.2579}}.

\bibitem{Diskin2012}
B.~Diskin, J.~L. Thomas, Comparison of node-centered and cell-centered
  unstructured finite-volume discretizations: inviscid fluxes, AIAA Journal 49
  (2012) 836--854.
\newblock \href {https://doi.org/10.2514/1.J050897}
  {\path{doi:10.2514/1.J050897}}.

\bibitem{nicoud_subgrid-scale_1999}
F.~Nicoud, F.~Ducros, Subgrid-scale stress modelling based on the square of the
  velocity gradient tensor, Flow, Turbulence and Combustion 62~(3) (1999)
  183--200.
\newblock \href {https://doi.org/10.1023/A:1009995426001}
  {\path{doi:10.1023/A:1009995426001}}.

\bibitem{sato_large_2017}
Y.~Sato, B.~Niceno, Large eddy simulation of upward co-current annular boiling
  flow using an interface tracking method, Nuclear Engineering and Design 321
  (2017) 69--81.
\newblock \href {https://doi.org/10.1016/j.nucengdes.2017.03.003}
  {\path{doi:10.1016/j.nucengdes.2017.03.003}}.

\bibitem{barbosa_high-speed_2003}
J.~R. Barbosa~Jr, G.~F. Hewitt, S.~M. Richardson, High-speed visualisation of
  nucleate boiling in vertical annular flow, International Journal of Heat and
  Mass Transfer 46 (2003) 5153--5160.
\newblock \href {https://doi.org/10.1016/S0017-9310(03)00231-X}
  {\path{doi:10.1016/S0017-9310(03)00231-X}}.

\bibitem{bottaro_note_1990}
A.~Bottaro, Note on open boundary conditions for elliptic flows, Numerical Heat
  Transfer, Part B: Fundamentals 18~(2) (1990) 243--256.
\newblock \href {https://doi.org/10.1080/10407799008944952}
  {\path{doi:10.1080/10407799008944952}}.

\bibitem{leonard_ultimate_1991}
B.~P. Leonard, The {ULTIMATE} conservative difference scheme applied to
  unsteady one-dimensional advection, Computer Methods in Applied Mechanics and
  Engineering 88~(1) (1991) 17--74.
\newblock \href {https://doi.org/10.1016/0045-7825(91)90232-U}
  {\path{doi:10.1016/0045-7825(91)90232-U}}.

\bibitem{Syrakos2023}
A.~Syrakos, O.~Oxtoby, E.~de~Villiers, S.~Varchanis, Y.~Dimakopoulos,
  J.~Tsamopoulos, A unification of least-squares and {Green--Gauss} gradients
  under a common projection-based gradient reconstruction framework,
  Mathematics and Computers in Simulation 205 (2023) 108--141.
\newblock \href {https://doi.org/10.1016/j.matcom.2022.09.008}
  {\path{doi:10.1016/j.matcom.2022.09.008}}.

\bibitem{Nishikawa2019}
H.~Nishikawa, Efficient gradient stencils for robust implicit finite-volume
  solver convergence on distorted grids, Journal of Computational Physics 386
  (2019) 486--501.
\newblock \href {https://doi.org/10.1016/j.jcp.2019.02.026}
  {\path{doi:10.1016/j.jcp.2019.02.026}}.

\end{thebibliography}
	
	\appendix
	\renewcommand{\thesection}{Appendix~\Alph{section}}       
	\section{CICSAM}
	\label{app:cicsam}
	The Compressive Interface Capturing Scheme for Arbitrary Meshes (CICSAM), developed by \cite{ubbink_method_1999}, is summarised here. Within the finite-volume framework, CICSAM integrates the transport equation over a control volume $P$ and a time interval $\Delta t$ (from time $t^n$ to $t^{n+1}$):
	\begin{equation}
		V_P\big( \alpha_P^{n+1} - \alpha_P^n \big)
		+ \frac{\Delta t}{2}\sum_{f} \Big[ \alpha_f\,(\mathbf{u}_f \cdot \mathbf{A}_f) \Big]^n + \Big[ \alpha_f\,(\mathbf{u}_f \cdot \mathbf{A}_f) \Big]^{n+1} = 0,
	\end{equation}
	where $V_P$ is the volume of cell $P$, $\mathbf{A}_f$ is the outward-pointing area vector of face $f$, and the summation is over all cell faces. Assuming the fluxes vary only weakly over $\Delta t$, we simplify the expression to obtain:
	\begin{equation}
		\alpha_P^{n+1} = \alpha_P^n - \frac{\Delta t}{V_P}\sum_{f} \alpha_f^*\,(\mathbf{u}_f \cdot \mathbf{A}_f),
	\end{equation}
	where the time-averaged face value is defined as:
	\begin{equation}
		\alpha_f^* = \frac{1}{2}\Big( \alpha_f^n + \alpha_f^{\,n+1} \Big).
	\end{equation}
	The key task is choosing $\alpha_f$ to keep the interface sharp and the solution bounded. For each face, the donor cell $D$ is the cell from which fluid is convected, the acceptor cell $A$ receives the fluid, and the upwind cell $U$ lies immediately upstream of $D$.
	
	To ensure boundedness, the normalised variable is introduced:
	\begin{equation}
		\tilde{\alpha} = \frac{\alpha - \alpha_U}{\alpha_A - \alpha_U},
	\end{equation}
	so that $\tilde{\alpha} = 0$ at the upwind cell and $\tilde{\alpha} = 1$ at the acceptor. Standard choices such as upwind differencing ($\alpha_f = \alpha_D$) are too diffusive, while downwind differencing ($\alpha_f = \alpha_A$) can violate the bounds.
	
	The convection boundedness criterion (CBC) requires that if $\tilde{\alpha}_D \in [0,1]$, then $\tilde{\alpha}_f$ must also lie in $[0,1]$. The most compressive bounded scheme is given by
	\begin{equation}
		\tilde{\alpha}_{f,\text{CBC-max}} =
		\begin{cases}
			\min\!\left\{\frac{\tilde{\alpha}_D}{\mathrm{Co}},\,1\right\}, & 0 \le \tilde{\alpha}_D \le 1,\\[1mm]
			\tilde{\alpha}_D, & \text{otherwise},
		\end{cases}
		\label{eq:cbc_max}
	\end{equation}
	where $\mathrm{Co}$ is the local Courant number.
	
	CICSAM blends two schemes depending on the local interface orientation: a compressive scheme (Hyper-C) for interfaces normal to the flow, and a higher-order scheme (ULTIMATE-QUICKEST) for interfaces nearly tangential to the flow.
	
	The \textbf{Hyper-C} scheme simply adopts the CBC upper bound (Eq.~\eqref{eq:cbc_max}), i.e.\ $\tilde{\alpha}_{f,\text{Hyper-C}} = \tilde{\alpha}_{f,\text{CBC-max}}$; this is the sharpest possible bounded interface representation.
	
	The \textbf{ULTIMATE-QUICKEST} scheme \cite{leonard_ultimate_1991} provides third-order accuracy in smooth regions:
	\begin{equation}
		\tilde{\alpha}_{f,\text{UQ}} =
		\begin{cases}
			\min\!\left\{\dfrac{8\,\mathrm{Co}\,\tilde{\alpha}_D + (1-\mathrm{Co})(6\tilde{\alpha}_D + 3)}{8},\,\tilde{\alpha}_{f,\text{CBC-max}}\right\}, & 0 \le \tilde{\alpha}_D \le 1,\\[2mm]
			\tilde{\alpha}_D, & \text{otherwise},
		\end{cases}
		\label{eq:uq}
	\end{equation}
	where $\tilde{\alpha}_{f,\text{CBC-max}}$ is the CBC upper bound from Eq.~\eqref{eq:cbc_max}. This scheme reduces numerical diffusion when the interface is aligned with the flow direction.
	
	The blending is performed in the normalised variable space:
	\begin{equation}
		\tilde{\alpha}_f = \gamma_f\,\tilde{\alpha}_{f,\text{Hyper-C}} + (1-\gamma_f)\,\tilde{\alpha}_{f,\text{UQ}},
	\end{equation}
	The weighting factor $\gamma_f$ depends on the angle $\theta_f$ between the interface normal $(\nabla \alpha)_D$ in the donor cell and the vector $\mathbf{d}_f$ connecting the donor and acceptor cell centres:
	\begin{equation}
		\gamma_f = \min\!\left\{k_\gamma\,\frac{\cos(2\theta_f) + 1}{2},\,1\right\},
	\end{equation}
	with the recommended parameter $k_\gamma=1$. When $\theta_f=0^\circ$ (interface normal aligned with the flow), $\gamma_f=1$, so the compressive scheme is fully applied; when $\theta_f=90^\circ$, $\gamma_f=0$, and the higher-order scheme dominates.
	
	The actual face value is recovered by inverting the normalisation:
	\begin{equation}
		\alpha_f = (1-\beta_f)\,\alpha_D + \beta_f\,\alpha_A,
	\end{equation}
	where the blending weight $\beta_f$ is
	\begin{equation}
		\beta_f = \frac{\tilde{\alpha}_f - \tilde{\alpha}_D}{\,1-\tilde{\alpha}_D\,}.
	\end{equation}
	This formulation guarantees that $\alpha_f$ remains bounded between the donor and acceptor values, and preserves the sharp interface.
	
	On unstructured meshes the upwind cell $U$ is not directly available, so CICSAM estimates the upwind value by extrapolation from the donor cell gradient $(\nabla \alpha)_D$. With $\mathbf{d}$ denoting the vector from the donor to the acceptor cell centre, the upwind value is approximated by
	\begin{equation}
		\alpha_U^* = \alpha_A - 2\,(\nabla \alpha)_D \cdot \mathbf{d}.
	\end{equation}
	The extrapolated value $\alpha_U^*$ is clamped to $[0,1]$. If the updated $\alpha$ slightly exceeds these bounds, a predictor--corrector step adjusts the flux weights to restore boundedness without sacrificing conservation.
	
	\section{Least Squares Gradient Reconstruction}
	\label{app:LeastSQGR}
	Gradient reconstruction on unstructured meshes is essential for computing diffusive fluxes and higher-order convective schemes. The least squares method, described in detail by \cite{ferziger_computational_2020}, provides a robust approach that is less sensitive to mesh irregularity than the simpler Green-Gauss method.
	
	Given a scalar field $\phi$ defined at cell centres, we estimate the gradient $\nabla \phi_P = (g_x, g_y, g_z)$ in each cell $P$ by enforcing a least squares fit to the known values in neighbouring cells. For a neighbour cell $Q$, define:
	\[
	\Delta \phi_{PQ} \;=\; \phi_Q \;-\; \phi_P,
	\quad
	\mathbf{r}_{PQ} \;=\; \mathbf{x}_Q - \mathbf{x}_P \;=\; (\,x_Q - x_P,\; y_Q - y_P,\; z_Q - z_P\,).
	\]
	We assume a linear variation of $\phi$ around cell $P$:
	\begin{equation}
		\phi_Q - \phi_P \;\approx\; \nabla \phi_P \,\cdot\, \mathbf{r}_{PQ}.
	\end{equation}
	To find the gradient that best fits all neighbours in the least squares sense, we minimise the sum of squared residuals:
	\[
	\sum_Q
	\Bigl( \Delta \phi_{PQ} \;-\; \nabla \phi_P \,\cdot\, \mathbf{r}_{PQ} \Bigr)^2
	\;\longrightarrow\; \min_{\nabla \phi_P}.
	\]
	
	\paragraph{Weighting.} In the general formulation, the weights $w_Q$ can be chosen as inverse distance ($1/|\mathbf{r}_{PQ}|$) or inverse distance squared ($1/|\mathbf{r}_{PQ}|^2$) to emphasise nearby cells on non-uniform meshes. In the present implementation, uniform weights ($w_Q = 1$) are used. Note that even with uniform weights, the outer product $\mathbf{r}_{PQ}\,\mathbf{r}_{PQ}^T$ naturally scales with $|\mathbf{r}_{PQ}|^2$, so nearby cells with shorter connection vectors inherently contribute smaller entries to the matrix $\mathbf{G}_P$.
	
	\paragraph{Interface modification.} For cells adjacent to the liquid--vapour interface, the distance vectors $\mathbf{r}_{PQ}$ are modified to account for the discontinuous nature of temperature and other fields across the interface. When a face $s$ between cells $P$ and $Q$ is crossed by the interface---detected by the sign change $(\alpha_P - 0.5)(\alpha_Q - 0.5) < 0$---the interface position $\mathbf{x}_f$ is estimated by linear interpolation of the volume fraction along the cell-centre connection (Eq.~\eqref{eq:front_interpolation}). The displacement vector for cell $P$ is then replaced by
	\begin{equation}
		\mathbf{r}_{Pf} \;=\; \mathbf{x}_f - \mathbf{x}_P,
		\label{eq:front_displacement}
	\end{equation}
	and the scalar value $\phi_Q$ is replaced by the saturation value $\phi_\mathrm{sat}$ (e.g.\ $T_\mathrm{sat}$ for the temperature field). Symmetrically, the neighbouring cell $Q$ uses $\mathbf{r}_{Qf} = \mathbf{x}_Q - \mathbf{x}_f$. Because $|\mathbf{r}_{Pf}| < |\mathbf{r}_{PQ}|$, the contribution of this face to the matrix $\mathbf{G}_P$ is reduced. On structured meshes, the degree of shortening depends on the angle between the face normal and the interface normal, which varies systematically with the cell's angular position relative to the bubble (see Section~\ref{sec:gradient_anisotropy}). On polyhedral meshes, the irregular face orientations distribute the shortening more uniformly, so that no preferential alignment develops.
	
	\paragraph{Solution procedure.} Setting the derivative of the objective function to zero gives the \emph{normal equations}:
	\[
	\sum_Q \mathbf{r}_{PQ}\,\mathbf{r}_{PQ}^T\;\nabla \phi_P
	\;=\;\sum_Q \mathbf{r}_{PQ}\,(\phi_Q - \phi_P).
	\]
	Defining the $3 \times 3$ matrix and right-hand side vector:
	\begin{equation}
		\mathbf{G}_{P}
		\;=\; \sum_Q \mathbf{r}_{PQ}\,\mathbf{r}_{PQ}^T,
		\quad
		\mathbf{b}_{P}
		\;=\; \sum_Q \mathbf{r}_{PQ}\, \Delta \phi_{PQ},
	\end{equation}
	we obtain the linear system:
	\begin{equation}
		\mathbf{G}_{P}\,\nabla \phi_P \;=\; \mathbf{b}_{P}.
	\end{equation}
	The gradient is found by direct inversion of the symmetric positive-definite $3 \times 3$ matrix:
	\begin{equation}
		\nabla \phi_P \;=\; \mathbf{G}_{P}^{-1}\,\mathbf{b}_{P}.
	\end{equation}
	Since $\mathbf{G}_P$ is only $3 \times 3$, direct inversion using Cramer's rule or explicit formulae is computationally efficient. For cells away from the interface, $\mathbf{G}_P$ depends only on the mesh geometry, so $\mathbf{G}_P^{-1}$ can be precomputed and stored during initialisation. For interface-adjacent cells, the interface modification (Eq.~\eqref{eq:front_displacement}) changes the displacement vectors at every time step as the interface moves, and $\mathbf{G}_P^{-1}$ must be recomputed accordingly. In the implementation, the six independent components of this symmetric inverse matrix are stored per cell, and the gradient is obtained at each time step by accumulating $\mathbf{G}_P^{-1}\,\mathbf{r}_{PQ}\,\Delta\phi_{PQ}$ over all face neighbours.
	
	\paragraph{Neighbour stencil.} The summation over $Q$ includes all cells sharing a face with cell $P$ (face neighbours). For improved accuracy on highly skewed meshes, the stencil can be extended to include vertex neighbours, though this increases computational cost and is not used in the present work.
	
	\paragraph{Boundary cells.} At domain boundaries, the neighbour stencil is incomplete. Symmetry boundaries are excluded from the stencil entirely. For all other boundary types, the boundary face value is included as a virtual neighbour located at the face centroid. For pressure gradients, an iterative procedure is used: the pressure at boundary faces is extrapolated from the interior using the current gradient estimate, and the gradient is then recomputed. This process is repeated for a prescribed number of iterations (four by default) to ensure consistent boundary values.
	
	\section{Gradient magnitude overestimate on hexahedral meshes}
	\label{app:gradient_magnitude}

\begin{figure}[htbp]
		\centering
		\includegraphics[width=\textwidth]{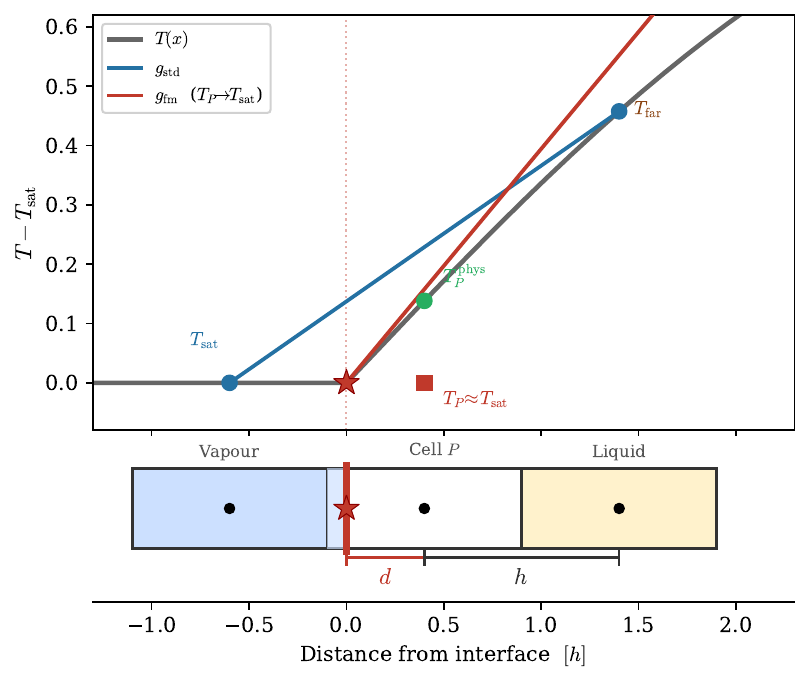}
		\caption{One-dimensional schematic of the gradient stencil for a cell~$P$ adjacent to the vapour--liquid interface. (a)~Standard gradient: symmetric central difference with equal spacing~$h$ on both sides; the cell-centre temperature~$T_P$ cancels. (b)~Interface-modified gradient: the vapour-side displacement vector is shortened to the interface intersection distance~$d < h$; when the mass transfer heat sink maintains $T_P \approx T_\mathrm{sat}$, the near-side contribution vanishes and the gradient is determined by the far neighbour alone.}
		\label{fig:gradient_schematic}
	\end{figure}

	As demonstrated in Section~\ref{sec:gradient_anisotropy}, the interface-modified least-squares gradient is essential for computing temperature gradients at the interface: without it, the standard gradient underestimates the bubble growth rate by $16\%$ (Figure~\ref{fig:scriven_sensitivity}(d)). However, on structured hexahedral meshes, the interplay between the modified stencil's inherent asymmetry and the temperature field produced by the mass transfer heat sink leads to a systematic overestimate of the mean gradient magnitude that drives the radius overshoot $R > R_\mathrm{Scriven}$. This appendix derives the mechanism responsible for this side effect. Figure~\ref{fig:gradient_schematic} illustrates the two stencil configurations analysed below.

	To assess whether the amplified gradient exceeds the true physical value at the interface, we model the local liquid temperature profile as
	\begin{equation}
		T(x) = T_\mathrm{sat} + \Delta T \, \operatorname{erf}\!\left(\frac{x}{\delta_T}\right),
		\label{eq:erf_profile}
	\end{equation}
	where $x$ is the distance from the interface into the liquid, $\Delta T = T_\infty - T_\mathrm{sat}$, and $\delta_T$ is the local thermal penetration depth. The true gradient at the interface is
	\begin{equation}
		g_\mathrm{true} = \frac{\mathrm{d}T}{\mathrm{d}x}\bigg|_{x=0} = \frac{2\,\Delta T}{\sqrt{\pi}\,\delta_T}.
		\label{eq:g_true}
	\end{equation}
	For the Scriven problem, matching this expression to the analytical interface gradient gives $\delta_T \approx 10$--$17~\mu$m over $t = 0.5$--$1.5$~ms---approximately half the pure-diffusion estimate $2\sqrt{\alpha_l t}$ because radial outflow compresses the boundary layer. On the $75^3$ mesh ($h = 4~\mu$m), the boundary layer spans only $2.5$--$4$ cell widths.

	Syrakos~et~al.~\cite{Syrakos2017} showed that the leading error term of the least-squares gradient vanishes when stencil points are symmetric about the cell centre. When this symmetry is broken, an uncancelled first-order term proportional to $\phi''$ appears. For a one-dimensional unweighted LSQ stencil with neighbours at $-h$ and $+\lambda h$ ($0 < \lambda < 1$), the truncation error is~\cite{Syrakos2017,Syrakos2023}
	\begin{equation}
		\epsilon_\mathrm{LSQ} = -\frac{h(1 - \lambda^3)}{2(1 + \lambda^2)}\,\phi''.
		\label{eq:lsq_truncation}
	\end{equation}
	When $\phi'' < 0$, i.e.\ when the temperature profile curves downward as it levels off toward the superheated bulk, this error is positive and the gradient is overestimated. This is precisely the shape of the temperature field near an evaporating interface, where $T$ rises steeply from $T_\mathrm{sat}$ and gradually flattens. The interface modification shortens the displacement toward the interface from $h$ to $d < h$ ($\lambda = d/h < 1$), which breaks the stencil symmetry and activates this error term. On a hexahedral mesh, each coordinate direction has only two stencil points, so there is no redundancy to mitigate the asymmetry~\cite{Nishikawa2019}.

	To expose this algebraically, consider the radial gradient component along a single coordinate axis (the hexahedral gradient matrix is diagonal, so each component is independent; in three dimensions, geometric projection effects reduce the predicted overestimation by ${\sim}7\%$ on average). With the standard stencil, the inner neighbour lies in the vapour at~$T_\mathrm{sat}$ and the outer at $T_\mathrm{far}$. Because both displacement vectors have equal magnitude~$h$, the cell-centre temperature cancels:
	\begin{equation}
		g_\mathrm{std} = \frac{T_\mathrm{far} - T_\mathrm{sat}}{2h}.
		\label{eq:g_std}
	\end{equation}
	This expression is independent of~$T_P$. With the interface modification, the vapour-side displacement vector is shortened from~$h$ to the distance $d < h$ from the cell centre to the interface intersection (Eq.~\eqref{eq:front_displacement}), and the scalar value is replaced by~$T_\mathrm{sat}$. The gradient becomes
	\begin{equation}
		g_\mathrm{fm} = \frac{(T_P - T_\mathrm{sat})\,d + (T_\mathrm{far} - T_P)\,h}{d^2 + h^2},
		\label{eq:g_fm}
	\end{equation}
	which now depends on~$T_P$. Setting $\theta_P = T_P - T_\mathrm{sat}$ and $\Delta T = T_\mathrm{far} - T_\mathrm{sat}$, the ratio of the two gradients is
	\begin{equation}
		\frac{g_\mathrm{fm}}{g_\mathrm{std}} = \frac{2h\bigl[\theta_P\,d + (\Delta T - \theta_P)\,h\bigr]}{\Delta T\,(d^2 + h^2)}.
		\label{eq:boost}
	\end{equation}

	The mass transfer acts as a heat sink at the interface, continuously extracting latent heat from the liquid cells adjacent to the interface and maintaining $T_P$ close to~$T_\mathrm{sat}$ ($\theta_P \ll \Delta T$). In the limit $\theta_P \to 0$, the near-side contribution vanishes (because $T_P \approx T_\mathrm{sat}$ and the interface value is also $T_\mathrm{sat}$), and the gradient is determined by the far neighbour alone:
	\begin{equation}
		\frac{g_\mathrm{fm}}{g_\mathrm{std}} \xrightarrow{\theta_P \to 0} \frac{2h^2}{d^2 + h^2}.
		\label{eq:boost_limit}
	\end{equation}
	For $d/h = 0.3$--$0.7$, this yields $g_\mathrm{fm}/g_\mathrm{std} = 1.34$--$1.83$; the limit $g_\mathrm{fm}/g_\mathrm{std} \to 2$ as $d \to 0$ corresponds to a pure one-sided difference over~$h$ rather than the symmetric difference over~$2h$.

	The erf profile (Eq.~\eqref{eq:erf_profile}) allows a direct comparison with the true interface gradient. For the $75^3$ mesh ($h = 4~\mu$m, $\delta_T \approx 10$--$17~\mu$m), the standard gradient gives $g_\mathrm{std}/g_\mathrm{true} \approx 0.68$--$0.69$: the central difference underestimates by ${\sim}30\%$ because the boundary layer spans only a few cells. With the analytical $T_P$ from the erf profile, the interface-modified gradient gives $g_\mathrm{fm}/g_\mathrm{true} \approx 0.93$--$0.95$: the stencil amplification compensates for most of the resolution deficit. However, in the coupled simulation, the mass transfer heat sink maintains $T_P \approx T_\mathrm{sat}$, and the model predicts $g_\mathrm{fm}/g_\mathrm{true} \approx 1.14$--$1.16$: the gradient exceeds the true value by $14$--$16\%$, which drives the radius overshoot. Physical overestimation thus requires both the stencil asymmetry and the heat sink feedback that suppresses the near-side contribution.

	The ablation test (Figure~\ref{fig:scriven_sensitivity}(d)) confirms this: the standard gradient gives $R/R_\mathrm{Scriven} \approx 0.84$, while the interface-modified gradient gives $R/R_\mathrm{Scriven} \approx 1.04$. On polyhedral meshes, the interface modification affects only 2--3 of the ${\sim}14$ faces per cell, so the amplification ($g_\mathrm{fm}/g_\mathrm{std} \approx 1.15$--$1.2$) is insufficient to overcome the ${\sim}30\%$ baseline underestimate, and no polyhedral mesh produces $R > R_\mathrm{Scriven}$.

	\paragraph{Behaviour under mesh refinement.}
	One might expect the overestimate to vanish on finer meshes, since the erf profile is better resolved and the standard gradient approaches $g_\mathrm{true}$. However, the boost factor $2h^2/(d^2 + h^2)$ depends only on the ratio $d/h$, which is approximately uniformly distributed within cells regardless of mesh size, so the amplification is essentially mesh-independent. At the same time, mesh refinement brings the cell centre closer to the interface: the physical cell-centre temperature from the erf profile is $T_P = T_\mathrm{sat} + \Delta T\,\mathrm{erf}(d/\delta_T)$, and since $d$ decreases proportionally to~$h$ while $\delta_T$ is a physical quantity independent of the mesh, $T_P$ naturally approaches $T_\mathrm{sat}$ on finer meshes. This means the near-side contribution $(T_P - T_\mathrm{sat})\,d$, which carries the local gradient information near the interface, becomes vanishingly small. The heat sink therefore needs to extract progressively less energy to push $T_P$ the remaining distance to~$T_\mathrm{sat}$, which makes the heat-sink limit ($\theta_P \to 0$) increasingly representative of the actual simulation state. Consequently, the gradient overestimate does not diminish with refinement: the boost factor remains constant while the physical temperature difference that would constrain it shrinks, so the overcompensation persists or even grows on finer meshes.

	\section{Computational cost}
	\label{app:computational_cost}

	Table~\ref{tab:scriven_cost} reports the profiled wall-clock time breakdown for the Scriven problem on the structured $75^3$ mesh (1800 time steps, 8 MPI processes, $\Delta t = 1~\mu$s). The pressure solver dominates ($24.5\%$) because the pressure Poisson equation requires the most iterations to converge. The energy equation ($15.6\%$) is the second largest cost, reflecting the interface-modified gradient computation at the interface. Preconditioning across all solvers accounts for $20.0\%$. The interface reconstruction (\texttt{isoap} front creation) and VOF advection together require less than $10\%$, confirming that the geometric interface reconstruction imposes a modest overhead relative to the overall solver cost.

	\begin{table}[htbp]
		\centering
		\caption{Computational cost breakdown for the Scriven problem on the structured $75^3$ mesh (421\,875 cells, 1800 time steps, 8 MPI processes, $\Delta t = 1~\mu$s). Percentages indicate the fraction of total wall-clock time (3.1~h).}
		\label{tab:scriven_cost}
		\begin{tabular}{lrc}
			\hline
			\textbf{Component} & \textbf{Time [s]} & \textbf{Fraction} \\
			\hline
			Pressure solver (CG)            & 2772 & 24.5\% \\
			Preconditioner (all solvers)     & 2264 & 20.0\% \\
			Energy equation                  & 1759 & 15.6\% \\
			Interface reconstruction (\texttt{isoap}) & 726 & 6.4\% \\
			Momentum solver (BiCG)           &  457 &  4.0\% \\
			Momentum equation                &  443 &  3.9\% \\
			Pressure gradient (Gauss)        &  415 &  3.7\% \\
			VOF advection (CICSAM)           &  354 &  3.1\% \\
			Advection term                   &  335 &  3.0\% \\
			Face interpolation               &  323 &  2.9\% \\
			Other (gradients, I/O, Rhie--Chow, \ldots) & 1462 & 12.9\% \\
			\hline
			\textbf{Total}                   & \textbf{11\,310} & \textbf{100\%} \\
			\hline
		\end{tabular}
	\end{table}

	Comparing mesh topologies, the polyhedral $75^3$ mesh (319\,156 cells) requires approximately $8.7$~s per time step versus $6.3$~s for the structured mesh (421\,875 cells), a ratio of $1.4$. Per cell, the polyhedral mesh is approximately $1.8$ times more expensive, consistent with the higher number of faces per cell (${\sim}12$--$14$ vs.\ $6$ for hexahedra), which increases the cost of face interpolation, gradient computation, and sparse matrix operations.

\end{document}